\documentclass[showpacs,aps,twocolumn]{revtex4-1}
\usepackage{epsfig}
\usepackage{graphicx}
\usepackage{amsmath,amssymb,amsfonts}
\usepackage{array}
\usepackage{url}
\usepackage{hyperref}
\usepackage{multirow}
\usepackage{float}
\usepackage{lineno}
\usepackage{xspace}
\usepackage{ulem}
\usepackage[usenames,dvipsnames]{color}

\begin{document}
\title{Impact of vorticity and viscosity on the hydrodynamic evolution of hot QCD medium}
 
\author{Bhagyarathi Sahoo}
\email{Bhagyarathi.Sahoo@cern.ch}

\author{Captain R. Singh}
\email{captainriturajsingh@gmail.com}

\author{Dushmanta Sahu}
\email{Dushmanta.Sahu@cern.ch}

\author{Raghunath Sahoo} 
\email{Corresponding Author: Raghunath.Sahoo@cern.ch}

\affiliation{Department of Physics, Indian Institute of Technology Indore, Simrol, Indore 453552, India}

\author{Jan-e Alam}
\email{jane@vecc.ac.in}
\affiliation{Variable Energy Cyclotron Centre, 1/AF, Bidhan Nagar, Kolkata, India}

\begin{abstract}

The strongly interacting transient state of quark-gluon plasma (QGP) medium created in ultra-relativistic collisions 
survives for a duration of a few fm/c. The spacetime evolution of QGP crucially depends on the equation of state (EoS),
vorticity, viscosity, and external magnetic field. In the present study, we obtain the lifetime of a vortical QGP
fluid within the ambit of relativistic second-order viscous hydrodynamics. We observe that the coupling of vorticity and 
viscosity significantly increases the lifetime of vortical QGP. The inclusion of a static magnetic field, vorticity, and viscosity makes the 
evolution slower. However, the static magnetic field slightly decreases the QGP lifetime by accelerating 
the evolution process for a non-rotating medium. We also report the rate of change of vorticity in the QGP, which will 
be helpful in studying the behavior of the medium in detail.

\end{abstract}
\date{\today}
\maketitle

\section{Introduction}
\label{intro}

It is reasonable to expect that angular momentum deposition in heavy-ion collisions can trigger a 
local vortical motion in the overlap region of the colliding species. 
The initial angular momentum ($L_0$) generated in a heavy-ion collision is directly proportional to the impact parameter ($b$)
of the collision and the center of mass energy ($\sqrt{s}$) as  $L_{0} \propto b\sqrt{s}$~\cite{Becattini:2007sr}. A fraction of the initial angular momentum 
is then transferred to the particles that are produced in the collisions. This can manifest as shear along the longitudinal momentum direction, 
creating vorticity in the system. The ultra-high magnetic field produced by the charged spectators in non-central heavy 
ion collisions can also generate vorticity. This generated vorticity, in turn, can affect the evolution of the hot and dense medium. 
From the global $\Lambda$ hyperon polarization measurement at Relativistic Heavy Ion Collider (RHIC), 
it has been estimated that a large vorticity ($\omega = (9\pm 1)\times 10^{21} \rm sec^{-1}$) is generated in 
the system produced in 
heavy-ion collisions~\cite{Adamczyk:2017}. This makes QGP the most vortical fluid found in nature so far.\\

There are several sources of vorticity besides the one mentioned above. One such example is the vorticity generated from the jet-like fluctuations in the fireball, which induces a smoke-loop type vortex around a fast-moving particle~\cite{Betz:2007kg}. This vorticity, however, does not contribute to global hyperon polarization. Another source of vorticity is the inhomogeneous expansion of the fireball. Due to the anisotropic flows in the transverse plane, a quadrupole pattern of the longitudinal vorticity along the beam direction is produced~\cite{Xia:2018tes,Jiang:2016woz, Wei:2018zfb,Becattini:2017gcx, Pang:2016igs, Voloshin:2017kqp}. On the other hand, the inhomogeneous transverse expansion produces transverse vorticity that circles the longitudinal axis. In addition, another source of vorticity can be due to the Einstein-de Haas effect~\cite{Einstein:1915}, where a strong magnetic field created by the fast-moving spectators magnetizes the QCD matter, and due to the magnetization, a rotation is induced. This leads to the generation of vorticity along the direction of the magnetic field. This effect is opposite to the Barnett effect, where a chargeless rotating system creates a non-zero magnetization \cite{Barnett}.\\

Vorticity formation in the ultra-relativistic heavy-ion collision has been studied by using hydrodynamic models such as ECHO-QGP, PICR, vHLLE, MUSIC, 3-FD, CLVisc in (3+1) dimensional model~\cite{ Becattini:2015ska,Csernai:2013bqa, Csernai:2014ywa, Ivanov:2019ern, Karpenko:2016jyx}. Event generators, such as AMPT, UrQMD, and HIJING, have also been used to estimate kinematic and thermal vorticity~\cite{Jiang:2016woz, Deng:2016gyh, Li:2017slc, Wei:2018zfb, Deng:2020ygd, Vitiuk:2019rfv}. Moreover, the non-zero local vorticity can help us to probe the chiral vortical effect (CVE), which is a non-trivial consequence of topological quantum chromodynamics~\cite{Rogachevsky:2010ys, Kharzeev:2007tn}. This effect is the vortical analog of the chiral magnetic effect (CME)~\cite{Kharzeev:2007jp, Fukushima:2008xe} and chiral separation effect (CSE) \cite{Son:2004tq, Metlitski:2005pr}. It represents the vector and axial currents generation along the vorticity ~\cite{Banerjee:2008th, Erdmenger:2008rm, Son:2009tf, Jiang:2015cva}. CVE is extremely important because it induces baryon charge separation along the direction of vorticity, which can be experimentally probed by two-particle correlations~\cite{Csernai:2013vda}.\\

Relativistic hydrodynamics govern the evolution of matter produced in ultra-relativistic collisions. 
Thus, relativistic hydrodynamics models with finite viscous correction become very useful in understanding the  
spacetime evolution of the system created in such collisions. From the AdS/CFT correspondence, the lower limit of shear viscosity ($\eta$) 
to entropy density($s$) ratio has been predicted, which is known as the KSS bound, given by $\eta/s \simeq 1/{4\pi}$~\cite{Kovtun:2004de}. 
Hydrodynamic models with $\eta/s \simeq 0.2$ explain the elliptic flow results from the RHIC experiments very well~\cite{Adam:2005}. 
Moreover, as observed in some recent studies~\cite{Fu:2021pok}, viscosity can generate some finite vorticity in the medium, 
even if initial vorticity is absent a priori. This makes the evolution dynamics of the viscous medium fascinating. \\

In the non-relativistic domain, the vorticity is defined as the curl of the velocity ($\omega$) field of the fluid as,
 \begin{align}
 \nonumber
     \vec{\omega} = \frac{1}{2} \vec{\nabla} \times \vec{v} 
\end{align}
Since high energy heavy-ion collision is a relativistic system, the generalized form of  vorticity  which is mostly used in the relativistic domain is 
thermal vorticity, which is defined as,

\begin{align}
\nonumber
    \omega_{\mu\nu} = -\frac{1}{2}\left(\partial_{\mu}\beta_{\nu}-\partial_{\nu}\beta_{\mu} \right) 
\end{align}
where $\beta_{\mu} = \frac{u_{\mu}}{T}$, with $u_{\mu}$ being the four-velocity of the fluid and $T$ is the temperature. 
Apart from thermal vorticity, there are several other kinds of vorticity; such as kinematic vorticity, temperature vorticity, 
and enthalpy vorticity in relativistic hydrodynamics, which have various implications as discussed in Ref.~\cite{Huang:2020dtn, Becattini:2015ska}. \\

In ref~\cite{Singh:2018bih}, the authors have used an ideal equation of state and estimated the 
time evolution of non-relativistic vorticity. They show that vorticity decreases as the system evolves with time. 
As mentioned earlier, the finite viscosity and vorticity of a rotational viscous fluid 
originate from several sources. 
In the present work, we study the evolution of QGP using second-order viscous hydrodynamics in the presence of vorticity. 
The effect of static magnetic field on evolution has also been 
included here. We obtain a set of coupled differential equations describing the evolution of the system. 
These coupled equations together describe the time evolution of temperature, viscosity, and vorticity.\\

This paper is organized as follows. In section~\Ref{formulation}, we briefly discuss the effects of viscosity and vorticity on the temperature through a set of 
non-linear coupled differential equations. In section~\Ref{res}, we discuss the results obtained from hydrodynamic equations, which describe the evolution of
temperature, viscosity, and vorticity and how much it is sensitive to initial hydrodynamic conditions. Finally, we summarize the essential findings in section~\Ref{sum}.

\section{Evolution of the system }
\label{formulation}
We first discuss the temperature profile for a simple relativistic ideal fluid. Secondly, we discuss temperature and viscosity evolution with proper time for a second-order relativistic viscous fluid. The following subsection discusses the evolution of temperature, viscosity, and vorticity for a relativistic rotational viscous fluid. Finally, we discuss the temperature, viscosity, and vorticity evolution of a rotating viscous fluid in a static magnetic field.

\subsection{Ideal fluid}
\label{ideal}
For an ideal fluid, the energy-momentum tensor ($T^{\mu\nu}$) does not contain a gradient of the hydrodynamic fields. 
This is called a $0^{th}$ order hydrodynamic model. 
The energy-momentum tensor for relativistic ideal hydrodynamics is,
\begin{equation}
\label{eq1}
     T^{\mu\nu}_{Ideal} = (\epsilon+P)u^{\mu}u^{\nu} -g^{\mu\nu}P 
\end{equation}
where
$\epsilon $ is energy density, $P$ is pressure, $ u^{\mu} =\gamma(1, \vec{v})$ is the four-velocity vector, with $\gamma = \frac{1}{\sqrt{1-\vec{v}^{2}}}$ being the Lorentz factor, and $g^{\mu\nu} =diag(1,-1,-1,-1)$ is the metric 
tensor.
The conservation of energy-momentum (in absences of external field) is given by,
\begin{equation}
\label{eq2}
\partial_{\mu}{T^{\mu\nu}} = 0
\end{equation}
Projecting Eq.~(\ref{eq2}) in the direction parallel to the fluid velocity, we get;

\begin{equation}
\label{eq3}
    u_{\nu}\partial_{\mu}{T^{\mu\nu}} = 0
\end{equation}

Simplification of Eq.~(\ref{eq3}) leads to,

\begin{equation}
\label{eq4}
    \partial_{\mu}({(\epsilon + P) u^{\mu}}) =  u_{\nu} g^{\mu \nu} \partial_{\mu} P
\end{equation}

Using the relation $D\equiv u^{\mu}\partial_{\mu}=\gamma \frac{d}{d\tau} $ in Eq.~(\ref{eq4}), 
produces the dissipation rate for energy density, 

\begin{equation}
\label{eq5}
    \frac{d \epsilon} {d \tau} =  - \frac{1}{\gamma}(\epsilon + P)\; \partial_{\mu} u^{\mu}
\end{equation}

For this study, we use a simple equation of state (EoS) describing an ideal plasma of massless u, d, s quarks, and gluons. The pressure is given by  $P = \epsilon/3= aT^{4} $ with zero baryon chemical potential, where $a$ is the constant, defined as ~\cite{Muronga:2001zk, Muronga:2003ta}

\begin{equation}
  a = \frac{\pi^{2}}{90} \left[ 16 + \frac{21}{2}N_{f} \right] \nonumber
 \end{equation}
where $N_{f} = 3$, is the number of flavours.
Using the EoS mentioned above, the equation governing the  cooling rate can be obtained as,

\begin{equation}
\label{eq6}
    \frac{dT}{d\tau} = - \frac{T}{3\gamma}\; \partial_{\mu} u^{\mu}    
\end{equation}

Equation~(\ref{eq6}) represents the cooling rate in $0^{th}$ order hydrodynamics or for ideal fluid.

\subsection{Viscous fluid}
\label{viscosity}
The viscosity in a medium originates due to the velocity gradient between
fluid cells which slows down the flow. 
Therefore, considering QGP as a viscous fluid modifies the medium evolution.  The
dissipative term ($\Pi^{\mu\nu}$)  needs to be added to the  energy-momentum tensor ($T^{\mu\nu}_{Ideal}$) 
representing the ideal fluid such that the total energy-momentum tensor is given by:

\begin{equation}
\label{eq7}
     T^{\mu\nu} =  T^{\mu\nu}_{Ideal} + \Pi^{\mu\nu}
\end{equation}
where $\Pi^{\mu\nu}$ is the viscous stress tensor, expressed as,
\begin{equation}
     \Pi^{\mu\nu}  = \pi^{\mu\nu} + \Delta^{\mu\nu}\Pi \nonumber
\end{equation}
where $\Delta^{\mu\nu} = g^{\mu\nu} -u^{\mu}u^{\nu}$ is the projection operator,
such that $\Delta^{\mu\nu}u_\nu=0$. The $\Pi^{\mu\nu}$ contains two parts; $\pi^{\mu\nu}$ accounts for the shear viscosity, and $\Delta^{\mu\nu}\Pi$ accounts for the bulk viscosity. For conformal fluids, the bulk viscous pressure does not contribute ($\Pi = 0$) \cite{Weinberg:1972}. For first-order hydrodynamic theory, the $ \pi^{\mu\nu}$ has the form;

\begin{equation}
\label{eq8}
    \pi^{\mu\nu} = \eta \bigtriangledown^{<\mu}u^{\nu>}
\end{equation}
where $\eta$ is the shear viscosity, with:
\begin{equation}
 \bigtriangledown^{<\mu}u^{\nu>} \equiv 2\bigtriangledown^{(\mu}u^{\nu)} - \frac{2}{3}\Delta^{\mu\nu}\bigtriangledown^{\alpha} u_{\alpha} 
\nonumber
\end{equation} 
where $\bigtriangledown^{(\mu}u^{\nu)}$ is defined as $A^{(\mu}B^{\nu)} = \frac{1}{2} \left(A^{\mu}B^{\nu} + A^{\nu}B^{\mu} \right)$.

For second-order hydrodynamic theory, the $ T^{\mu\nu}$ contains both the first and second-order gradient of the hydrodynamic fields. In M\"{u}ller-Israel-Stewart (MIS) second-order theory, the $\pi^{\mu\nu}$ is given by ~\cite{Romatschke:2009im},

\begin{equation}
\label{eq9} 
\pi^{\mu\nu} = \eta \bigtriangledown^{<\mu}u^{\nu>} +  \tau_{\pi} \left[\Delta_{\alpha}^{\mu}\Delta_{\beta}^{\nu}D\pi^{\alpha\beta}....\right] + O(\delta^{2})
\end{equation}
where
$\tau_{\pi}$ is the relaxation time. 
Inclusion of the viscous term in energy density evolution changes  Eq.~(\ref{eq5}) to the following form\cite{Muronga:2001zk, Muronga:2003ta};

\begin{equation}
\label{eq10}
    \frac{d\epsilon}{d\tau} = - \frac{1}{\gamma}( \epsilon+P - \Phi )\; \partial_{\mu} u^{\mu}
\end{equation}

Here $\Phi = \pi^{00} - \pi^{zz}$ is the difference between temporal and spatial components of the shear viscosity tensor 
representing the viscous term. \\

For first-order theory, the viscous shear term $ \Phi  = \frac{4 \eta}{3 \tau}$. The second-order MIS relaxation equation using Grad's  14 moments methods for shear viscosity has the following form \cite{Muronga:2001zk, Muronga:2003ta};
 
 \begin{align}
     D\pi^{\mu\nu} = - \frac{1}{\tau_{\pi}}\pi^{\mu\nu}- \frac{1}{2\beta_{2}}\pi^{\mu\nu}\left[\beta_{2}\theta + TD\left(\frac{\beta_{2}}{T}\right)\right] \nonumber \\  + \frac{1}{\beta_{2}} \bigtriangledown^{<\mu}u^{\nu>}
      \label{eq11}
 \end{align}
where $\theta \equiv \partial_{\mu}u^{\mu}$ represents  the expansion 
of the system, $\tau_{\pi} = 2\eta\beta_{2}$ is the relaxation time, $\beta_{2}$ is the relaxation coefficient given as; $\beta_{2} = 3/4P$.  
Here we take shear viscosity $\eta = bT^{3}$, where $b$ is defined as;
 
\begin{equation}
 b = (1 + 1.70 N_{f})\frac{0.342}{(1 + N_{f}/6)\alpha_{s}^{2}\ln(\alpha_{s}^{-1})} \nonumber
\end{equation}
where  $\alpha_s = 0.5$, is the strong coupling.\\

Now, the evolution of shear viscosity can be obtained from the  Eq.~(\ref{eq11}) as a viscous shear tensor,
\begin{eqnarray}
    \frac{d\Phi}{d\tau} = - \frac{\Phi}{\gamma\tau_{\pi}} - \frac{\Phi}{2\gamma}\left( \partial_{\mu}u^{\mu} + \frac{\gamma}{\beta_{2}}T\frac{d}{d\tau}\left(\frac{\beta_{2}}{T}\right) \nonumber \right)\\
    +\frac{1}{\gamma\beta_{2}}\left(\bigtriangledown^{<0}u^{0>} - \bigtriangledown^{<z}u^{z>}\right)
    \label{eq12}
\end{eqnarray}

Using the equation of state, $ P = \epsilon/3= aT^{4} $ to the Eq.~(\ref{eq10}) and Eq.~(\ref{eq12}), we have, 

\begin{equation}
    \frac{dT}{d\tau} = -\frac{1}{\gamma} \bigg( \frac{T}{3} - \frac{T^{-3}\Phi}{12a} \bigg)\;\partial_{\mu}u^{\mu}
    \label{eq13}
\end{equation}
 
\begin{eqnarray}
    \frac{d\Phi}{d\tau} = - \frac{2aT\Phi}{3b\gamma} - \frac{\Phi}{2\gamma}\left(\partial_{\mu}u^{\mu} - \frac{5\gamma}{T}\frac{dT}{d\tau}\right) \nonumber \\ +\frac{4aT^{4}}{3\gamma}\left(\bigtriangledown^{<0}u^{0>} - \bigtriangledown^{<z}u^{z>}\right)
\label{eq14}
\end{eqnarray}

Thus, Eq.~(\ref{eq13}) and Eq.~(\ref{eq14}) represent the space-time evolution of temperature and viscous 
term ($\Phi$) with the proper time, 
which cumulatively affects the temperature evolution in the second-order theory. Furthermore, 
by putting $\Phi$ = 0 in Eq.~(\ref{eq13}) and Eq.~(\ref{eq14}), one gets the equation of motion for the ideal fluid.

\subsection{Rotational viscous fluid}
\label{vorticity}
Next, we consider a viscous medium with non-zero vorticity, which can
couple with the spin of the particles and  gives rise to spin polarization
in the system. Here spin polarization tensor is obtained using a tensor decomposition with the help of Ref.~\cite{Florkowski:2017ruc}. The antisymmetric spin polarization tensor is given as; 

\begin{equation}
 \omega_{\mu\nu} = k_{\mu}u_{\nu} - k_{\nu}u_{\mu} + \epsilon_{\mu\nu\alpha\beta}u^{\alpha}\omega^{\beta}
  \label{adeq1}
\end{equation}
where, $k_{\mu}$ and $\omega_{\mu}$ are defined in terms of spin polarization tensor;
\begin{equation}
  k_{\mu} = \omega_{\mu\nu}u^{\nu}, \;\;\;\;\;\;\;\;\; \omega_{\mu} = \frac{1}{2} \epsilon_{\mu\nu\alpha\beta}\omega^{\nu\alpha}u^{\beta}
    \label{adeq2}
\end{equation}

To hold the relation $k^{\mu}u_{\mu} = \omega^{\mu}u_{\mu} = 0$, the $\omega_{\mu}$ and  $k_{\mu}$ are set to orthogonal to the fluid velocity $u_{\mu}$.  Here $\epsilon_{\mu\nu\alpha\beta} $ is the Levi Civita antisymmetric four tensor, $\epsilon^{0123}= - \epsilon_{0123} = 1$. Considering the rotation in the $x$-$z$ plane, i.e. $\omega_{\mu} = (0,0,\omega,0)$, one needs to solve Eq.~(\ref{adeq1}) and Eq.~(\ref{adeq2}) self-consistently to obtain the spin polarization tensor, $\omega_{\mu\nu}$; 

\begin{equation}
  \omega_{\mu\nu} =
  \left[ {\begin{array}{cccc}
    0 & 0 & 0 & 0 \\
    0 & 0 & 0 & \frac{\omega}{T} \\
    0 & 0 & 0 & 0 \\
    0 & -\frac{\omega}{T} & 0 & 0 \\
  \end{array} } \right]
  \label{adeq3}
\end{equation}
For this work, we have considered the velocity profile  $ u^{\mu} =\gamma(1, v_{x}, 0, v_{z} )$. 
The velocity profile is chosen in such a way that the transverse component of velocity depends upon the 
longitudinal component and the longitudinal component of velocity  develops a transverse component. The chosen velocity profiles are~\cite{Singh:2018bih};

\begin{equation}
\label{eq15}
    v_{x} =  \frac{\omega z}{2}
\end{equation}

\begin{equation}
\label{eq16}
    v_{z} =  \frac{z}{\tau} - \frac{\omega x}{2}
\end{equation}
where $x$ and $z$ are the position coordinates. Here it is important to note that vorticity is the cause of inducing the velocity along $x$-direction, i.e., $v_{x}$. We have introduced the vorticity into the system through the modified Euler's thermodynamic relation~\cite{Becattini:2010, Florkowski:2017ruc, Singh:2018bih}, we have;
\begin{equation}
    \epsilon + P = Ts + \mu n + \Omega \rm w
    \label{eq17}
\end{equation}
Here, $\Omega$ is the chemical potential corresponding to rotation, and \rm w is the rotation density. Further, one can define $\Omega = \frac{T}{2\sqrt{2}}\sqrt{\omega_{\mu\nu}\omega^{\mu\nu}} $ and $\rm w = 4cosh(\xi)n_{0}$, where $\xi = \frac{\omega}{2T}$ and $n_{0} = \frac{T^{3}}{\pi^{2}}$ is the number density of the particles in the massless limit. Thus, the rotation density becomes $\rm w = 4\frac{T^{3}}{\pi^{2}}\rm cosh\left(\frac{\omega}{2T}\right)$
\cite{Singh:2018bih}.\\
 
Thus, taking all the above inputs at zero 
baryonic chemical potential, Eq.~(\ref{eq17}) can be modified as,
\begin{equation}
\label{eq18}
    \epsilon + P = Ts + \frac{2\omega T^{3}}{\pi^{2}} \cosh\left(\frac{\omega}{2T}\right)
\end{equation}
Under the ideal limit, $\epsilon = 3P$. Hence the above equation becomes,
\begin{equation}
\label{eq19}
    \epsilon = \frac{3}{4}\bigg[ Ts + \frac{2\omega T^{3}}{\pi^{2}} \cosh\left(\frac{\omega}{2T}\right)\bigg]
\end{equation}
Differentiating the above equation with respect to proper time $\tau$,
\begin{equation}
\label{eq20}
    \frac{d\epsilon}{d\tau} = \frac{3}{4}\bigg[ \frac{Tds}{d\tau}+\frac{sdT}{d\tau} + \frac{2}{\pi^{2}}\frac{d}{d\tau}\left(\omega T^{3} \cosh\left(\frac{\omega}{2T}\right)\right)\bigg]
\end{equation}

 We  use the standard form of entropy,
 $s = c+dT^{3}$, where c and d are constants to obtain, 
 
 \begin{equation}
\label{eq21}
    \frac{d\epsilon}{d\tau} = \frac{3}{4}\bigg[ \bigg(s+3dT^{3}+\frac{2F}{\pi^{2}}\bigg)\frac{dT}{d\tau} + \frac{2G}{\pi^{2}}\frac{d\omega}{d\tau}\bigg]
\end{equation}
where, $F= 3T^{2}\omega \cosh\left(\frac{\omega}{2T}\right) - \frac{1}{2} \omega^{2}T \sinh\left(\frac{\omega}{2T}\right)$ and
$G = T^{3} \cosh\left(\frac{\omega}{2T}\right) + \frac{1}{2} \omega T^{2}  \sinh\left(\frac{\omega}{2T}\right)$. Now, using Eq.~(\ref{eq18}) in Eq.~(\ref{eq10}), we get,
 
\begin{equation}
\label{eq22}
    \frac{d\epsilon}{d\tau} = -\frac{1}{\gamma} \left(Ts + \frac{2\omega T^{3}}{\pi^{2}} \cosh\left(\frac{\omega}{2T}\right) - \Phi \right)\; \partial_{\mu} u^{\mu}
\end{equation}

Comparing Eq.~(\ref{eq21}) and Eq.~(\ref{eq22}) we get,

\begin{align}
\label{eq23}
    \frac{d\omega}{d\tau} = \frac{-\pi^{2}}{2G}\bigg[ \frac{4T}{3\gamma}\bigg( s + \frac{2T^{2}\omega}{\pi^{2}} \rm cosh\left(\frac{\omega}{2T}\right) - \frac{\Phi}{T}\bigg)\partial_{\mu} u^{\mu} \nonumber\\
    + \bigg(s+3dT^{3}+\frac{2F}{\pi^{2}}\bigg)\frac{dT}{d\tau}\bigg]
\end{align}

The temperature evolution equation can be obtained from the energy evolution Eq.~(\ref{eq22}) using the aforementioned EoS. The modified temperature cooling rate 
is presented as;

\begin{equation}
\label{eq24}
    \frac{dT}{d\tau} = \frac{1}{\gamma}\left[- \frac{T}{3}\bigg( 1 + \frac{2\omega T^{2}}{s\pi^{2}}\rm \cosh\left(\frac{\omega}{2T}\right) \bigg) +  \frac{\Phi T^{-3}} {12 \mathit a} \right]\partial_{\mu} u^{\mu}
\end{equation}

Thus, vorticity can also generate viscosity in the medium. In this work, we have taken the direct 
contribution of vorticity in viscosity evolution through MIS equation~\cite{Muronga:2003ta}. Here we have incorporated the viscous and vorticity coupling term  $\pi^{(\mu}_{\alpha}\omega^{\nu)\alpha}$ through a second order transport coefficient $\lambda$ \cite{Song:2008si}.

\begin{align}
\label{eq25}
    D\pi^{\mu\nu} = - \frac{1}{\tau_{\pi}}\pi^{\mu\nu}- \frac{1}{2\beta_{2}}\pi^{\mu\nu}\left[\beta_{2}\theta + TD\left(\frac{\beta_{2}}{T}\right)\right] \nonumber\\
    +\frac{1}{\beta_{2}} \bigtriangledown^{<\mu}u^{\nu>} + \lambda \pi^{(\mu}_{\alpha} \omega^{\nu)\alpha}
\end{align}
Starting with Eq.~(\ref{eq18}), the coupling of shear stress tensor with vorticity 
can be  written as: 


\begin{align}
 \frac{d\Phi}{d\tau} &= - \frac{2aT\Phi}{3b\gamma} - \frac{\Phi}{2\gamma}\left( \partial_{\mu}u^{\mu} - \frac{5\gamma}{T}\frac{dT}{d\tau} \right) \nonumber \\ &+\frac{4aT^{4}}{3\tau\gamma}\left(\bigtriangledown^{<0}u^{0>} - \bigtriangledown^{<z}u^{z>}\right) - \frac{\omega \Phi}{\gamma T \tau} 
 \label{eq26}
\end{align}

In the present context the detailed expressions for $u^{\mu}\partial_{\mu}$, $\partial_{\mu}u^{\mu}$, 
$\bigtriangledown^{<0}u^{0>} - \bigtriangledown^{<z}u^{z>}$, and $\pi^{(\mu}_{\alpha}\omega^{\nu)\alpha}$  
are derived in Appendix~(\ref{appe1}),~(\ref{appe2}),~(\ref{appe3}) and~(\ref{appe4}), respectively. 
Finally, we get the three non-linear coupled differential Eqs.~(\ref{eq23}),~(\ref{eq24}), and~(\ref{eq26}) describing 
the medium evolution in terms of vorticity, temperature, and viscosity, respectively.  
If we take  $\omega$ = 0, then it reduces to the second-order viscous hydrodynamics, and further, if we take  
$\Phi =0 $, then it gives us a solution corresponding to the ideal fluid.

\subsection{Rotational viscous fluid in the presence of magnetic field}
\label{mag}
Next, we consider the evolution of charged fluids rotating in a viscous medium in the presence 
of the magnetic field. In such a case, the energy-momentum tensor for rotating, 
viscous and magnetized fluid is given by ~\cite{Roy:2015kma, Biswas:2020rps};

\begin{equation}
\label{eq27}
     T^{\mu\nu} =  \left( \epsilon+P+B^{2}\right) u^{\mu}u^{\nu} -g^{\mu\nu}\left(P+\frac{B^{2}}{2}\right) - B^{\mu}B^{\nu}  + \pi^{\mu\nu}
\end{equation}
where $B^{\mu} = \frac{1}{2}\epsilon^{\mu\nu\alpha\beta}F_{\nu\alpha}u_{\beta}$ is the magnetic field in the 
fluid,  $F_{\nu\alpha}$ is the field strength tensor. The magnetic field four vector $B^{\mu}$ 
is  space-like four vector with modulus $B^{\mu}B_{\mu} = -1$ and orthogonal to $u^\mu$ that 
is $B^{\mu}u_{\mu} = 0$, where $B = |\vec{\textbf{B}}|$, and  $|\vec{\textbf{B}}|$ is the magnetic three vector.

The energy density evolution equation for a viscous medium in the presence of a magnetic field can be 
obtained from the energy-momentum conservation, Eq.~(\ref{eq2}) is given by \cite{Pu:2016ayh, Roy:2015kma};

\begin{equation}
\label{eq28}
    \frac{d\epsilon}{d\tau} = - \frac{1}{\gamma}\left( \epsilon+P+B^2 - \Phi \right)\;\partial_{\mu} u^{\mu} - B\frac{dB}{d\tau}
\end{equation}

Proceeding in the same way as Sec.~\ref{vorticity}, using the modified Euler equation $\epsilon + P = Ts + \mu n + \Omega \rm w + eBM $, where  $M = \chi_{m} B$ is the magnetization of the fluid, $\chi_{m}$ being the magnetic susceptibility, we have;

\begin{align}
\label{eq29}
    &\frac{d\omega}{d\tau} = \frac{-\pi^{2}}{2G}\bigg[ \bigg(s+3dT^{3}+\frac{2F}{\pi^{2}}\bigg)\frac{dT}{d\tau} + \left(\frac{4}{3}+2e\chi_{m}\right)B\frac{dB}{d\tau} \nonumber \\
   & +\frac{4T}{3\gamma}\bigg(s + \frac{2T^{2} \omega}{\pi^{2}} \rm cosh(\frac{\omega}{2T}) + (1+e\chi_m)\frac{B^{2}}{T}- \frac{\Phi}{T}\bigg)\; \partial_{\mu} u^{\mu} \bigg]
\end{align}

The changing magnetic field induces the electric field, making the medium evolution more complex. 
Therefore,  to reduce the complexity, we have considered a  static magnetic field for our calculation, 
i.e., $ \frac{dB}{d\tau}=0 $.  In such a situation, 
Eq.~(\ref{eq29}) reduces to:
 
\begin{align}
\label{eq30}
& \frac{d\omega}{d\tau} = \frac{-\pi^{2}}{2G}\bigg[\bigg(s+3dT^{3}+\frac{2F}{\pi^{2}}\bigg)\frac{dT}{d\tau}\nonumber \\
& + \frac{4T}{3\gamma}\bigg(s + \frac{2T^{2}\omega}{\pi^{2}}\cosh\left(\frac{\omega}{2T}\right) + (1+e\chi_m)\frac{B^{2}}{T}- \frac{\Phi}{T}\bigg)\partial_{\mu} u^{\mu}  \bigg]
\end{align}

The  temperature evolution equation in the presence of spin vorticity and magnetic field coupling is given by,

\begin{align}
\label{eq31}
    \frac{dT}{d\tau} =&  - \frac{T}{3\gamma}\bigg( 1 + \frac{2\omega T^{2}}{s\pi^{2}}\cosh\left(\frac{\omega}{2T}\right)  +\frac{\chi_{m}eB^{2}}{Ts}\bigg) \partial_{\mu} u^{\mu}\nonumber \\
    &+\frac{\Phi T^{-3}}{12a\gamma}\partial_{\mu} u^{\mu}
\end{align}

We have used Eq.~(\ref{eq26}) to include the viscous effect in this case as well.\\

The following section presents the interplay among vorticity, viscosity, and temperature on their dissipation using the above-discussed formalism.   

\section{Results and Discussion}
\label{res}

 
\begin{figure*}[ht!]
\centering
\includegraphics[scale = 0.32]{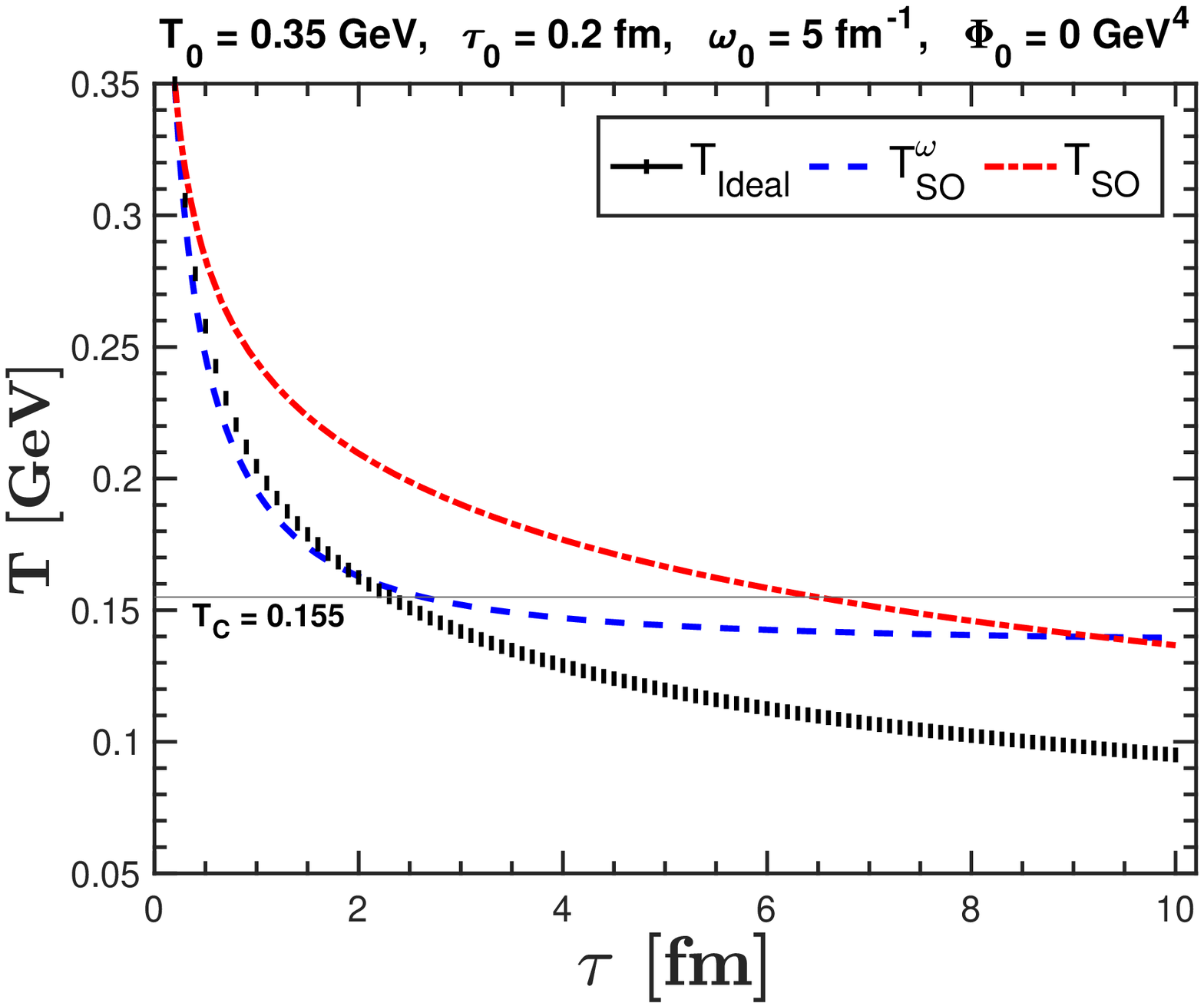}
\includegraphics[scale = 0.32]{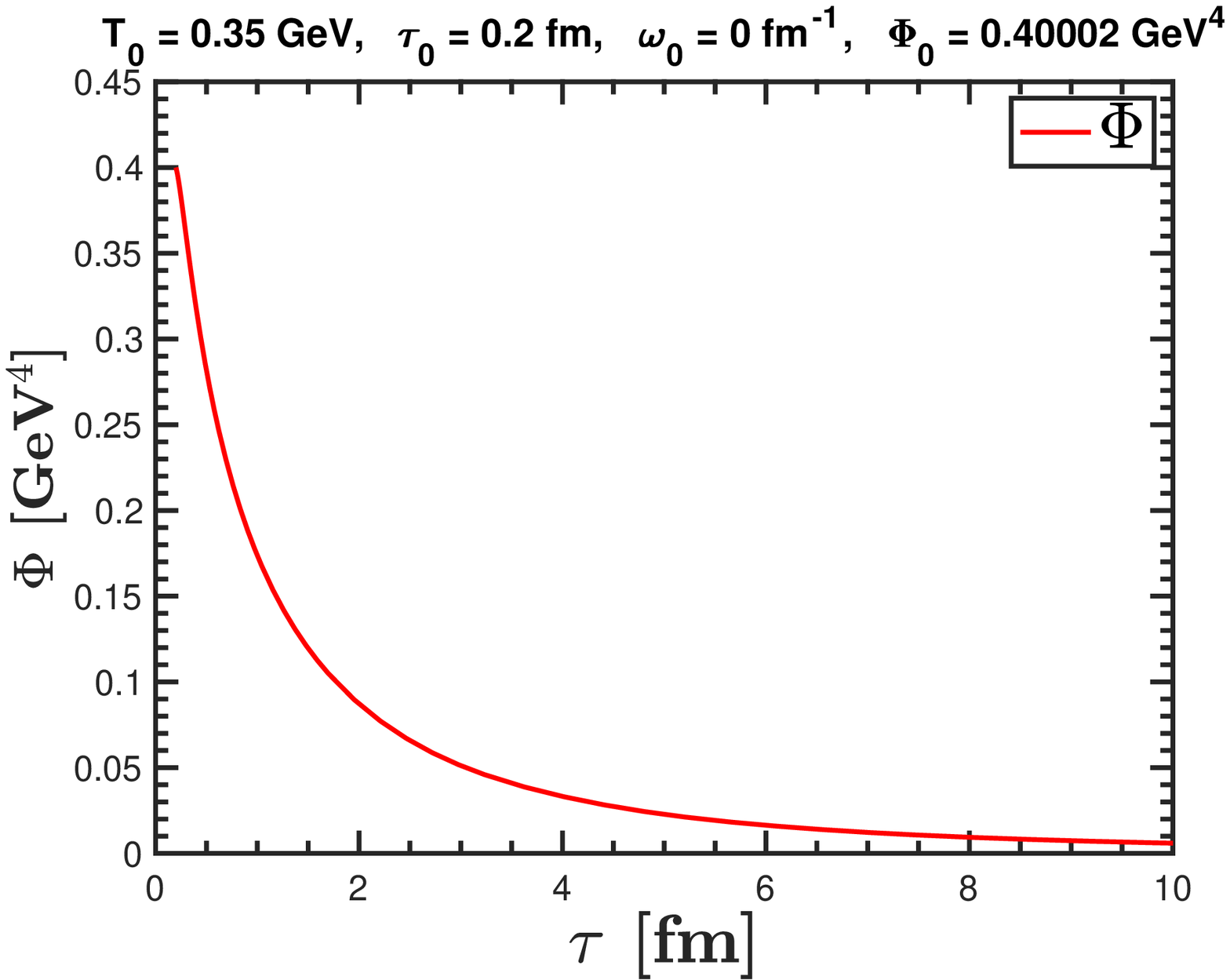}
\includegraphics[scale = 0.32]{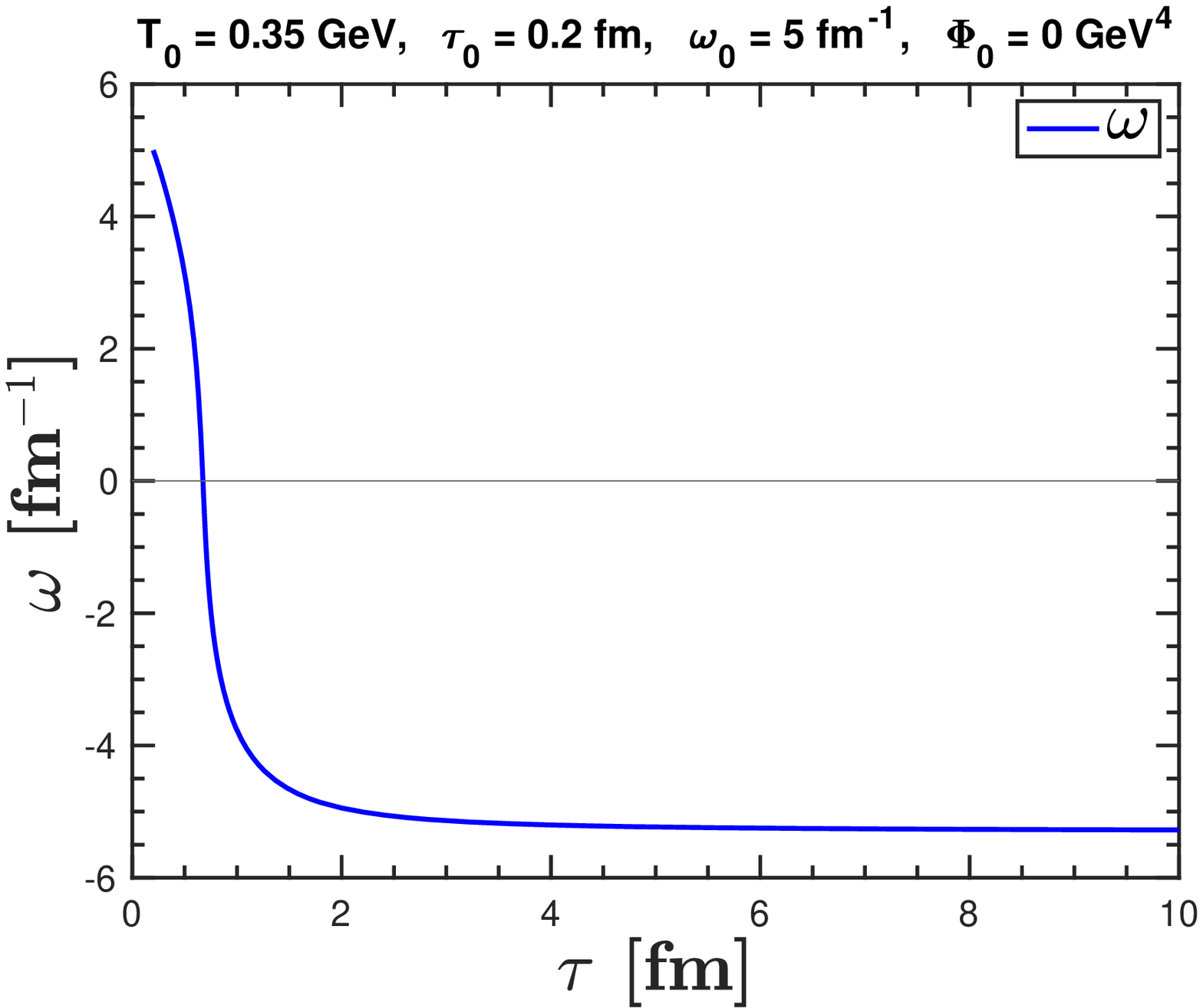}
\caption{(Color Online) {\bf Left to Right:} Temperature (T), viscous term ($\Phi$) and vorticity ($\omega$) are plotted, respectively, against time $\tau$ with the initial conditions: {\bf T = 0.35 GeV, $\tau_{0}$ = 0.2 fm, $\omega_{0}$ = 0.1 fm$^{-1}$, $\Phi_{0}$ = 0.40002 GeV$^4$}. For T$_{\text{Ideal}}$; $\omega = 0$ and $\Phi = 0$. For T$_{\text{SO}}$; $\omega = 0$ but $\Phi \ne 0$. For T$^{\omega}_{SO}$; $\omega \ne 0$ but $\Phi = 0$. In $\Phi$ plot,  $\omega = 0$ and in $\omega$ plot,  $\Phi = 0$.} 
\label{fig1}
\end{figure*}
\begin{figure*}[ht!]
\centering
\includegraphics[scale = 0.32]{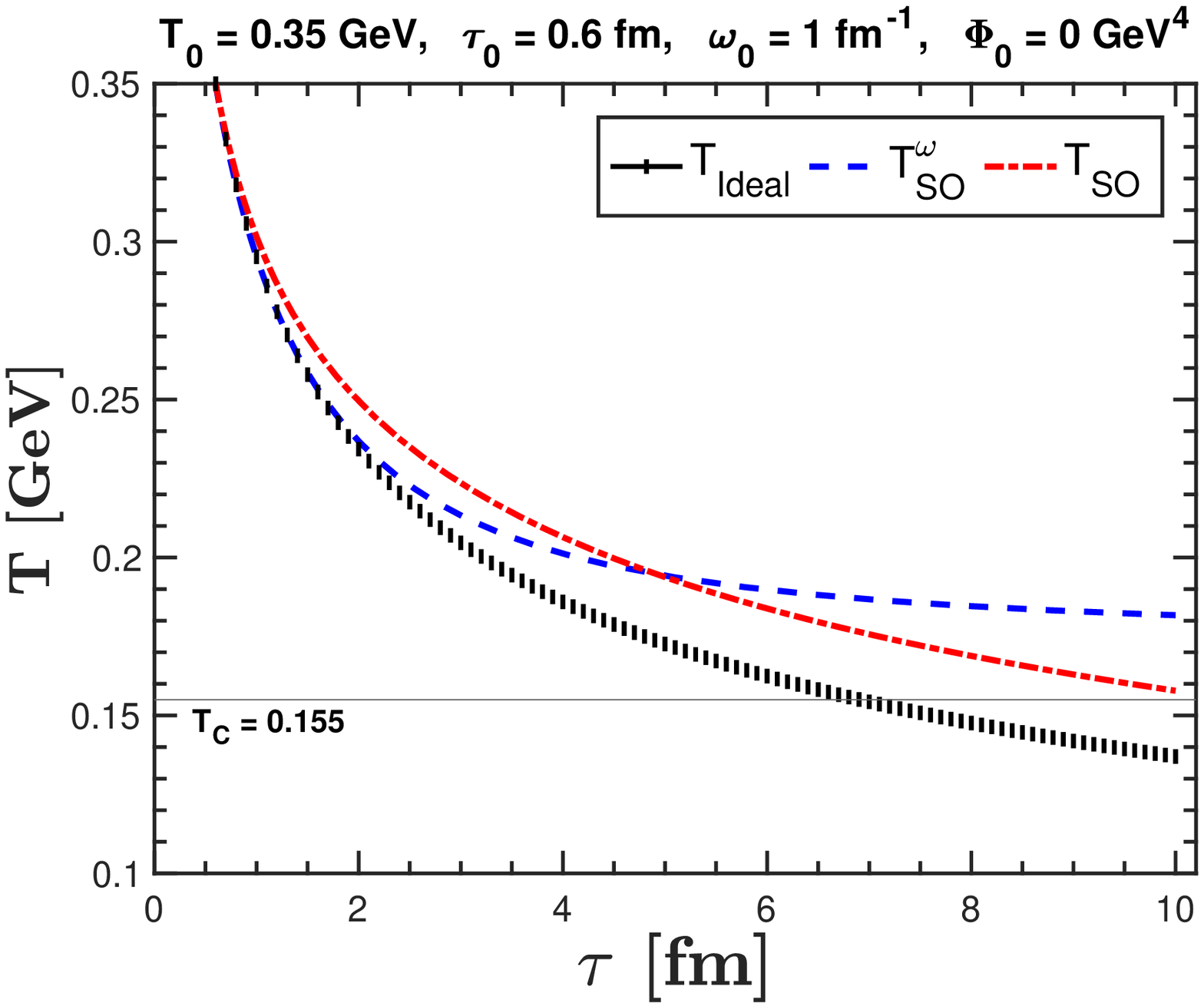}
\includegraphics[scale = 0.32]{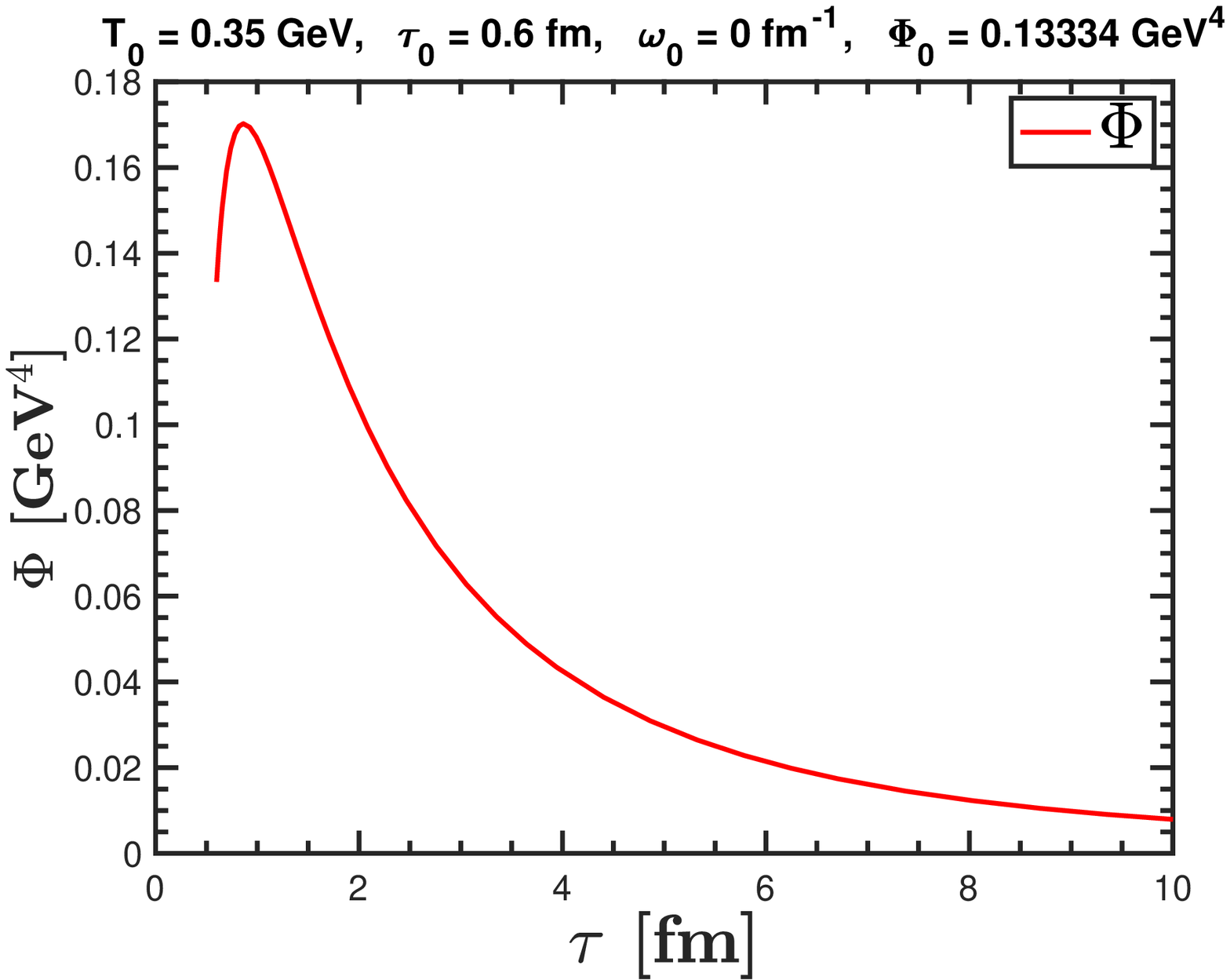}
\includegraphics[scale = 0.32]{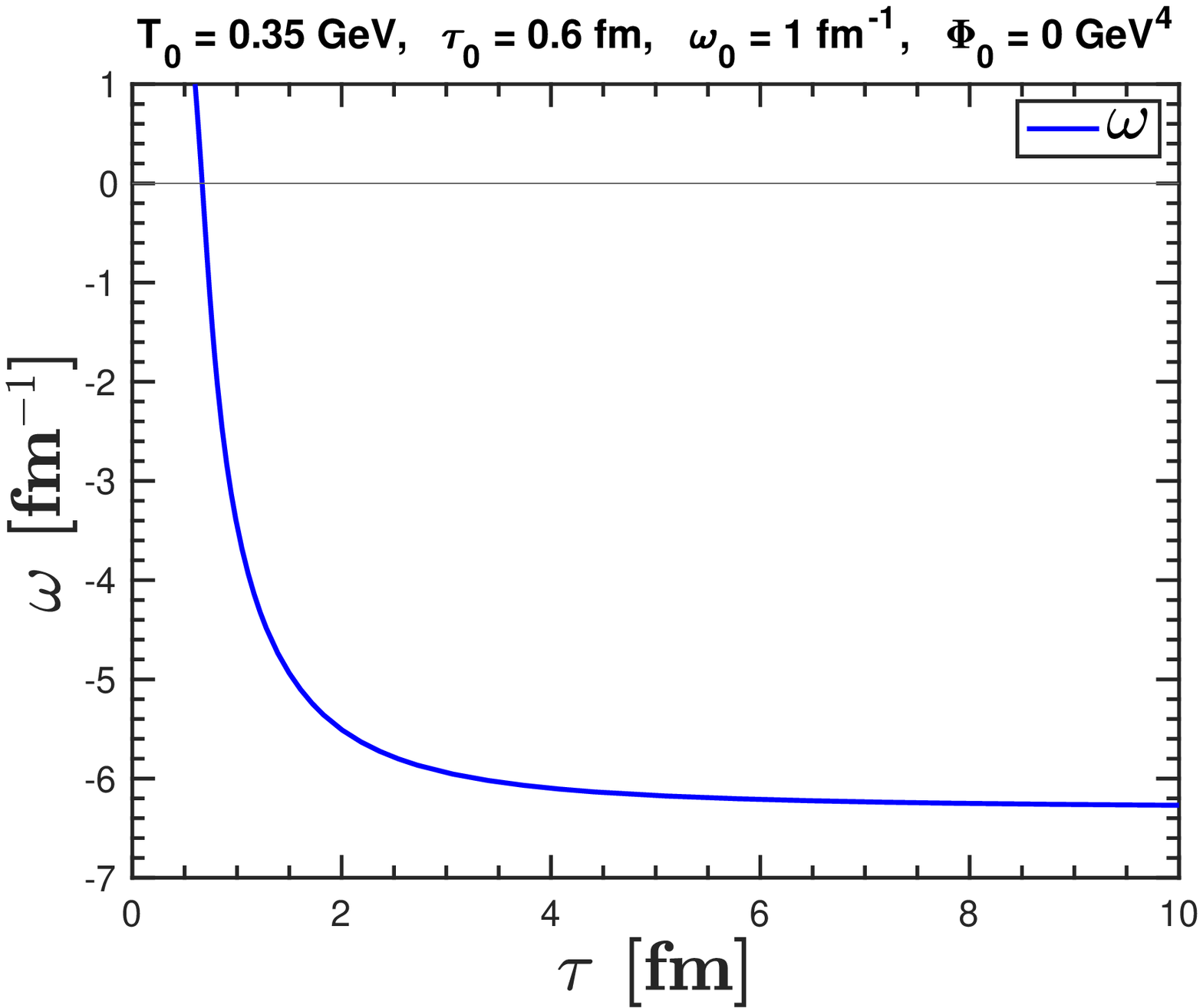}
\caption{(Color Online) {\bf Left to Right:} Temperature (T), viscous term ($\Phi$) and vorticity ($\omega$) are plotted, respectively, against time $\tau$ with the initial conditions: {\bf T = 0.35 GeV, $\tau_{0}$ = 0.6 fm, $\omega_{0}$ = 1.0 fm$^{-1}$, $\Phi_{0}$ = 0.13334 GeV$^4$}. For T$_{\text{Ideal}}$; $\omega = 0$ and $\Phi = 0$. For T$_{\text{SO}}$; $\omega = 0$ but $\Phi \ne 0$. For T$^{\omega}_{SO}$; $\omega \ne 0$ but $\Phi = 0$. In $\Phi$ plot,  $\omega = 0$ and in $\omega$ plot,  $\Phi = 0$.}
\label{fig2}
\end{figure*}
\begin{figure*}[ht!]
\centering
\includegraphics[scale = 0.319]{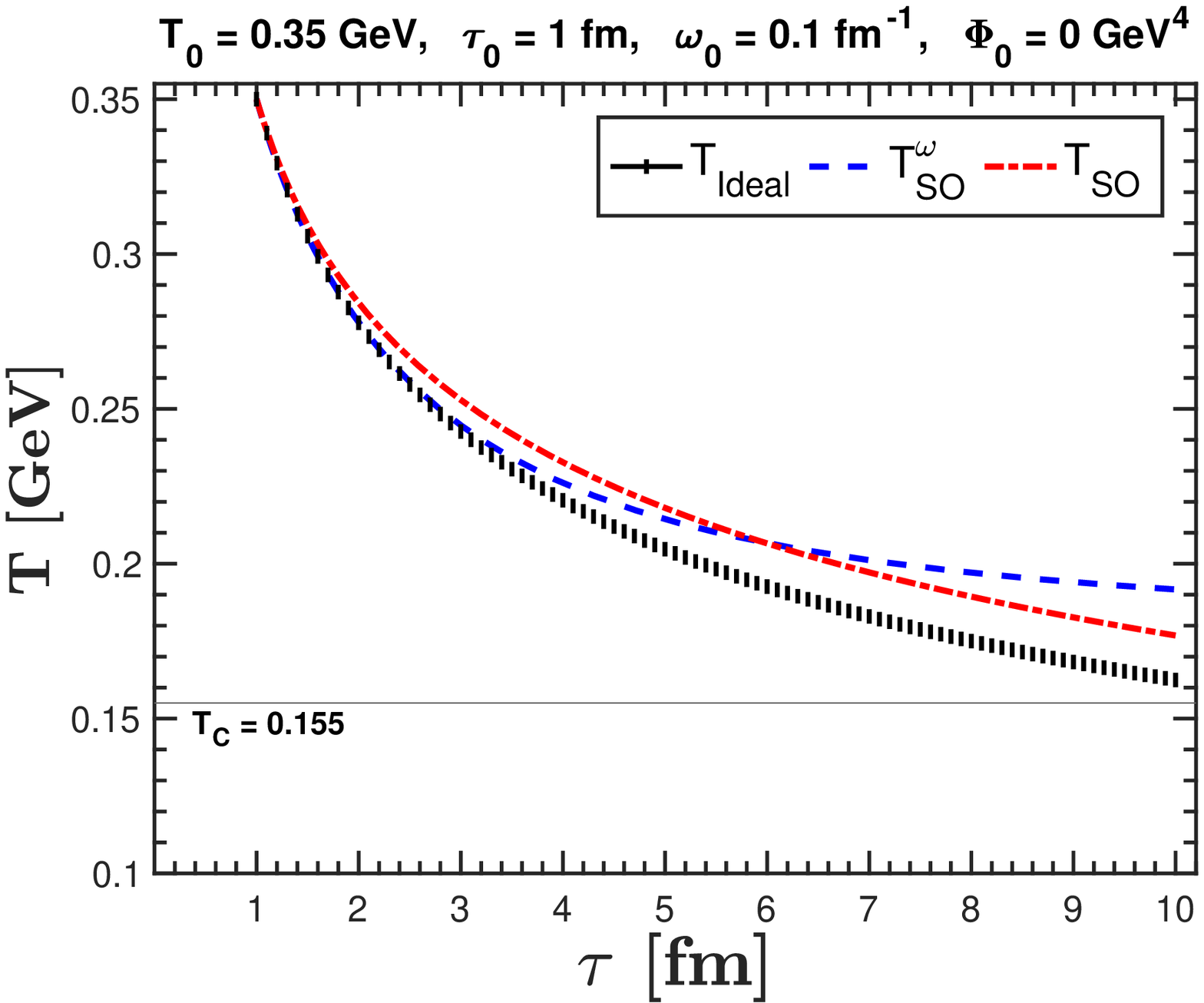}
\includegraphics[scale = 0.319]{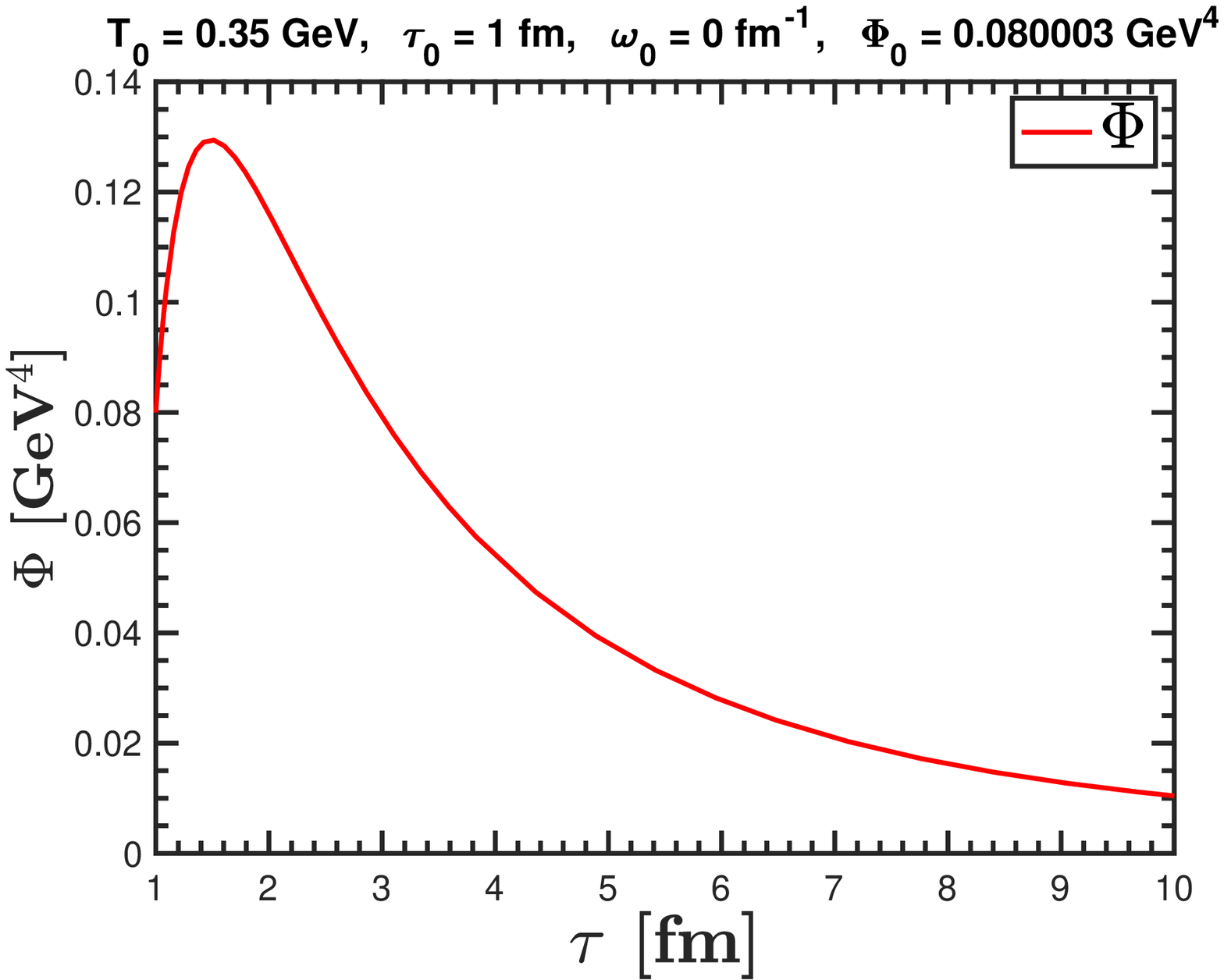}
\includegraphics[scale = 0.319]{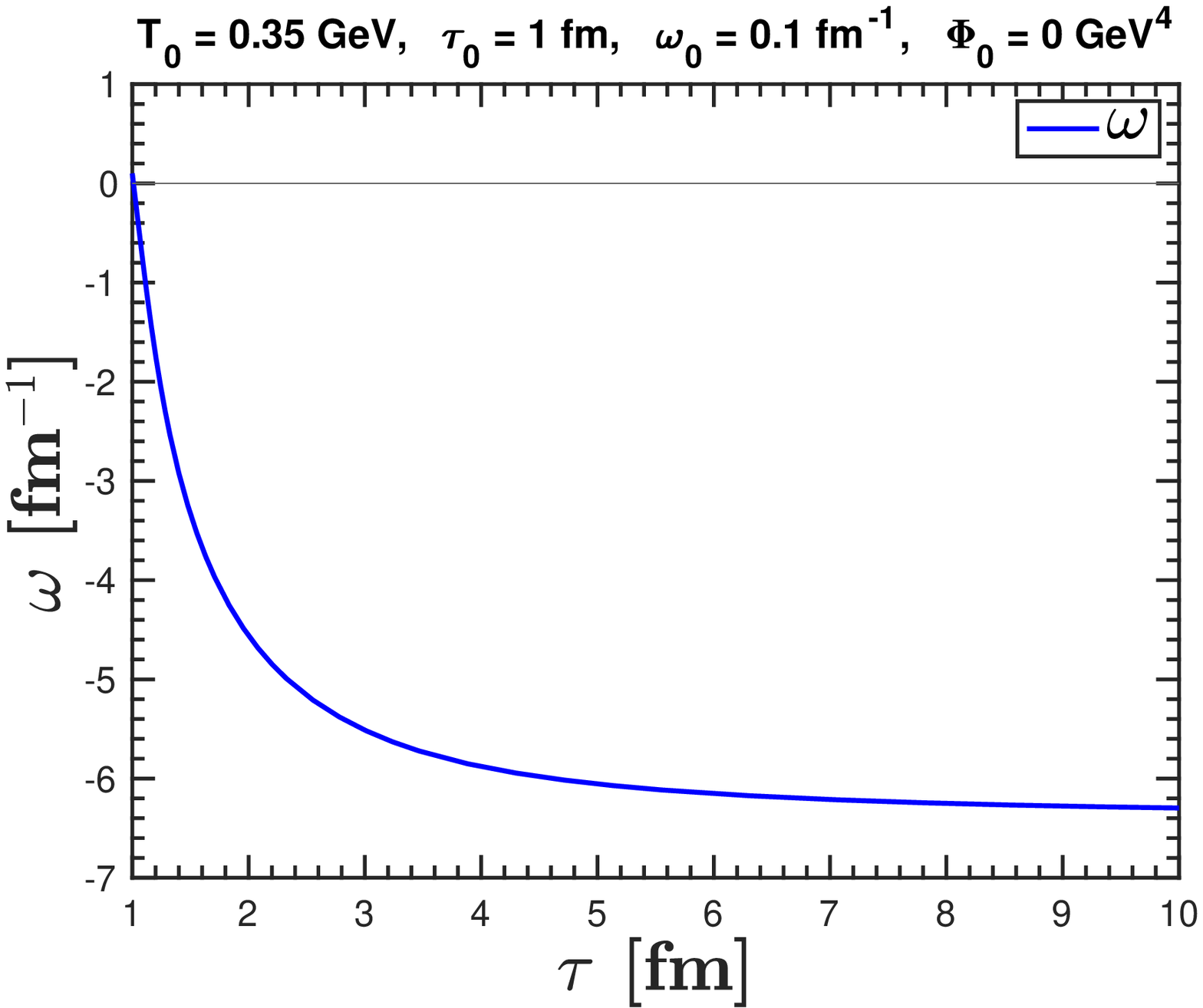}
\caption{(Color Online) {\bf Left to Right:} Temperature (T), viscous term ($\Phi$) and vorticity ($\omega$) are plotted, respectively, against time $\tau$ with the initial conditions: {\bf T = 0.35 GeV, $\tau_{0}$ = 1.0 fm, $\omega_{0}$ = 0.1 fm$^{-1}$, $\Phi_{0}$ = 0.080003 GeV$^4$}. For T$_{\text{Ideal}}$; $\omega = 0$ and $\Phi = 0$. For T$_{\text{SO}}$; $\omega = 0$ but $\Phi \ne 0$. For T$^{\omega}_{SO}$; $\omega \ne 0$ but $\Phi = 0$. In $\Phi$ plot,  $\omega = 0$ and in $\omega$ plot,  $\Phi = 0$.}
\label{fig3}
\end{figure*}

This section explores the effect of rotation and viscous forces on the evolution of the QGP. Their individual and combined roles in the evolution of temperature 
are discussed. The vorticity, viscosity, and temperature evolution are governed respectively by the three coupled 
equations Eq.~(\ref{eq23}), Eq.~(\ref{eq26}), and Eq.~(\ref{eq24}). The solution of these coupled differential
 equations is very sensitive to the initial conditions, i.e., $T_{0}$, $\tau_{0}$, $\omega_{0}$ and $\Phi_{0}$.  
We have considered the initial viscosity $\Phi_{0} = \frac{1}{3\pi}\frac{s_{0}}{\tau_{0}}$ at
$\tau=\tau_{0}$, where $s_{0} = c + dT_{0}^{3}$ is the initial entropy density~\cite{crs}. While the initial condition for vorticity is chosen in such a way that 
the speed of rotation does not violate the causality. Therefore, $\omega_{0}$ is taken as $\omega \propto \frac{1}{\tau_0}$ 
to preserve the causality. Given these conditions, we have chosen three sets of initial conditions for $T_{0}$, $\tau_{0}$, and $\omega_{0}$. Each set of initial conditions corresponds to a completely new evolving system. 
We have solved the coupled differential equation corresponding to $T$, $\omega$, and $\Phi$ using these initial conditions. First, we illustrate how vorticity, viscosity, and temperature change with $\tau$ when there is no coupling between viscosity and vorticity.  Next, we explore the scenario when viscosity contributes to the vorticity and their combined effect on temperature variation. Further, the direct contribution of vorticity in viscosity will be shown. 
It is to be noted that T$_{\text{Ideal}}$ stands for the case when  $\omega = 0$ and  $\Phi =  0$ in Eq.~(\ref{eq27}). $T_{SO}$ ($T^{\omega}_{SO}$) 
stands for temperature obtained by solving the second-order hydrodynamic equations for $\omega=0$ ($\omega\neq 0$). It is noteworthy to mention that for irrotational fluid ($\omega = 0$), the longitudinal boost invariant velocity profile is assumed, and the evolution is similar to a Bjorken-like flow. However, for the rotational fluid ($\omega \neq 0$) we consider the velocity profile mentioned in Eq.~(\ref{eq15}) and Eq.~(\ref{eq16}).

\begin{figure*}[ht!]
\centering
\includegraphics[scale = 0.32]{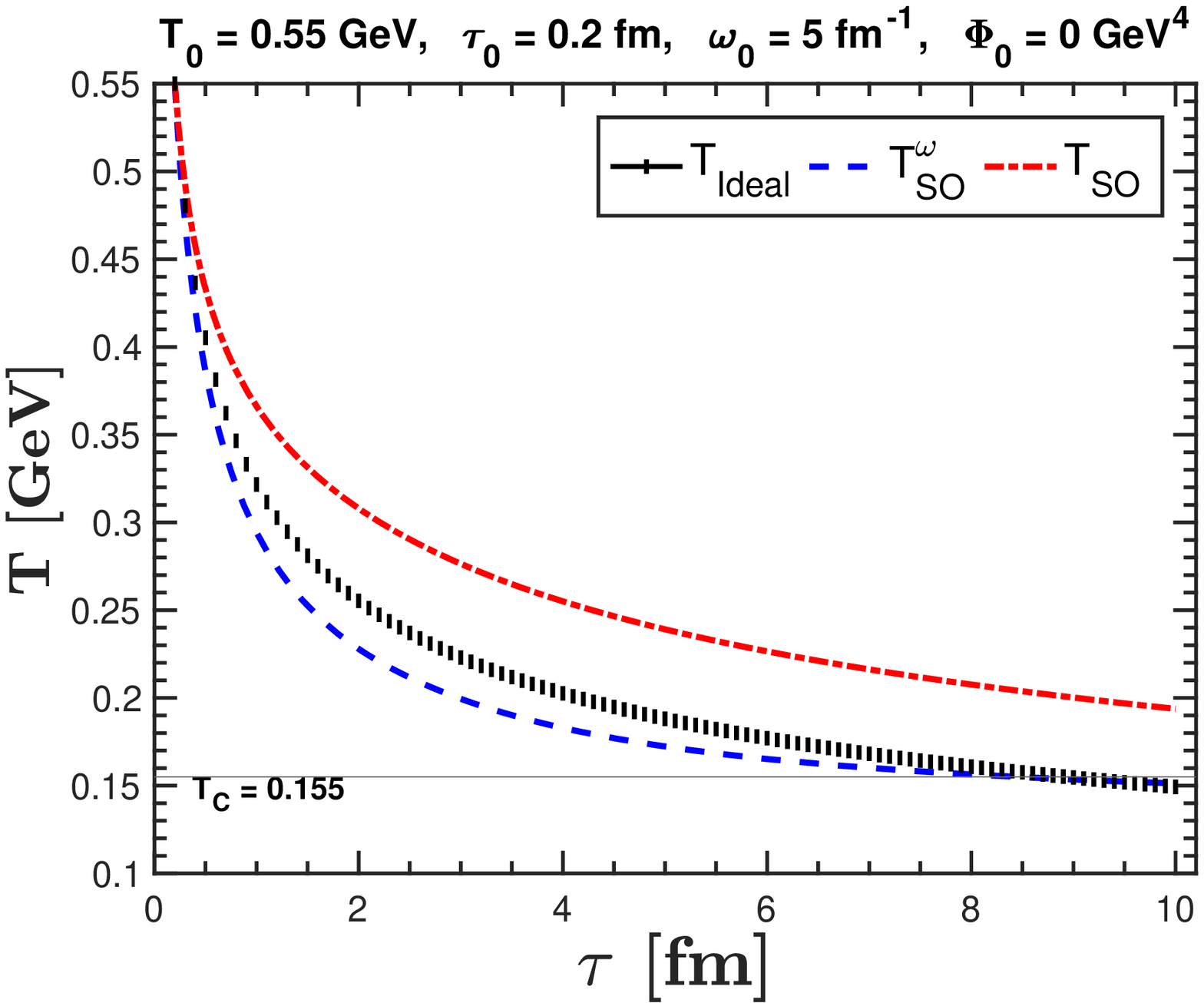}
\includegraphics[scale = 0.32]{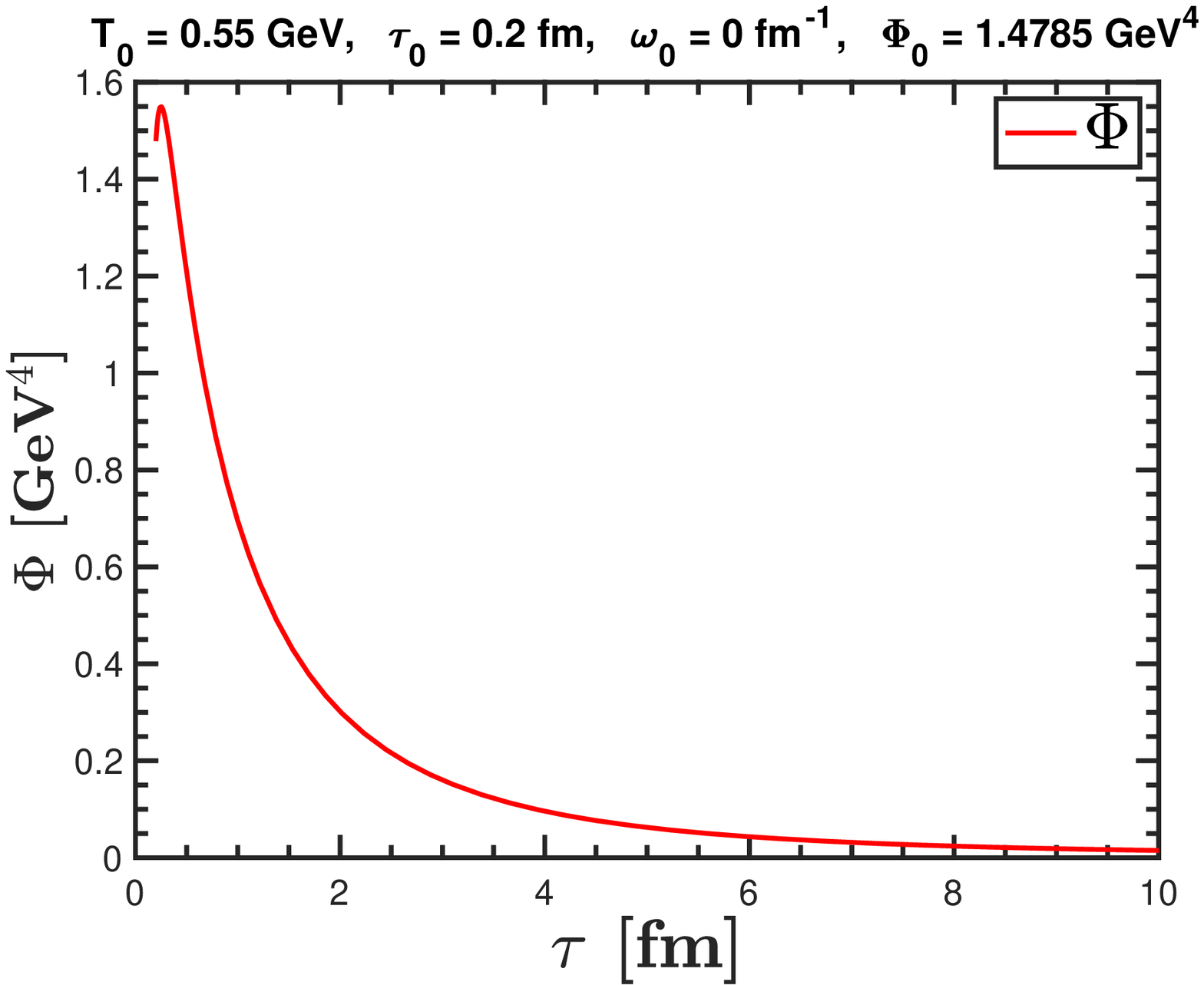}
\includegraphics[scale = 0.32]{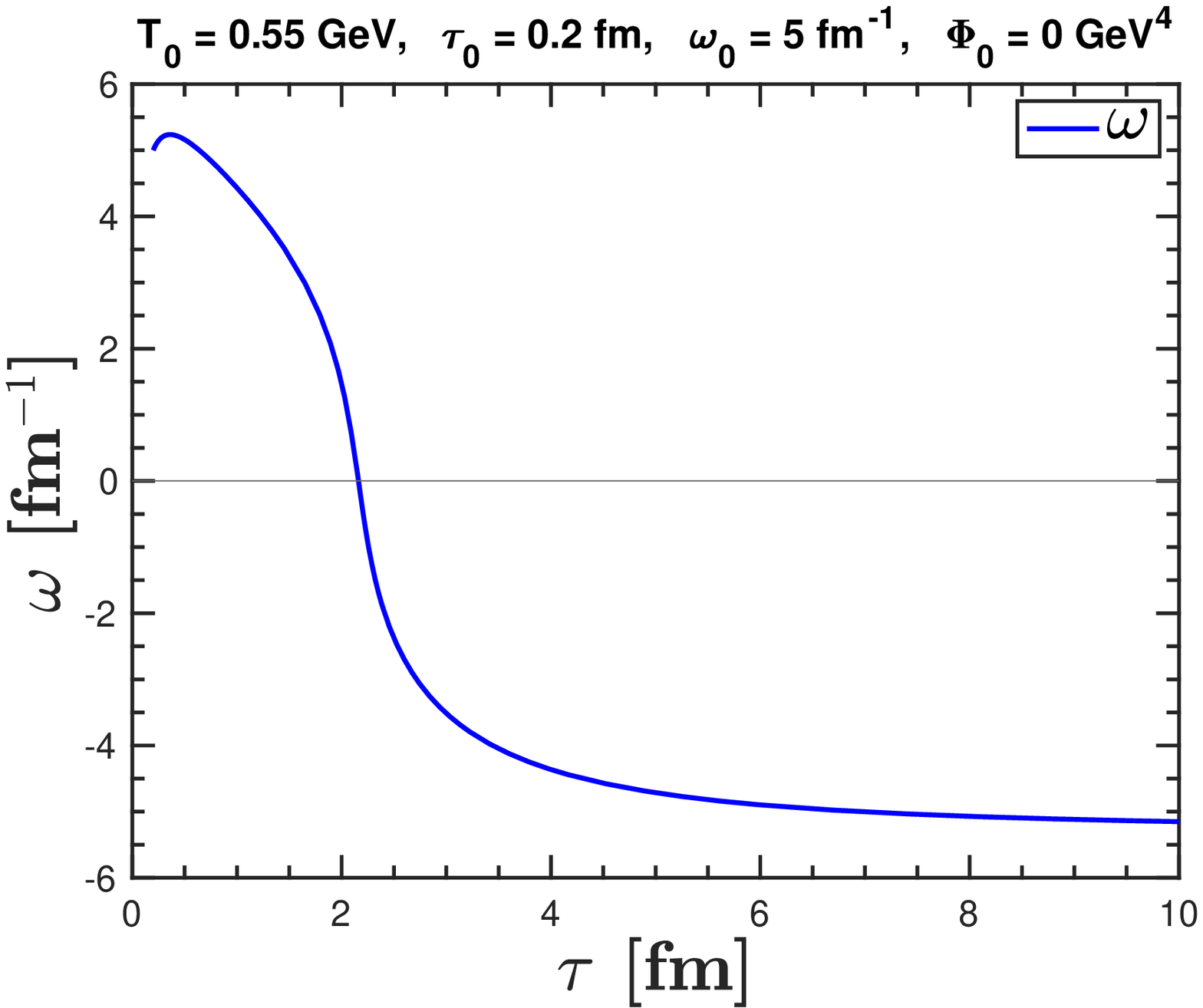}
\caption{(Color Online) {\bf Left to Right:} Temperature (T), viscous term ($\Phi$) and vorticity ($\omega$) are plotted, respectively, against time $\tau$ with the initial conditions: {\bf T = 0.55 GeV, $\tau_{0}$ = 0.2 fm, $\omega_{0}$ = 5 fm$^{-1}$, $\Phi_{0}$ = 1.4785 GeV$^4$}.  For T$_{\text{Ideal}}$; $\omega = 0$ and $\Phi = 0$. For T$_{\text{SO}}$; $\omega = 0$ but $\Phi \ne 0$. For T$^{\omega}_{SO}$; $\omega \ne 0$ but $\Phi = 0$. In $\Phi$ plot,  $\omega = 0$ and in $\omega$ plot,  $\Phi = 0$.}
\label{fig4}
\end{figure*}
\begin{figure*}[ht!]
\centering
\includegraphics[scale = 0.32]{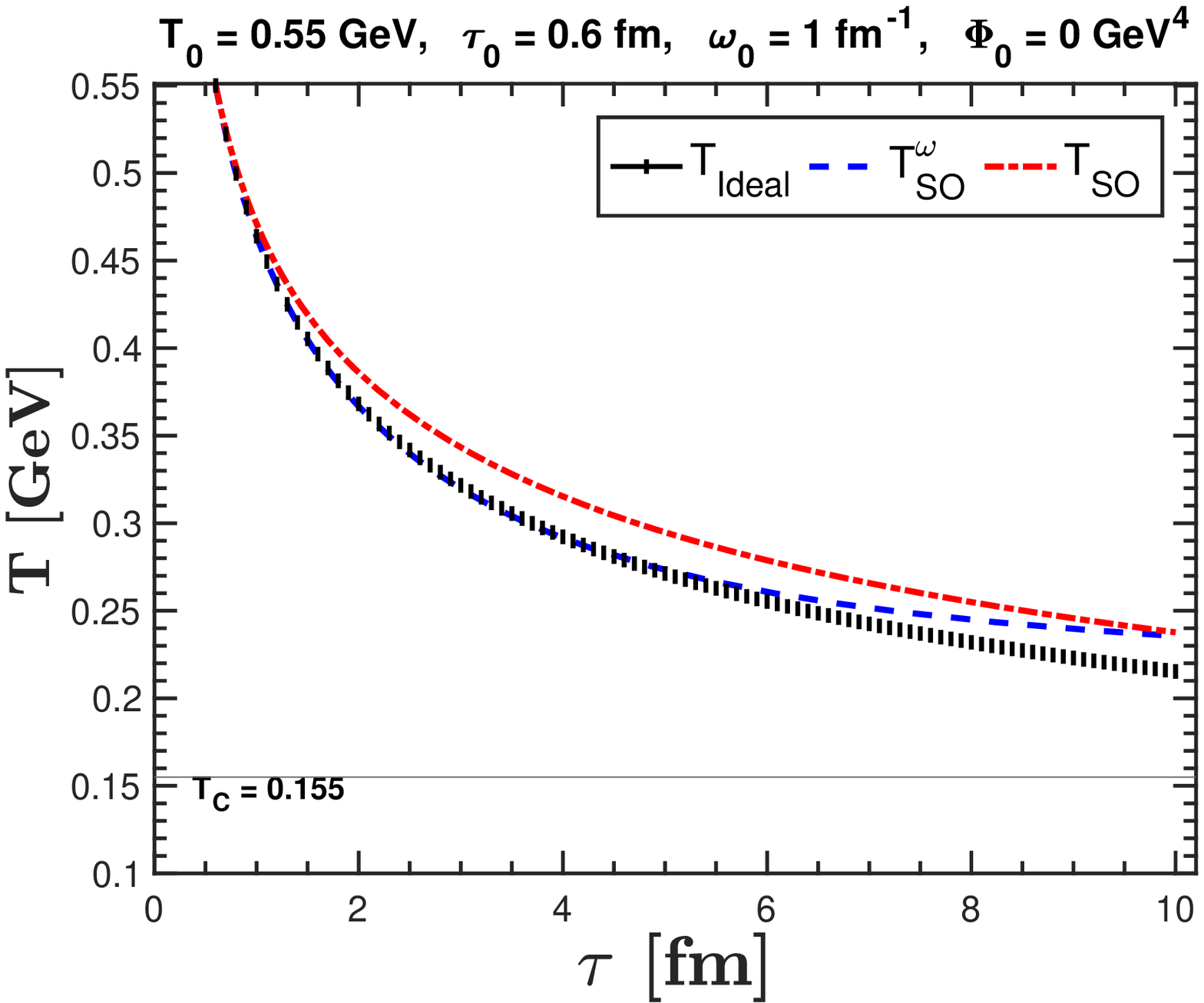}
\includegraphics[scale = 0.32]{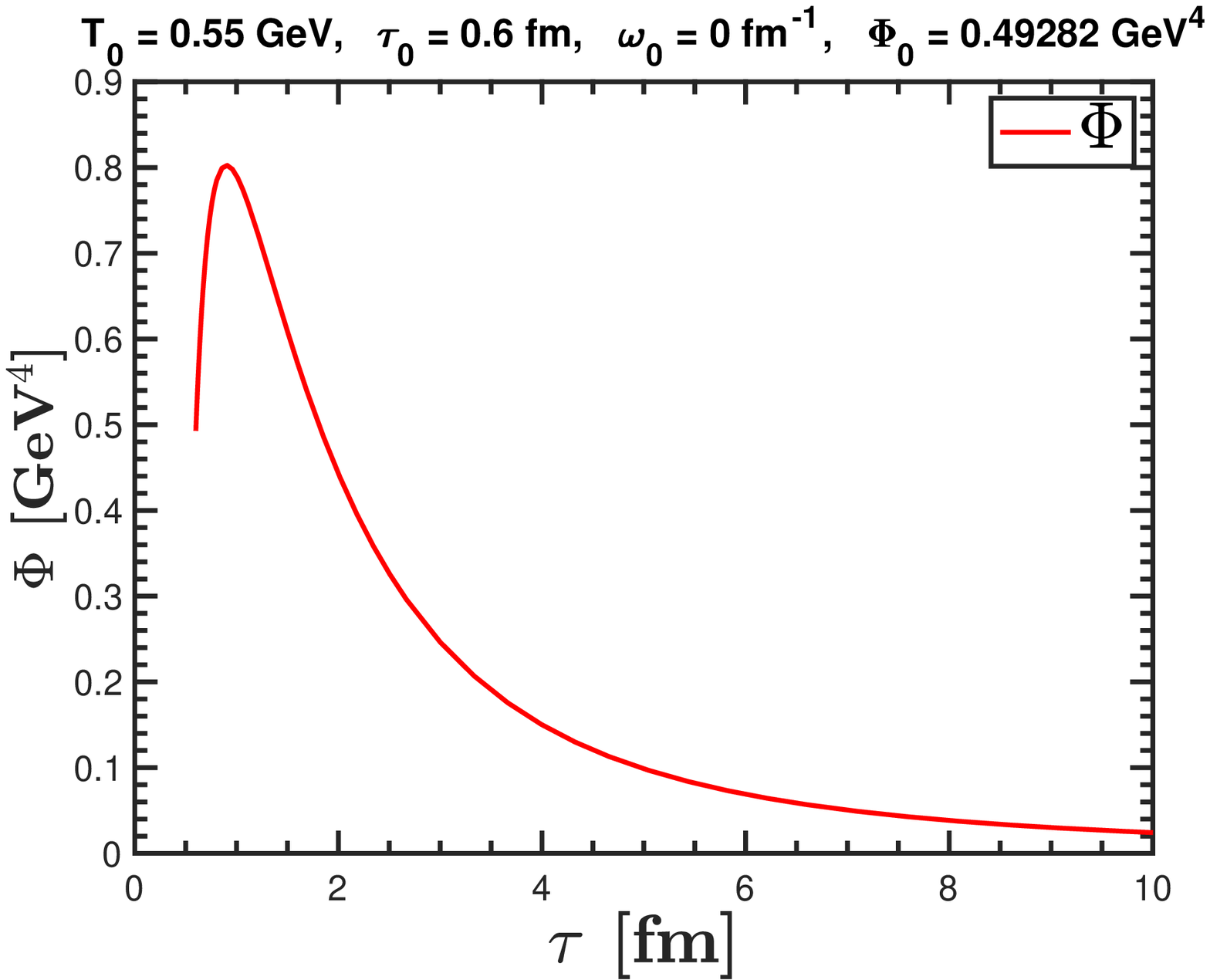}
\includegraphics[scale = 0.32]{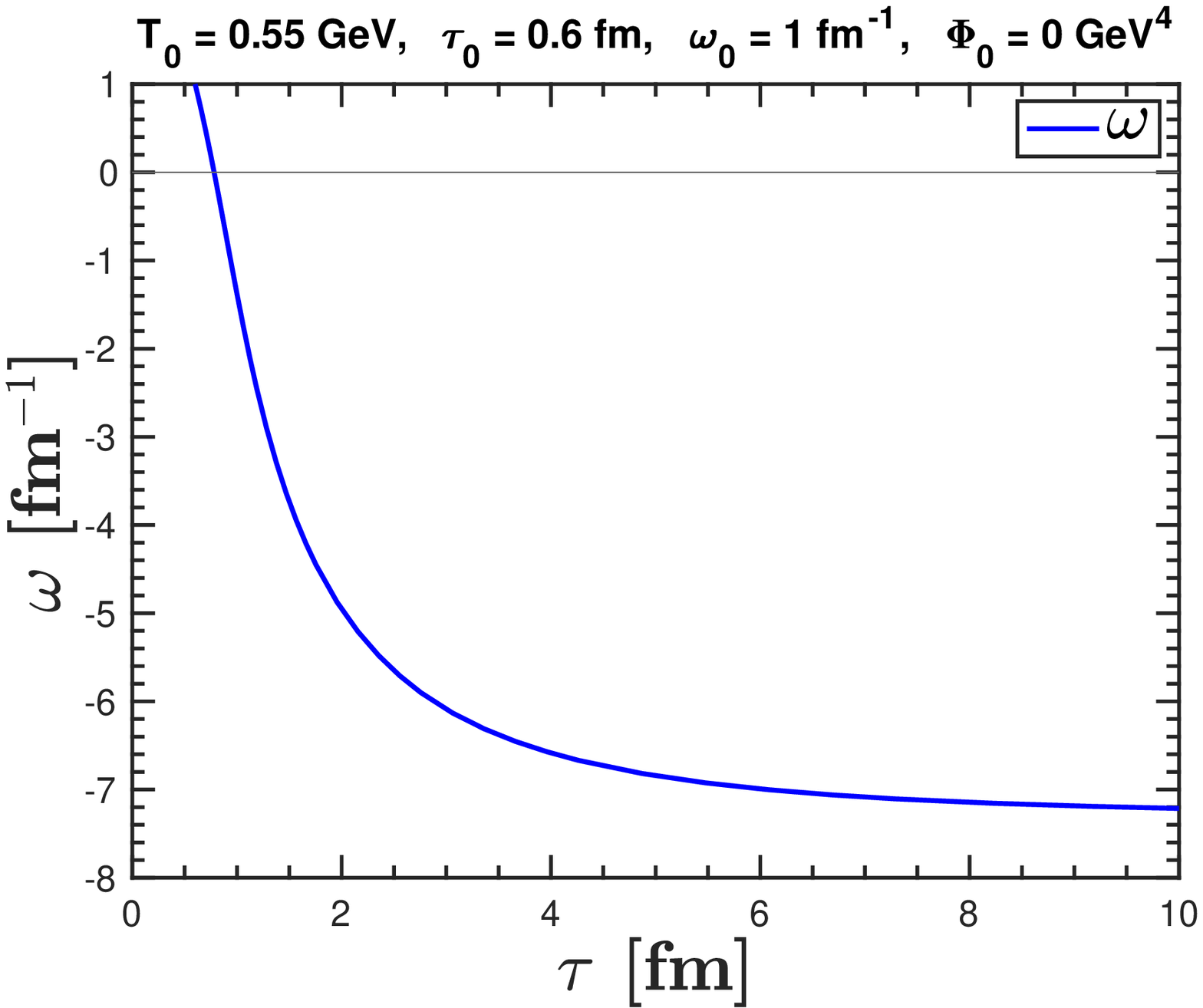}
\caption{(Color Online) {\bf Left to Right:} Temperature (T), viscous term ($\Phi$) and vorticity ($\omega$) are plotted, respectively, against time $\tau$ with the initial conditions: {\bf T = 0.55 GeV, $\tau_{0}$ = 0.6 fm, $\omega_{0}$ = 1 fm$^{-1}$, $\Phi_{0}$ = 0.49282 GeV$^4$}. For T$_{\text{Ideal}}$; $\omega = 0$ and $\Phi = 0$. For T$_{\text{SO}}$; $\omega = 0$ but $\Phi \ne 0$. For T$^{\omega}_{SO}$; $\omega \ne 0$ but $\Phi = 0$. In $\Phi$ plot,  $\omega = 0$ and in $\omega$ plot,  $\Phi = 0$.}
\label{fig5}
\end{figure*}
\begin{figure*}[ht!]
\centering
\includegraphics[scale = 0.32]{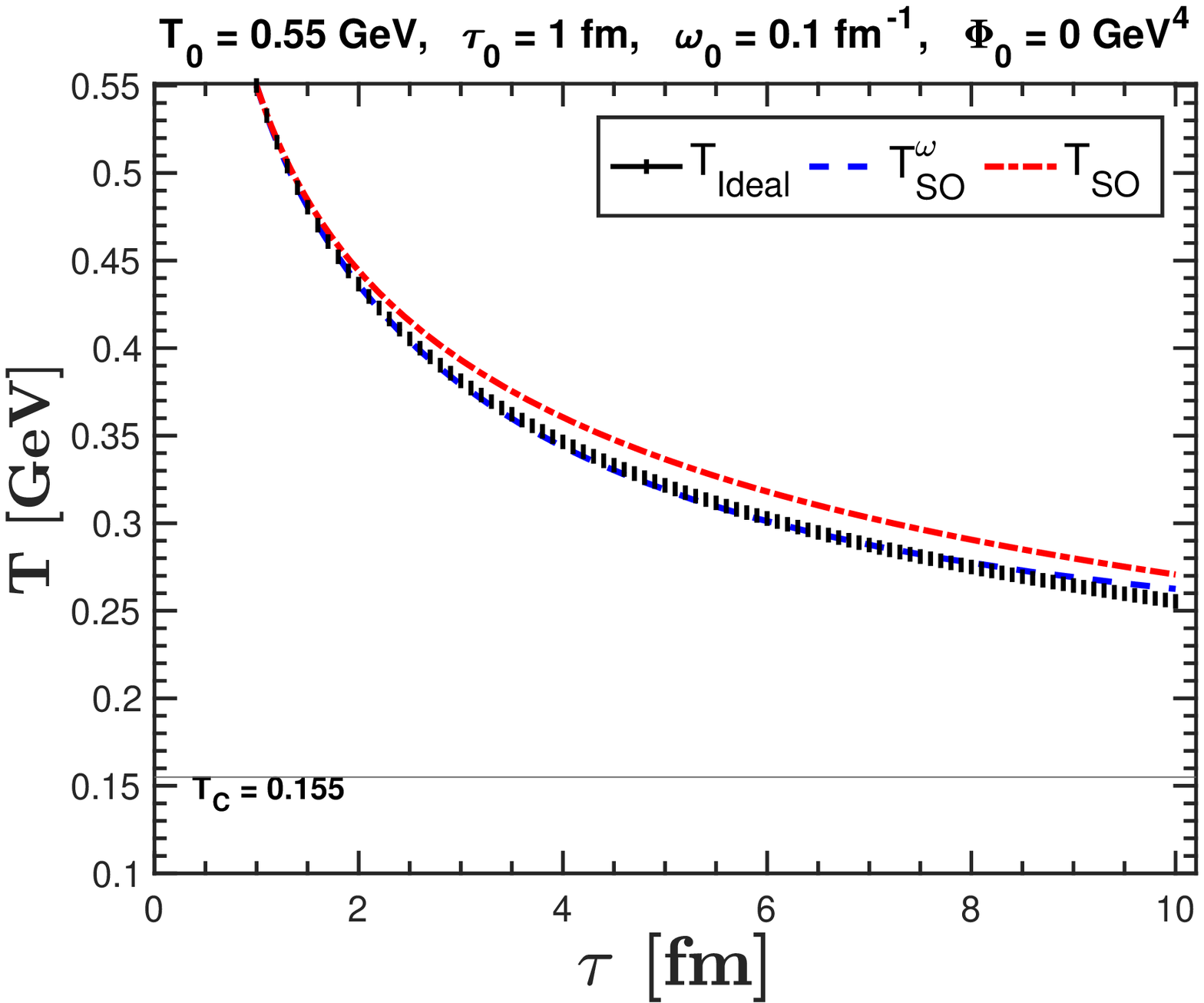}
\includegraphics[scale = 0.32]{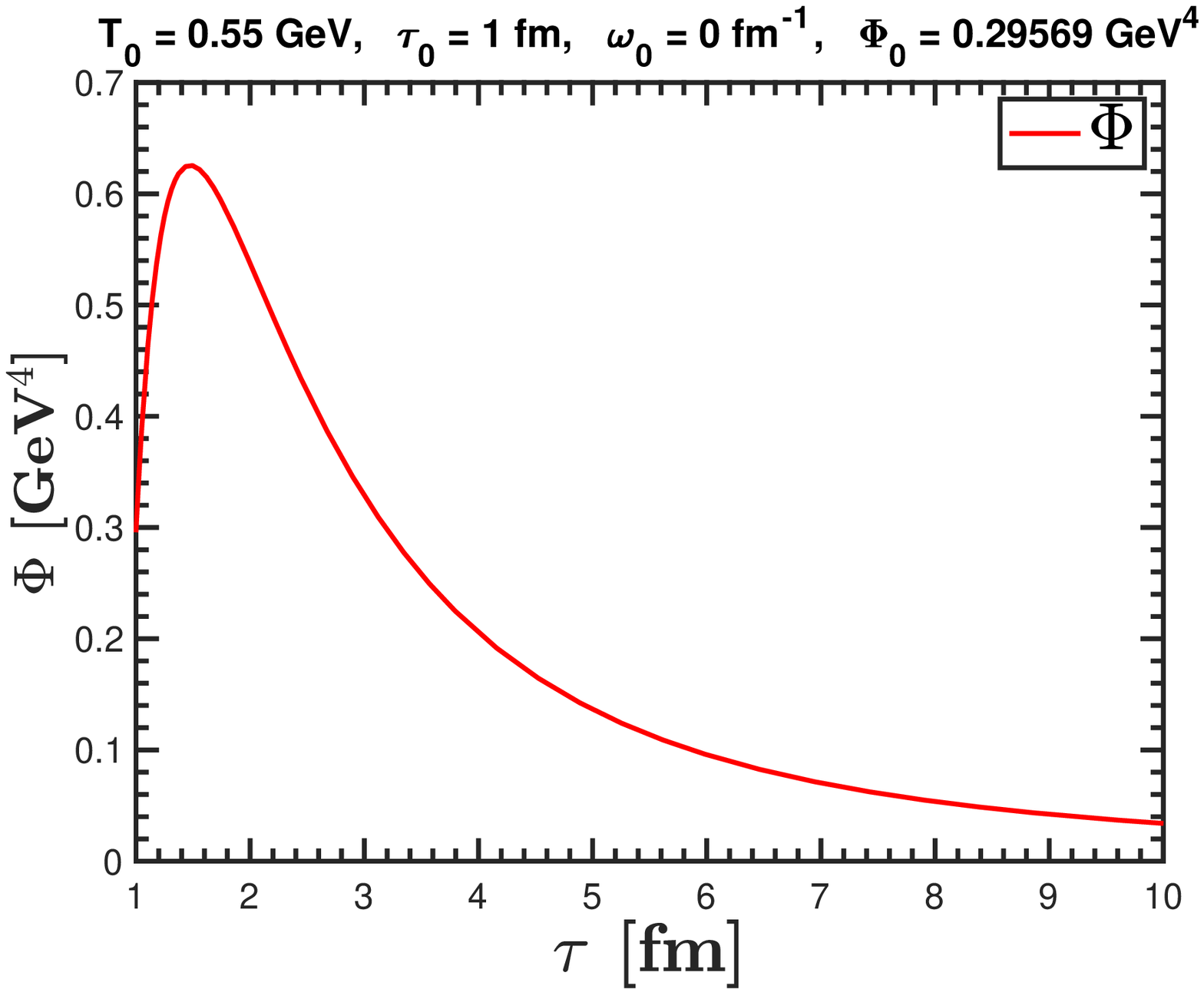}
\includegraphics[scale = 0.32]{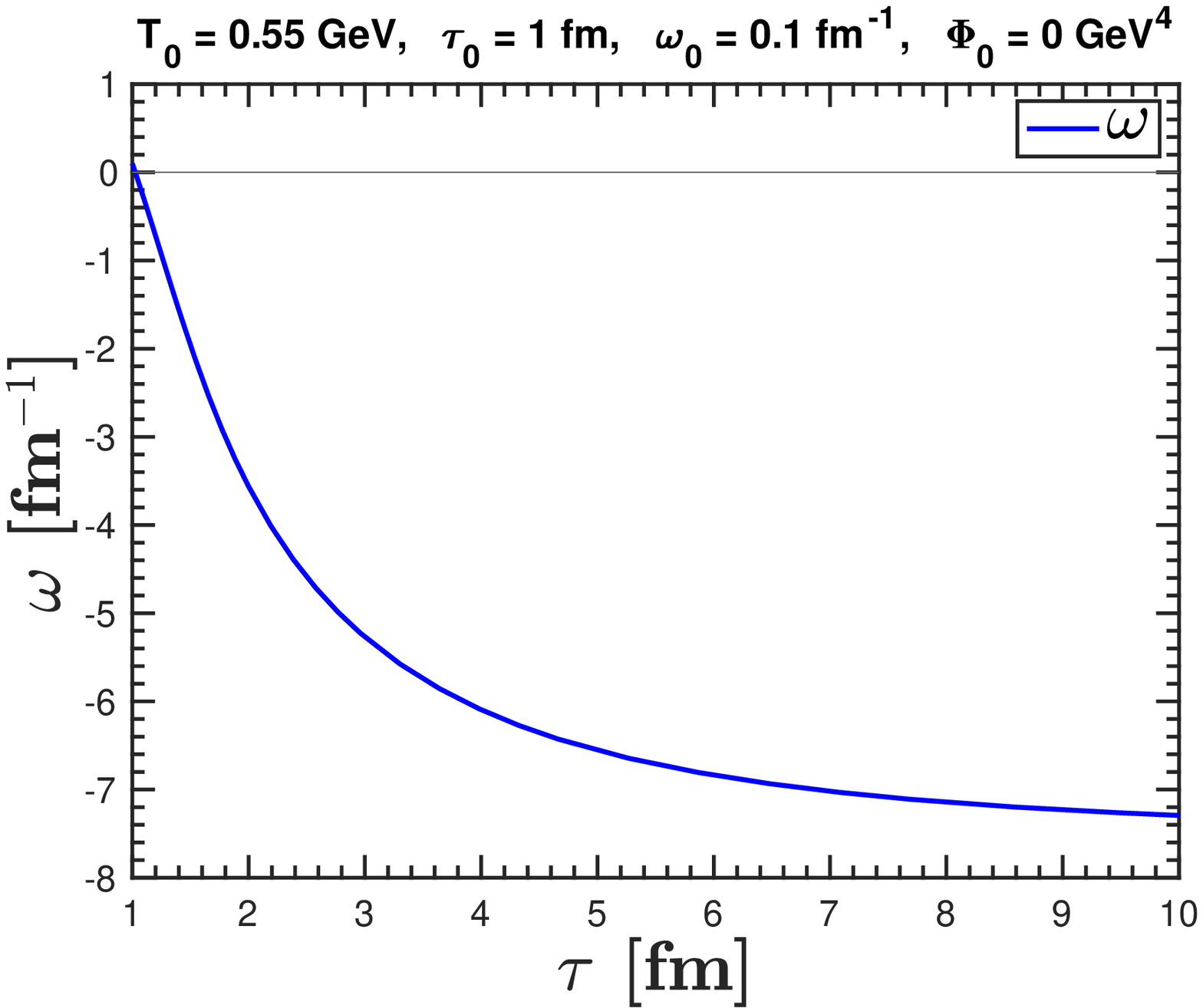}
\caption{(Color Online) {\bf Left to Right:} Temperature (T), viscous term ($\Phi$) and vorticity ($\omega$) are plotted, respectively, against time $\tau$ with the initial conditions: {\bf T = 0.55 GeV, $\tau_{0}$ = 1 fm, $\omega_{0}$ = 0.1 fm$^{-1}$, $\Phi_{0}$ = 0.29569 GeV$^4$}. For T$_{\text{Ideal}}$; $\omega = 0$ and $\Phi = 0$. For T$_{\text{SO}}$; $\omega = 0$ but $\Phi \ne 0$. For T$^{\omega}_{SO}$; $\omega \ne 0$ but $\Phi = 0$. In $\Phi$ plot,  $\omega = 0$ and in $\omega$ plot,  $\Phi = 0$.}
\label{fig6}
\end{figure*}


\subsection*{Case I: No coupling between $\Phi$ and $\omega$} 
The individual effect of $\omega$ and $\Phi$ on medium cooling is explored in this section, The corresponding differential equations for the cooling  are:

\begin{equation}
 \frac{dT}{d\tau} = - \frac{T}{3 \gamma }\left[ 1 + \frac{2\omega T^{2}}{s\pi^{2}}\rm \cosh\left(\frac{\omega}{2T}\right)\right] \partial_{\mu} u^{\mu} \nonumber  
\end{equation}

\begin{equation}
   \frac{dT}{d\tau} = \frac{1}{\gamma}\left[ - \frac{T}{3 }+  \frac{\Phi T^{-3}}{12a } \right]\partial_{\mu} u^{\mu} \nonumber
\end{equation}

 
The solutions of these two equations are $T^{\omega}_{SO}$ and $T_{SO}$ respectively.

In Fig.~\ref{fig1}, rapid cooling is observed  for ideal fluid in the absence of any dissipation.  
In presence of viscosity additional heat production reduces the cooling. Similar to viscosity, vorticity too affects the cooling.  Vortical motion present in the QGP medium imposes a constraint on the medium cooling. During the first moments of evolution, the rotation speed is almost equal to the medium evolution rate; due to this, it does not affect the cooling rate much. Therefore, as shown in  Fig.~\ref{fig1},  initially upto $\tau \sim$2 fm the cooling rate of T$^{\omega}_{SO}$ $\approx$ T$_{\text{Ideal}}$. Afterward, the system tries to hold back the evolution process when the rotation speed becomes smaller than the fluid velocity. The $\omega$ diffusion  and $\Phi$ dissipation with time is plotted in Fig.~\ref{fig1},  initially $\omega$ changes with a high rate, but at a later stage, it becomes almost 
 constant in the absence of any other external force while $\Phi$ approaches zero at large $\tau$.
The negative value of $\omega$ in the plot depicts the change in the direction of the rotation. 
This change in the rotation happens due to the initial fast expansion of the medium and the restriction imposed on it by 
the rotational motion of the fluid. 
This means medium evolution induces the rotation opposite to the initial vorticity. As time increases, 
vorticity also grows/diffuses in the opposite direction and gets saturated when medium evolution becomes static. 
Results displayed in Fig.~\ref{fig1} also suggest that cooling becomes almost independent of the vorticity if fluid is rotating close to the speed of light and, therefore, 
the cooling rate at $\omega_{0} = 5$ fm$^{-1}$ becomes almost the same as the ideal one, i.e., T$^{\omega}_{SO}$ $\approx$ T$_{\text{Ideal}}$.\\

Fig.~\ref{fig2} and Fig.~\ref{fig3} depicts the cooling rate change with changing initial conditions.
The variation of $T^{\omega}_{SO}$ cooling shown in Fig.~\ref{fig2} and Fig.~\ref{fig3} is the implication 
of low speed of the rotation along with the change in other initial conditions as compared to Fig.~\ref{fig1}.  
Depending on the speed of the medium rotation, the $\omega$ evolution is shown in Fig.~\ref{fig2} and Fig.~\ref{fig3}. 
The large value of $\tau_{0}$ reduces the $\Phi_{0}$, which leads to a faster cooling for T$_{SO}$. 
When the temperature cooling is faster than the $\Phi$ dissipation rate, it induces the medium viscosity, 
which causes a smooth rise in $\Phi$  as seen in the viscous evolution displayed in Fig.~\ref{fig2} and Fig.~\ref{fig3}.  \\

Now we take $T_0= 0.550$ GeV and keep the same initial conditions for $\Phi_{0}$ and $\omega_{0}$ as earlier and
evaluate $T$, $\Phi$, and $\omega$ to check the sensitivity of the results on the value
of initial temperature. The results in such cases are shown in  Fig.~\ref{fig4} to Fig.~\ref{fig6}.
The results show that the high initial vorticity effect almost vanishes at a relatively high initial temperature. 
As a result, the cooling for non-viscous rotating fluid behaves like ideal fluid. The dissipation of $\omega$ with proper time plotted 
in Fig.~\ref{fig4}, shows that temperature and vorticity coupling dominate when both are very large at the initial stage (T$_0$ = 0.550 GeV and  
$\omega_0$ = 5.0 fm$^{-1}$).  The short thermalization time and large initial temperature provide a large initial viscosity, reducing the cooling 
for T$_{SO}$  respective to Fig.~\ref{fig1}. Fig.~\ref{fig4} depicts that the vorticity diffusion rate is slow until a certain time; thereafter, 
vorticity increases in the opposite direction and gets saturated with time.  Fig.~\ref{fig5} and Fig.~\ref{fig6} follow similar trend
as Fig.~\ref{fig3} and Fig.~\ref{fig4} with higher initial temperature. \\


\subsection*{Case-II: $\Phi$ coupling with $\omega$}
In this case, we have considered both non-zero viscosity and vorticity in determining the
cooling rate as given in Eq.~(\ref{eq24}). 
T$_{SO-\Phi}^{\omega}$ in the figures represents the 
cooling rate corresponding to Eq.~(\ref{eq24}) and $\omega^{\Phi}$ stands for the vorticity obtained by solving Eq.~(\ref{eq23}).  Here we see how viscosity in a rotating fluid modifies the cooling rate and vorticity evolution ($\omega^{\Phi}$).\\

\begin{figure*}[!htp]
\centering
\includegraphics[scale = 0.32]{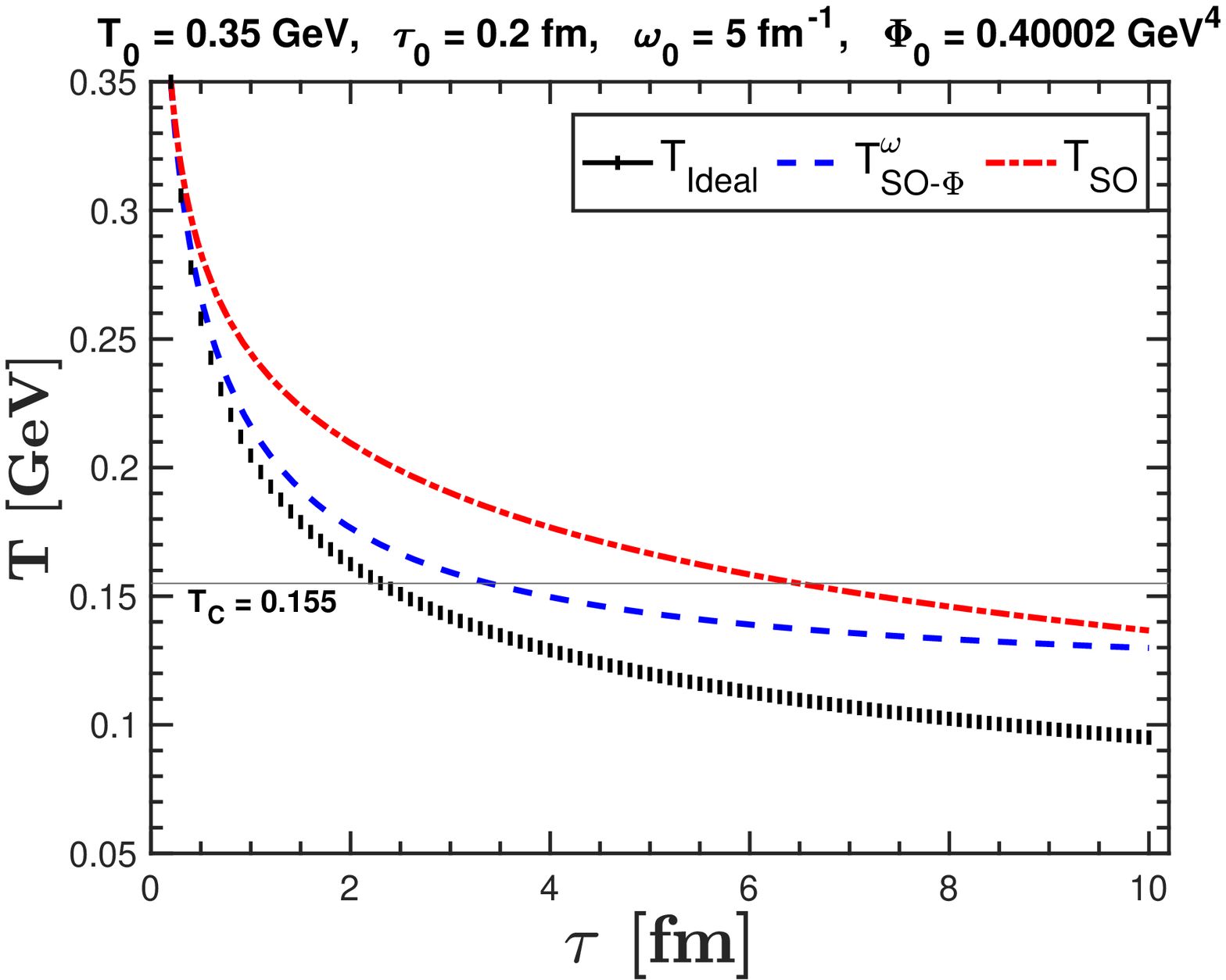}
\includegraphics[scale = 0.32]{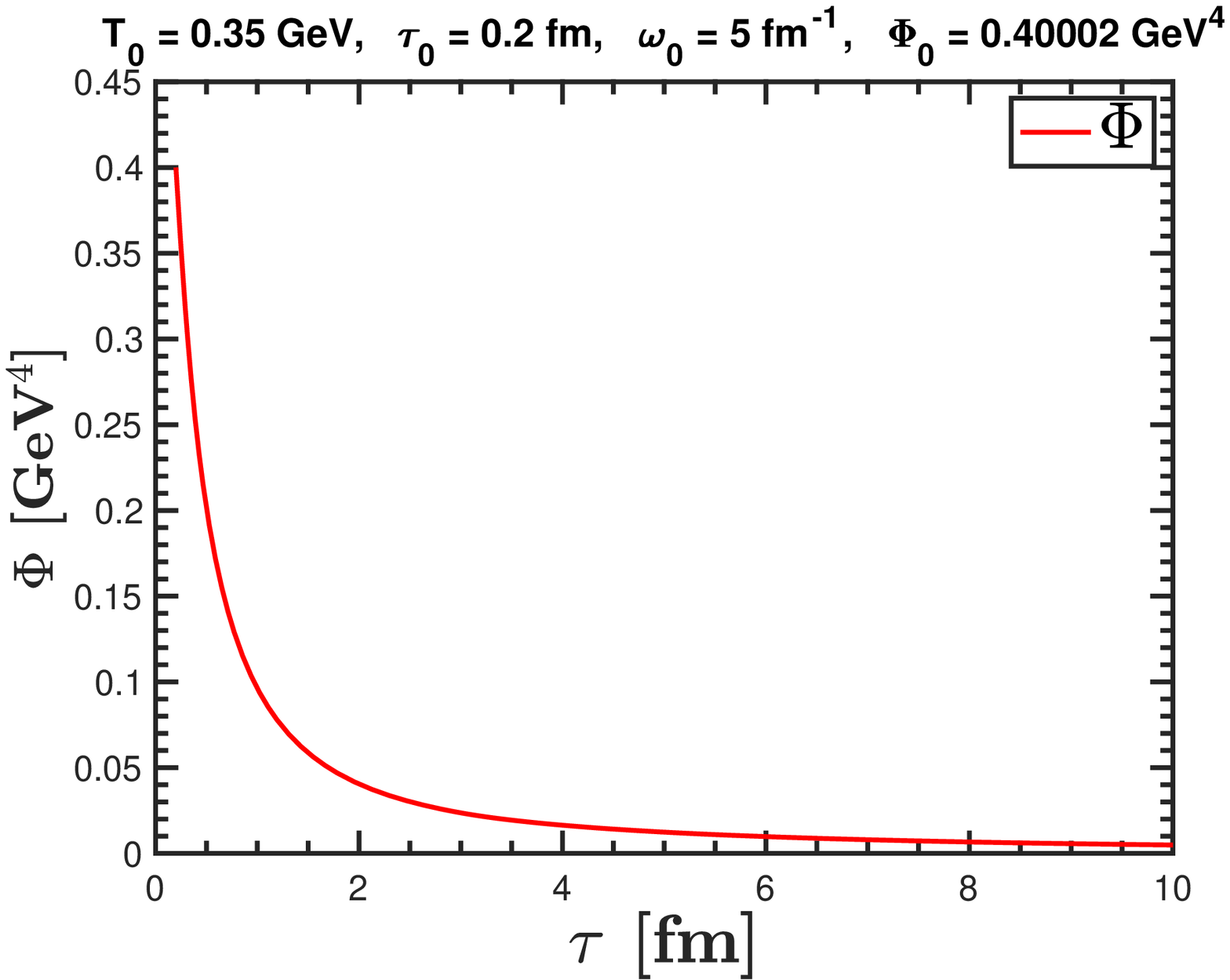}
\includegraphics[scale = 0.32]{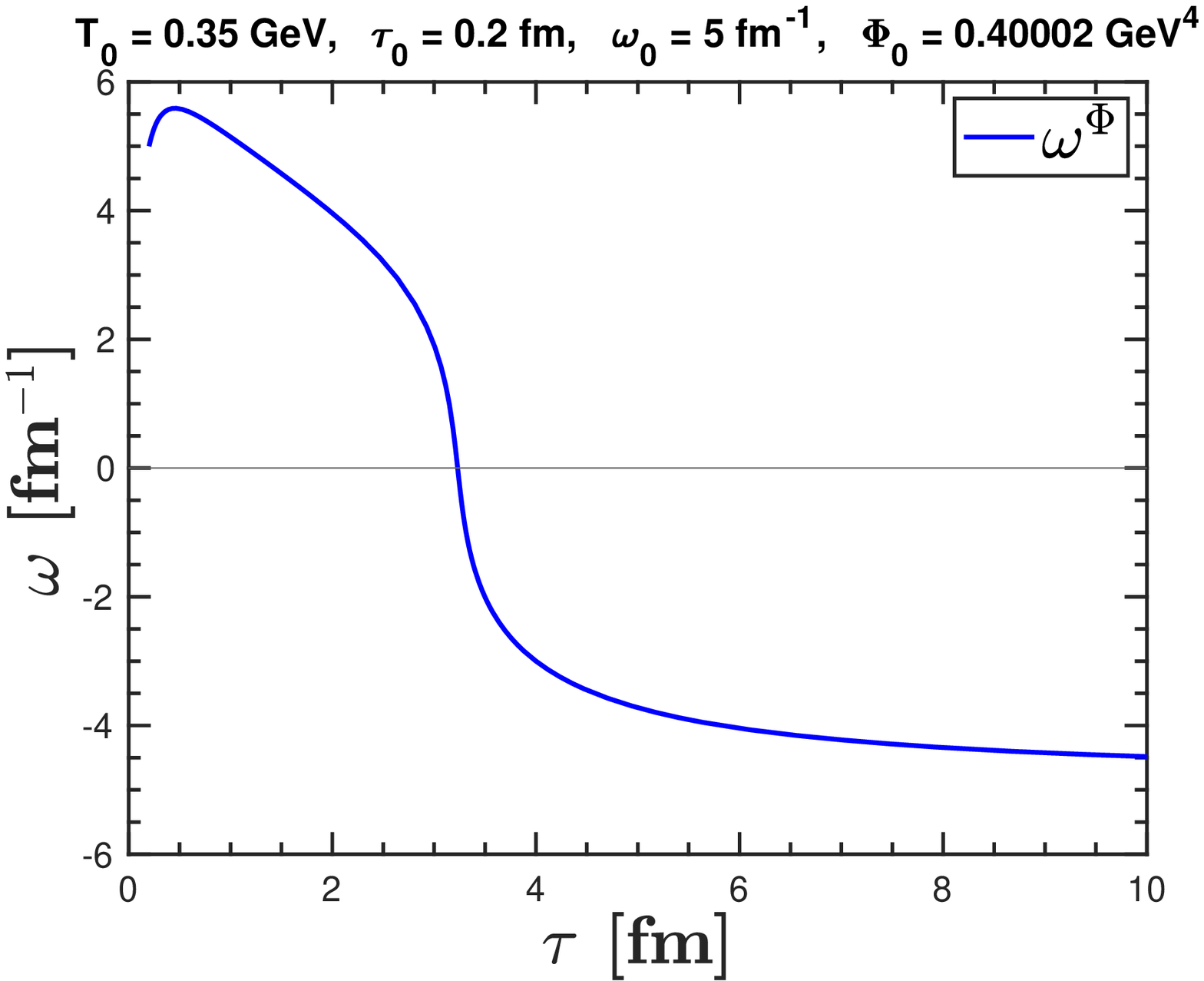}
\caption{(Color Online) {\bf Left to Right:} Temperature (T), viscous term ($\Phi$) and vorticity ($\omega$) are plotted, respectively, against time $\tau$ with the initial conditions: {\bf T = 0.35 GeV, $\tau_{0}$ = 0.2 fm, $\omega_{0}$ = 5.0 fm$^{-1}$, $\Phi_{0}$ = 0.40002 GeV$^4$}.}
\label{fig7}
\end{figure*}

\begin{figure*}[ht!]
\centering
\includegraphics[scale = 0.32]{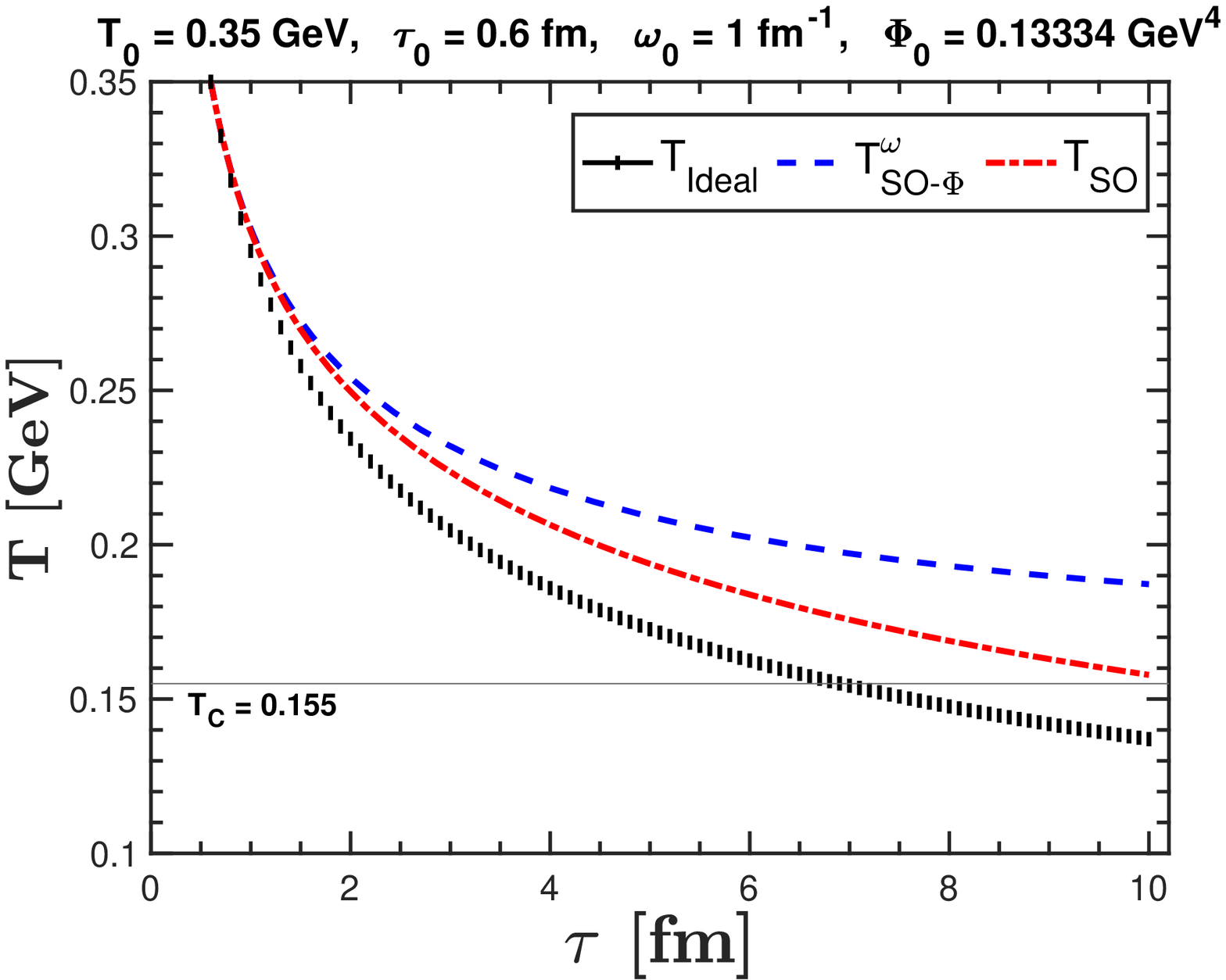}
\includegraphics[scale = 0.32]{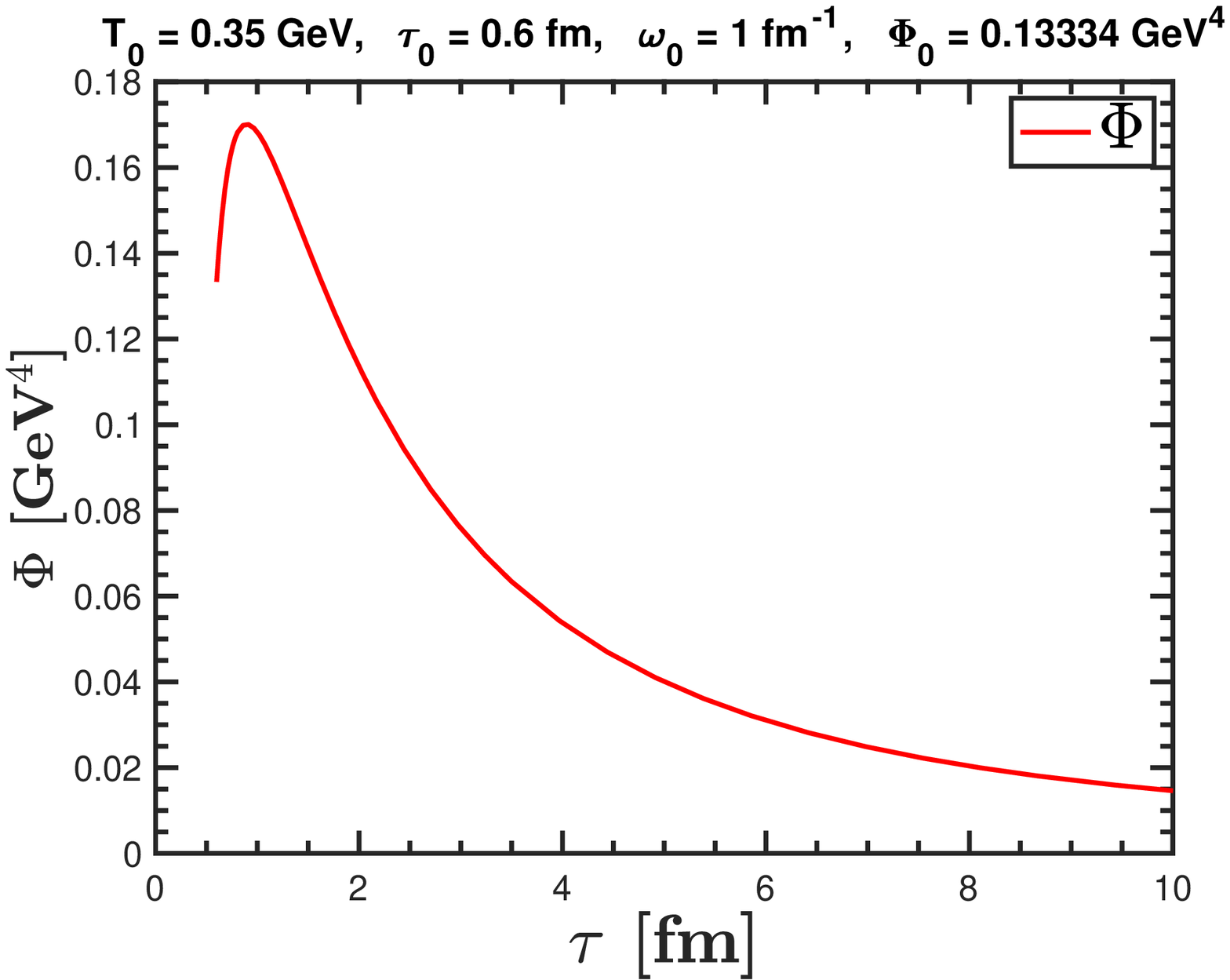}
\includegraphics[scale = 0.32]{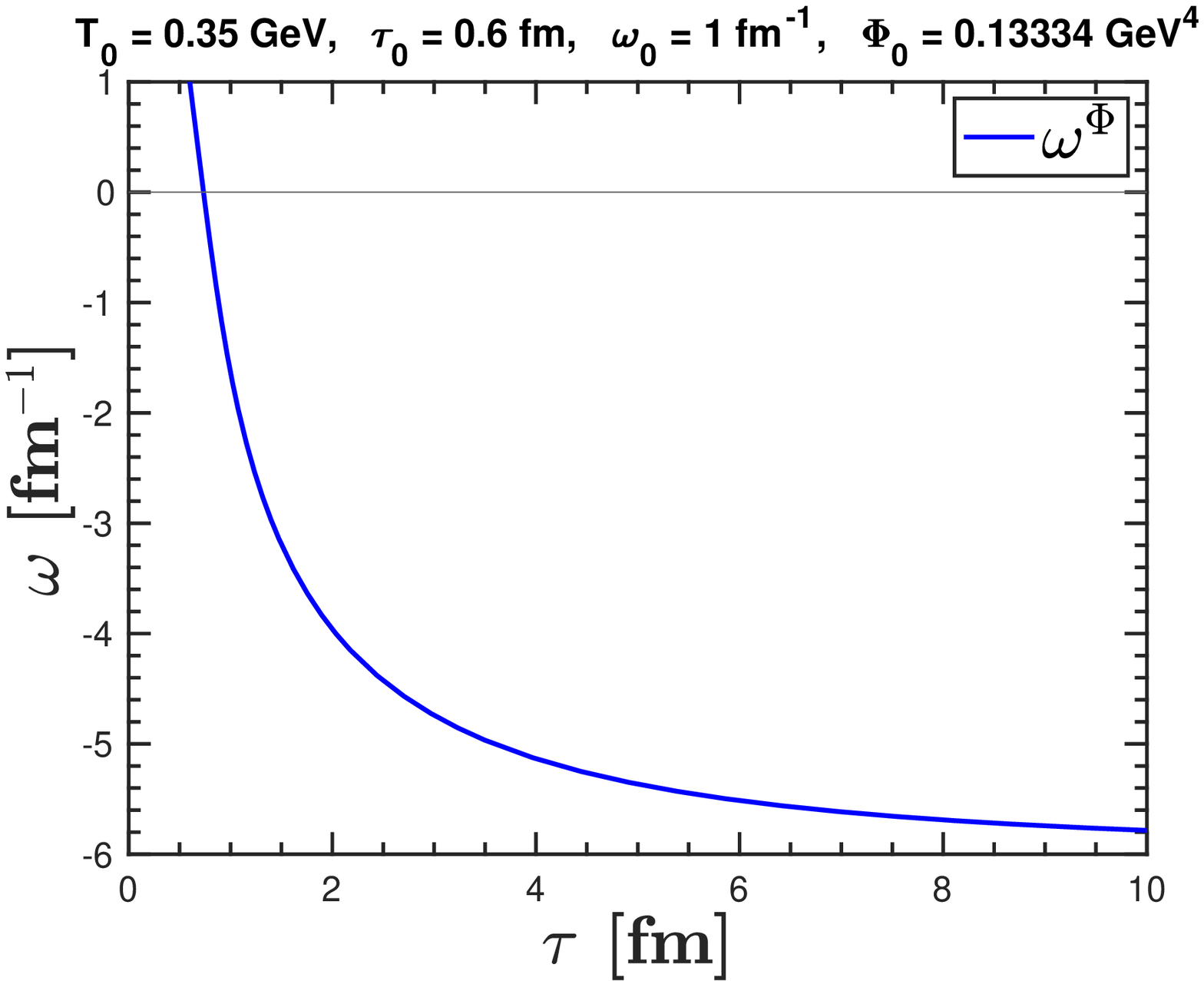}
\caption{(Color Online) {\bf Left to Right:} Temperature (T), viscous term ($\Phi$) and vorticity ($\omega$) are plotted, respectively, against time $\tau$ with the initial conditions: {\bf T = 0.35 GeV, $\tau_{0}$ = 0.6 fm, $\omega_{0}$ = 1.0 fm$^{-1}$, $\Phi_{0}$ = 0.13334 GeV$^4$}.}
\label{fig8}
\end{figure*}

\begin{figure*}[ht!]
\centering
\includegraphics[scale = 0.32]{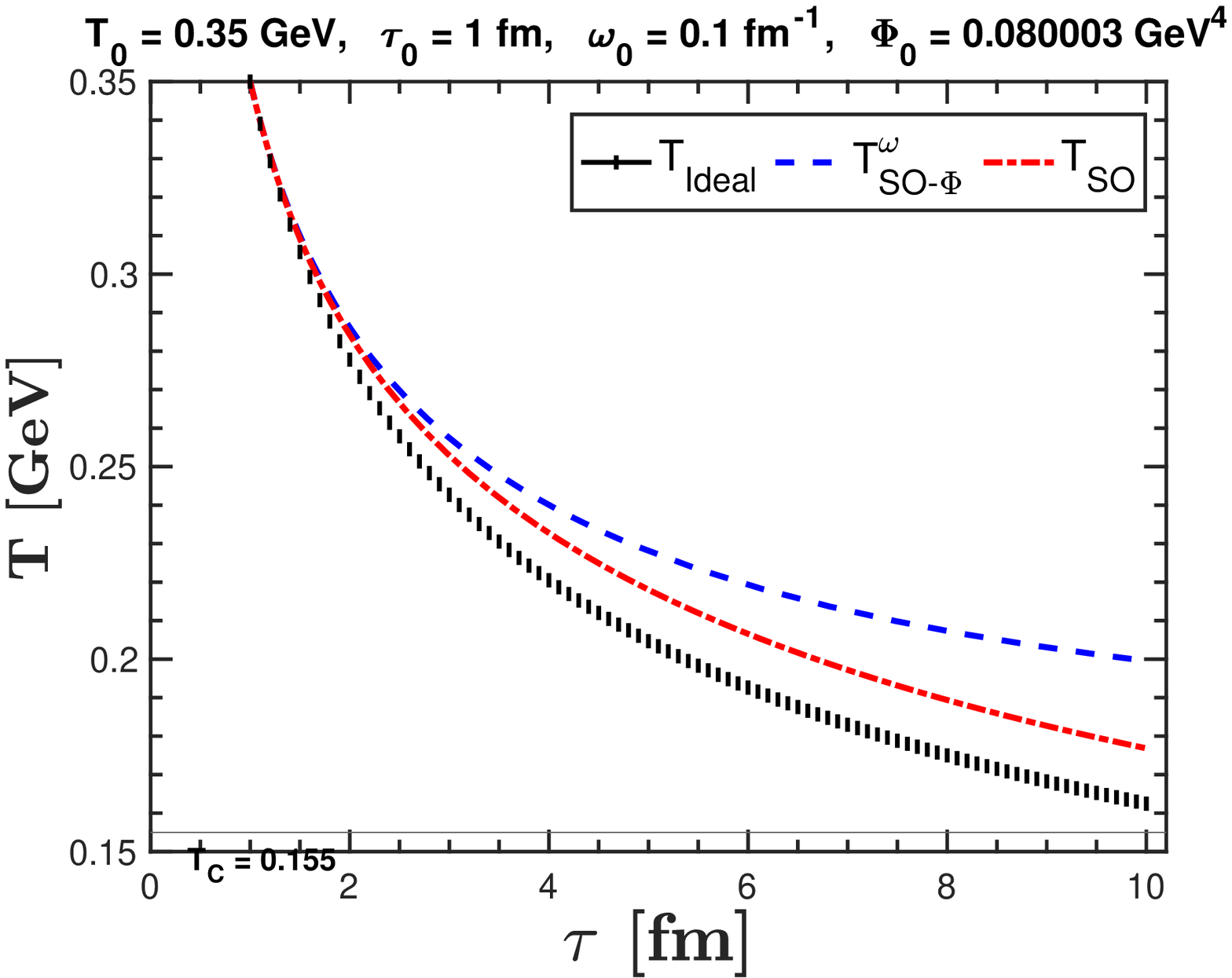}
\includegraphics[scale = 0.32]{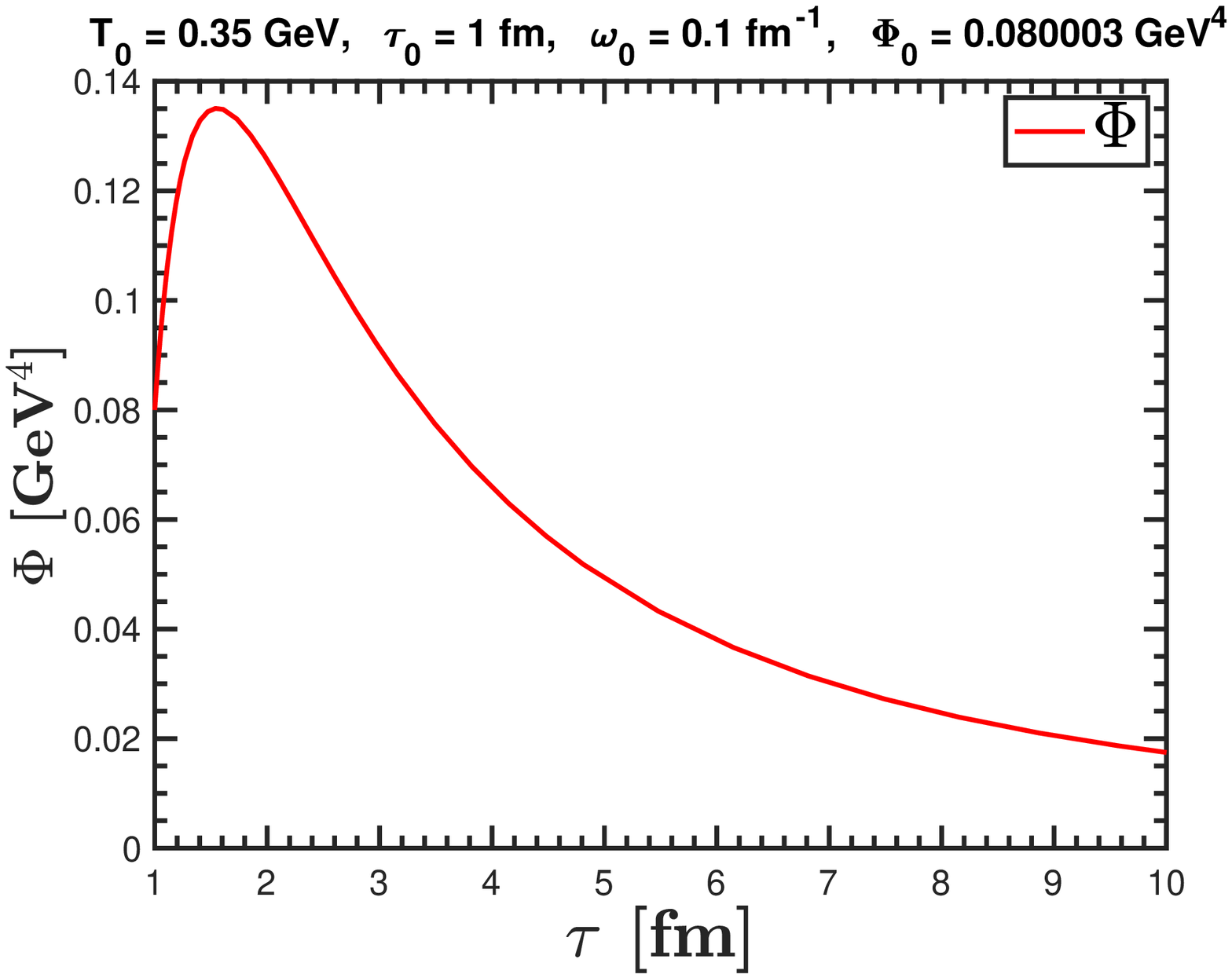}
\includegraphics[scale = 0.32]{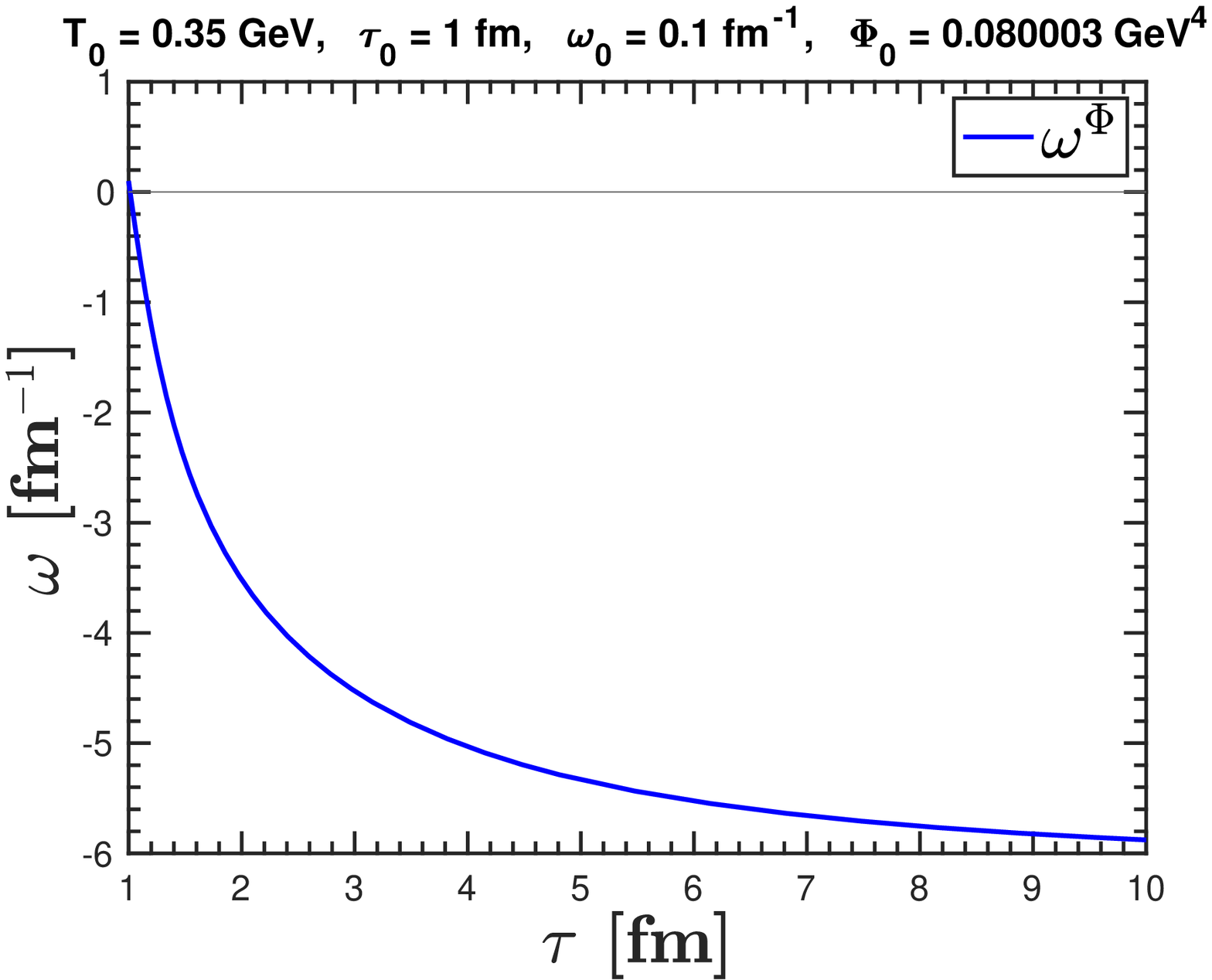}
\caption{(Color Online) {\bf Left to Right:} Temperature (T), viscous term ($\Phi$) and vorticity ($\omega$) are plotted, respectively, against time $\tau$ with the initial conditions: {\bf T = 0.35 GeV, $\tau_{0}$ = 1.0 fm, $\omega_{0}$ = 0.1 fm$^{-1}$, $\Phi_{0}$ = 0.080003 GeV$^4$}.}
\label{fig9}
\end{figure*}

In Fig.~\ref{fig7}, the combination of vorticity and viscosity is shown for large initial vorticity and a small thermalization time. T$_{SO-\Phi}^{\omega}$ cools down slightly faster than T$_{SO}$ due to the opposition of viscosity to changes in vorticity direction. Positive initial vorticity results in faster cooling, almost following the ideal rate. However, the presence of viscosity in the rotating fluid slows down T$_{SO-\Phi}^{\omega}$ cooling compared to T$_\text{Ideal}$. 
The impact of viscosity on vorticity change can be observed in the variation of $\omega^{\Phi}$ as shown in Fig.~\ref{fig7}. The $\Phi$ evolution
shown  in Fig.~\ref{fig7} follows similar  pattern as the corresponding result shown  in Fig.~\ref{fig1}. 
In Fig.~\ref{fig8}, the combined dynamics of $\omega$ and $\Phi$ are shown for an initial vorticity value of $\omega_{0}$ = 1.0 fm$^{-1}$ and a thermalization time of $\tau_0$ = 0.6 fm. A relatively smaller initial viscosity value is insufficient to resist rapid changes in vorticity. Smaller vorticities easily adapt to the changes imposed by the evolving medium. Negative vorticity slows down the cooling rate, resulting in T$_{SO-\Phi}^{\omega}$ cooling at a slower rate than T$_{SO}$. In Fig.~\ref{fig8}, the coupling between $\Phi$ and $\omega$ leads to the activation of a saturation point in the diffusion rate of $\omega^{\Phi}$. The $\Phi$ evolution plot in Fig.~\ref{fig8} follows a similar pattern to its corresponding plot in Fig.~\ref{fig2}. Fig.~\ref{fig9} follows similar explanation as Fig.~\ref{fig8}, with cooling becoming even slower for T$_{SO-\Phi}^{\omega}$ due to a very small initial vorticity and a large thermalization time.\\


\begin{figure*}[ht!]
\centering
\includegraphics[scale = 0.32]{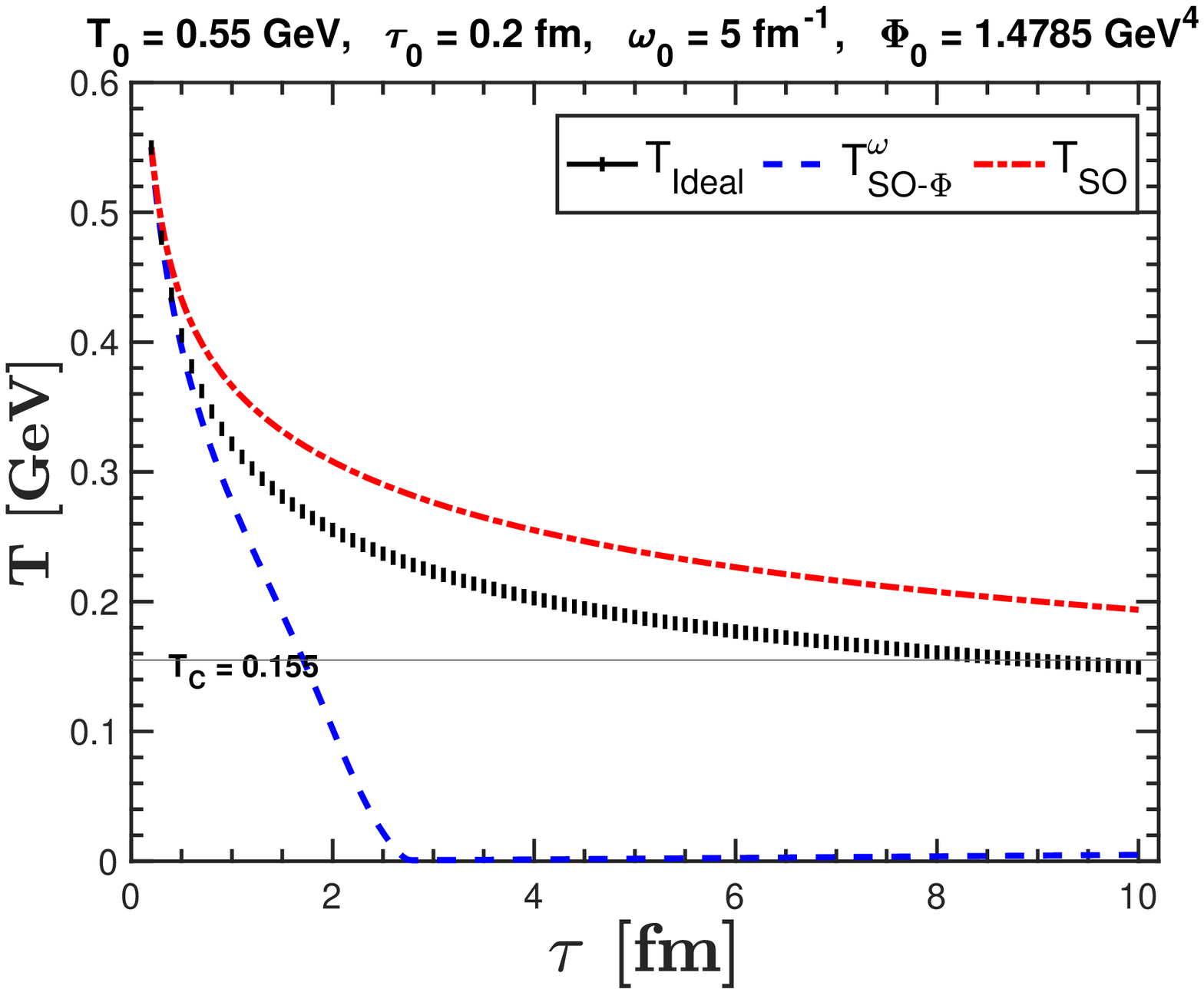}
\includegraphics[scale = 0.32]{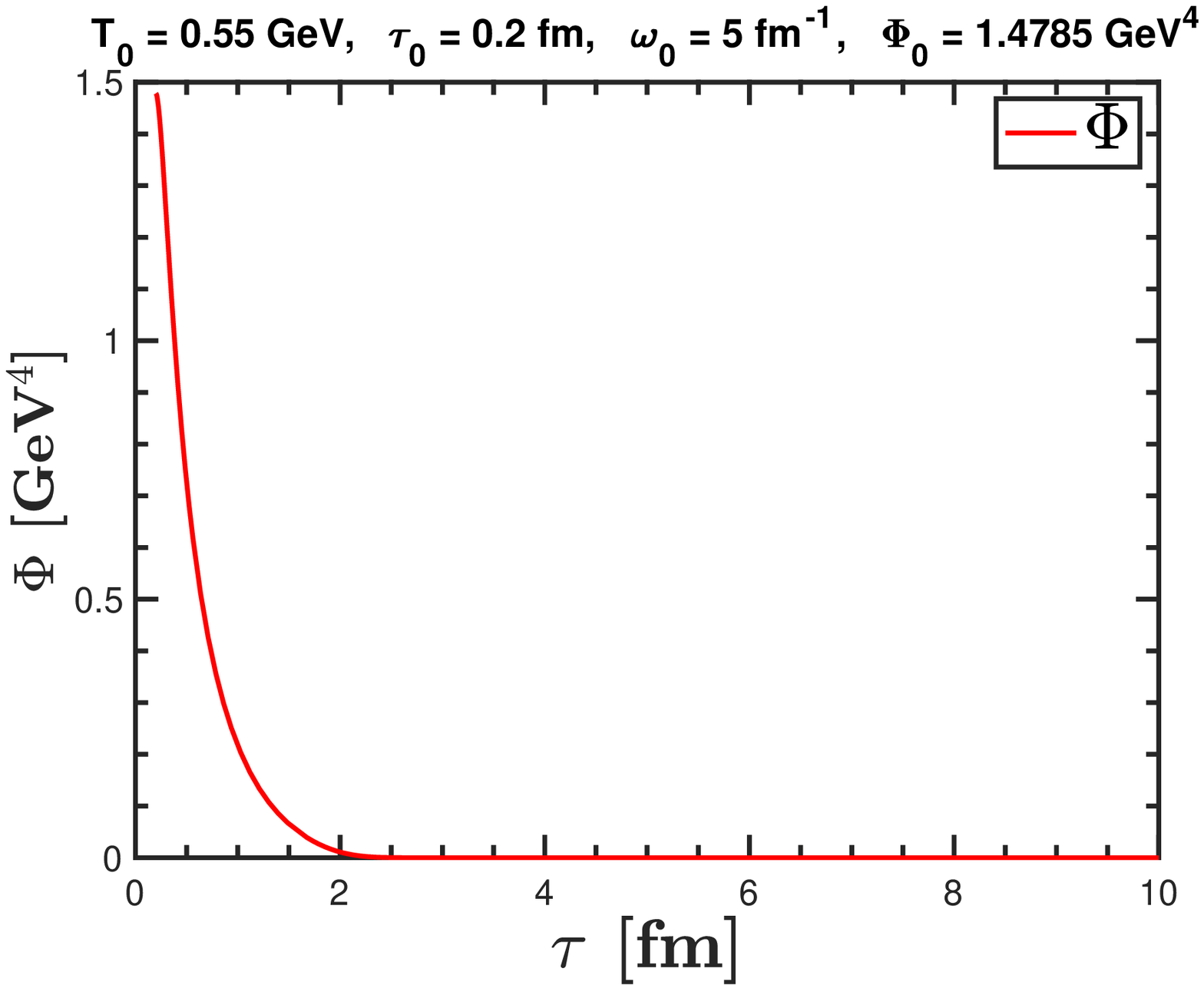}
\includegraphics[scale = 0.32]{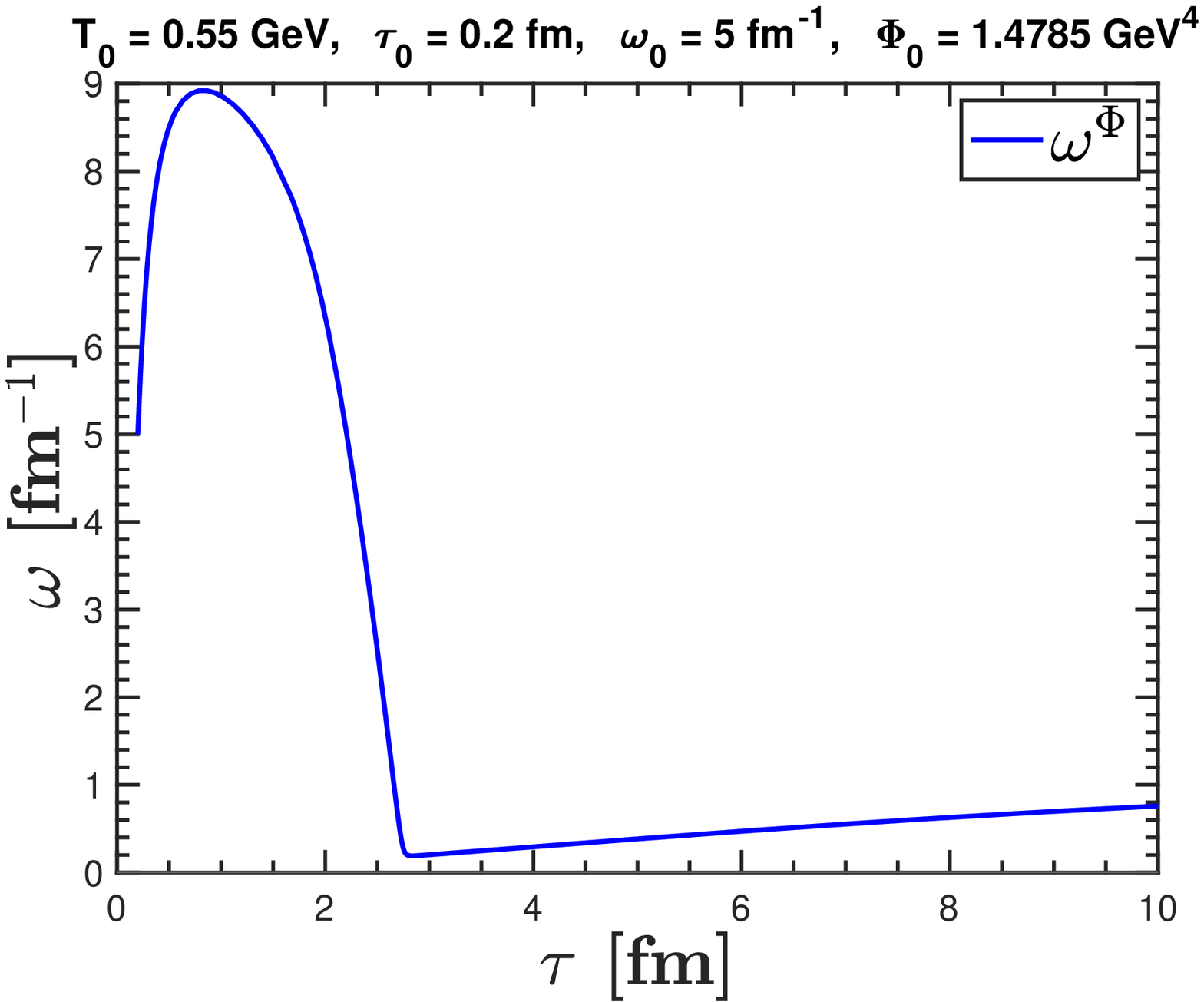}
\caption{(Color Online) {\bf Left to Right:} Temperature (T), viscous term ($\Phi$) and vorticity ($\omega$) are plotted, respectively, against time $\tau$ with the initial conditions: {\bf T = 0.55 GeV, $\tau_{0}$ = 0.2 fm, $\omega_{0}$ = 5.0 fm$^{-1}$, $\Phi_{0}$ = 1.4785 GeV$^4$}.}
\label{fig10}
\end{figure*}

\begin{figure*}[ht!]
\centering
\includegraphics[scale = 0.32]{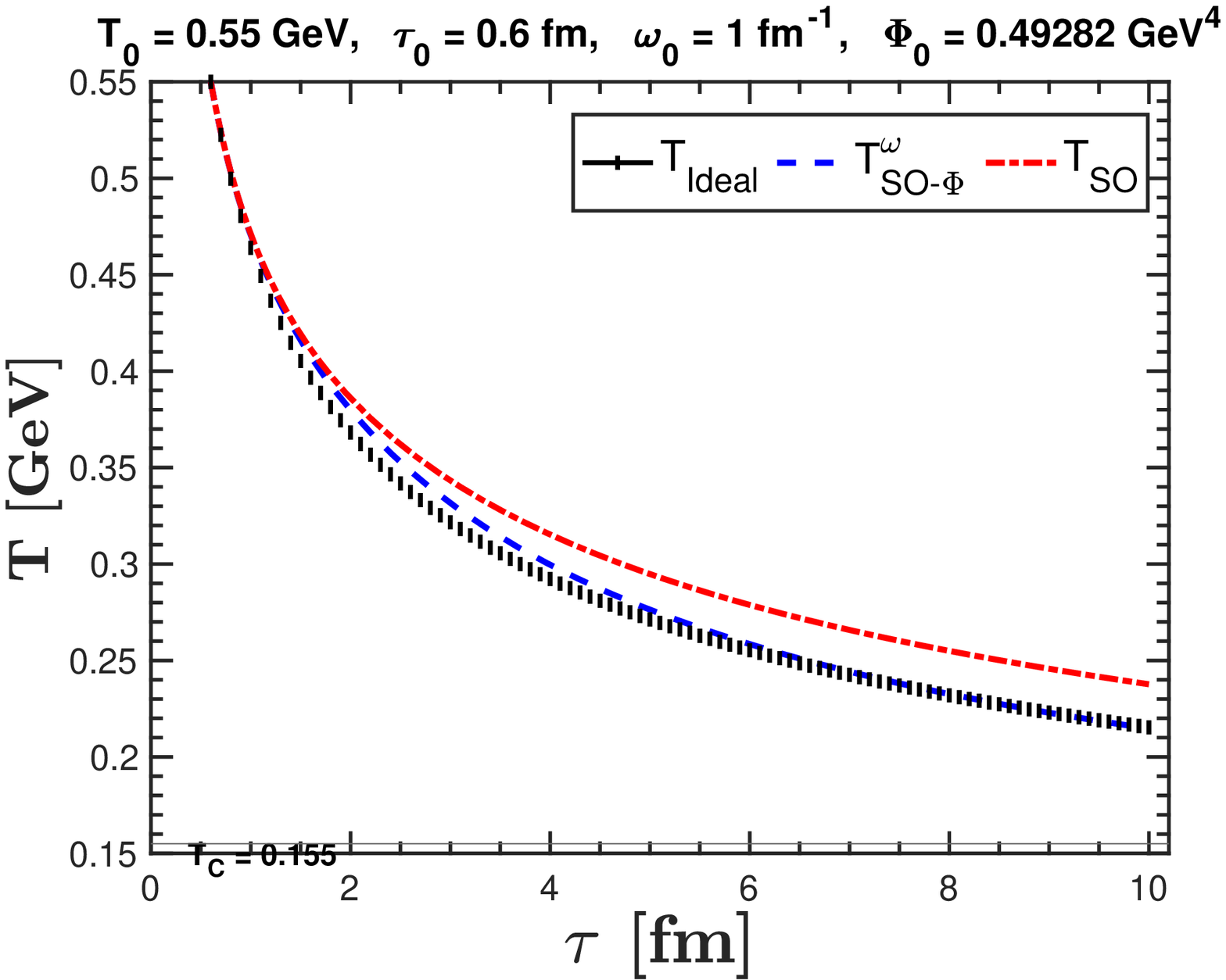}
\includegraphics[scale = 0.32]{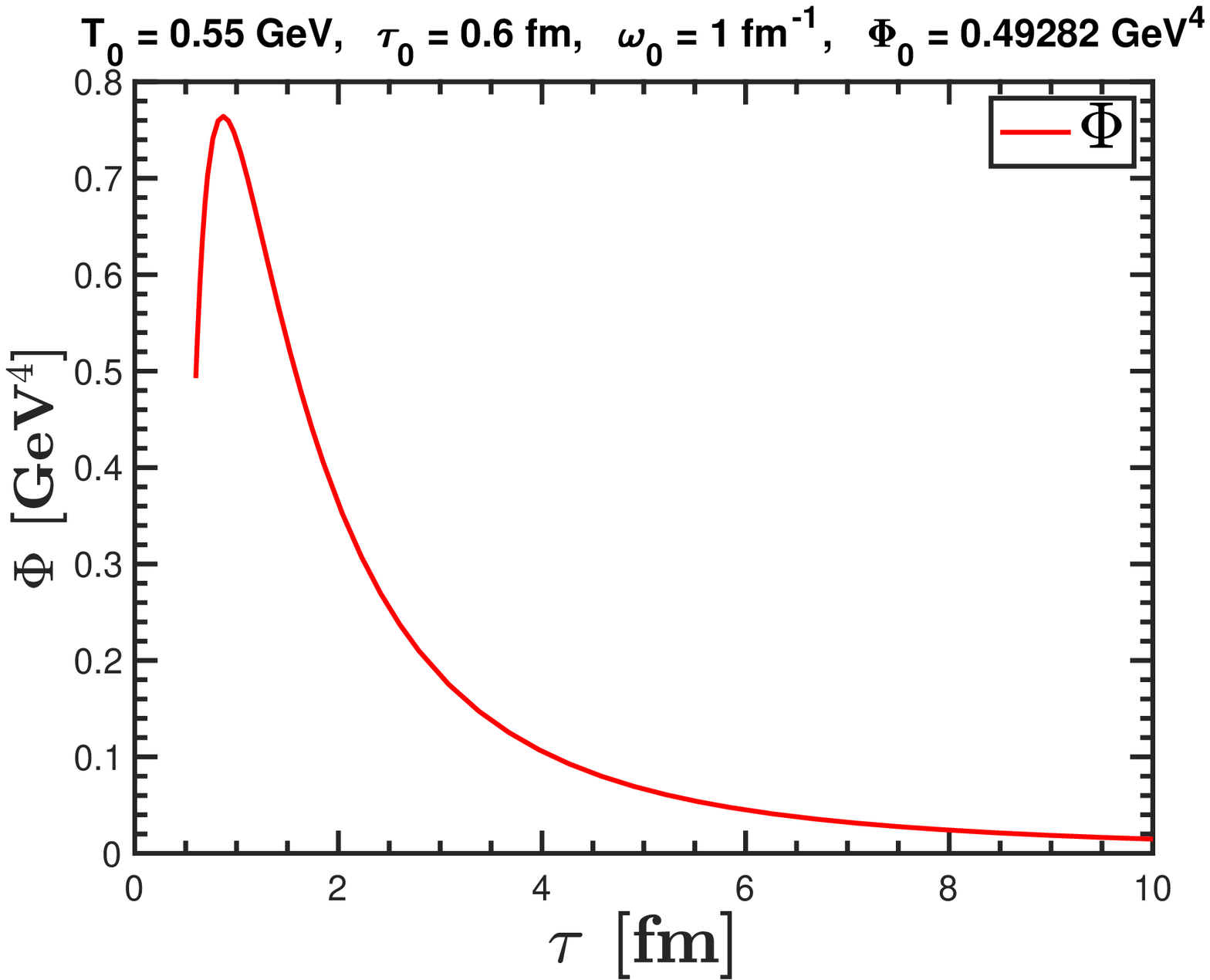}
\includegraphics[scale = 0.32]{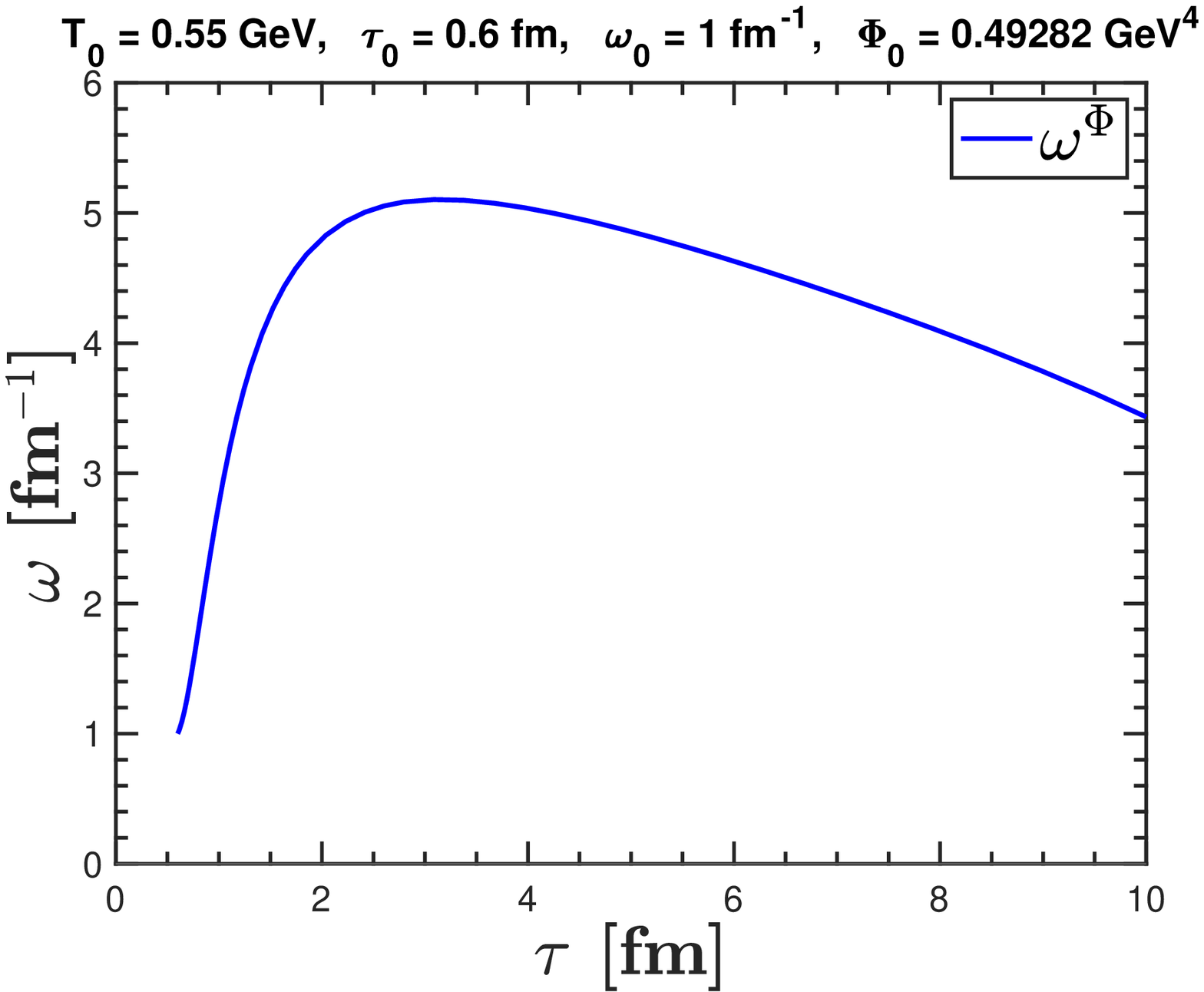}
\caption{(Color Online) {\bf Left to Right:} Temperature (T), viscous term ($\Phi$) and vorticity ($\omega$) are plotted, respectively, against time $\tau$ with the initial conditions: {\bf T = 0.55 GeV, $\tau_{0}$ = 0.6 fm, $\omega_{0}$ = 1.0 fm$^{-1}$, $\Phi_{0}$ = 0.49282 GeV$^4$}.}
\label{fig11}
\end{figure*}

\begin{figure*}[ht!]
\centering
\includegraphics[scale = 0.32]{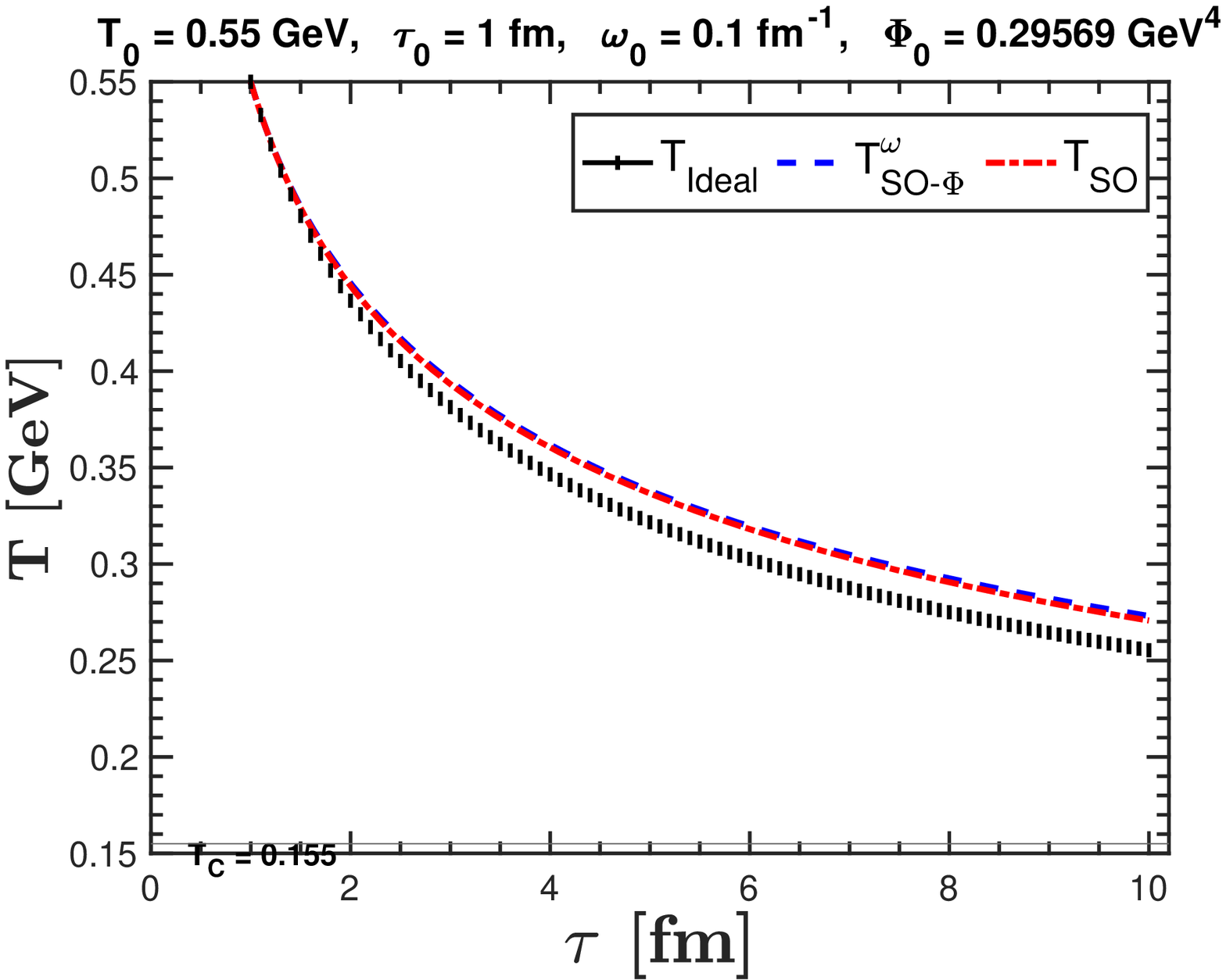}
\includegraphics[scale = 0.32]{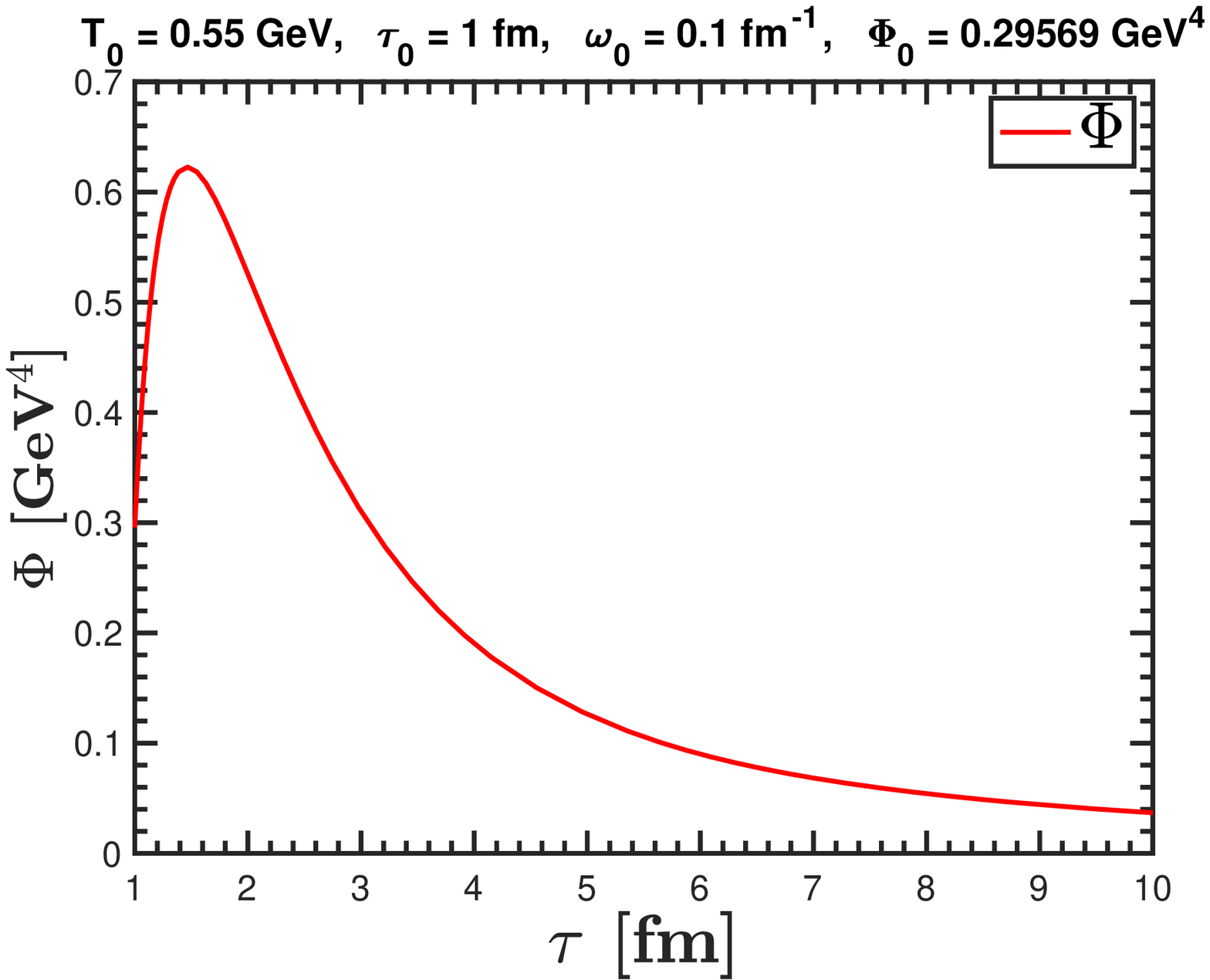}
\includegraphics[scale = 0.32]{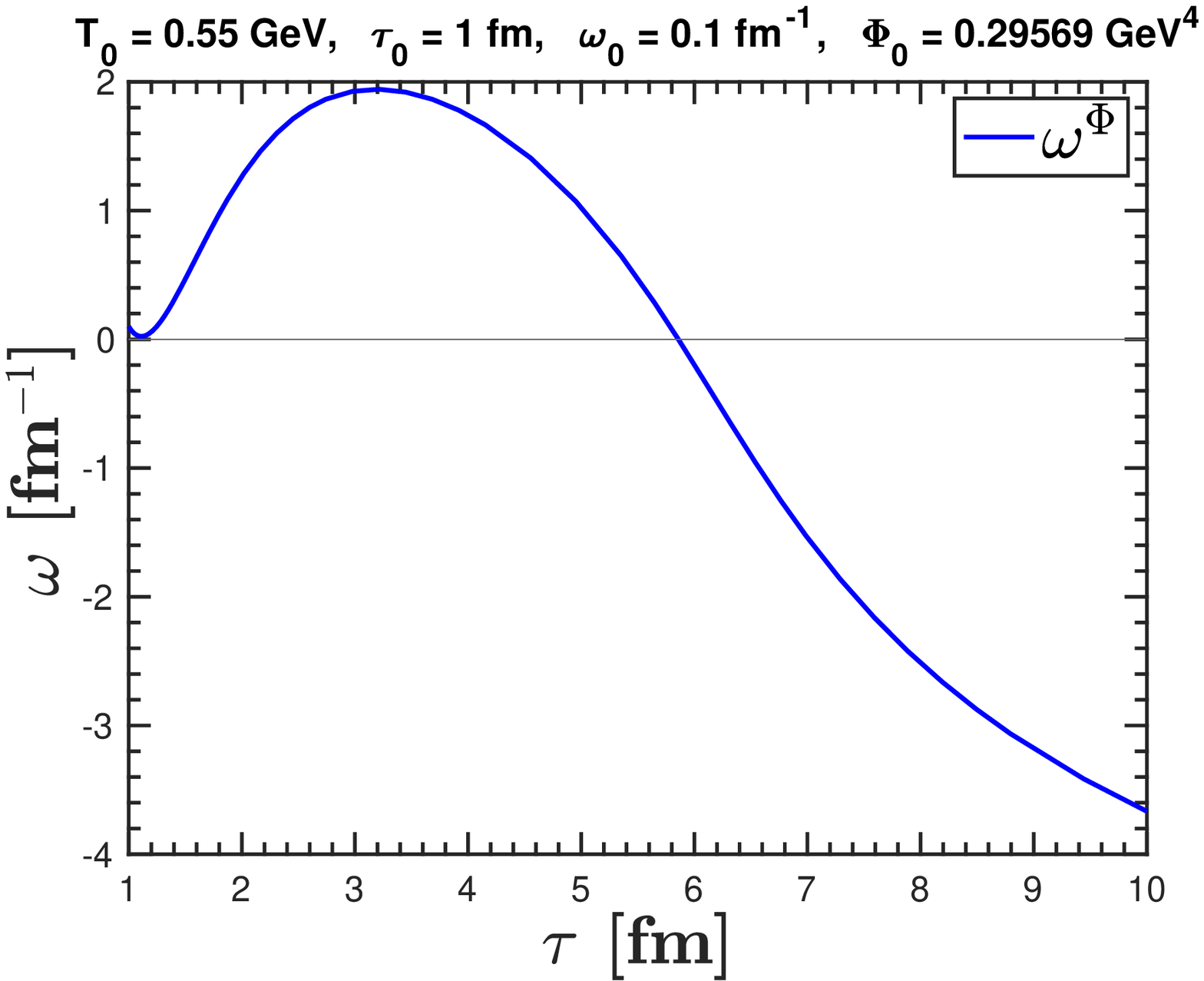}
\caption{(Color Online) {\bf Left to Right:} Temperature (T), viscous term ($\Phi$) and vorticity ($\omega$) are plotted, respectively, against time $\tau$ with the initial conditions: {\bf T = 0.55 GeV, $\tau_{0}$ = 1.0 fm, $\omega_{0}$ = 0.1 fm$^{-1}$, $\Phi_{0}$ = 0.29569 GeV$^4$}.}
\label{fig12}
\end{figure*}

What happens if, along with initial vorticity, the initial temperature is large and the thermalization time is short? The answer to this question is given in Fig.~\ref{fig10}; 
it implies that at low $\tau_{0}$ and high $T_{0}$, the initial viscosity is very high. 
As discussed earlier, due to viscosity coupling with vorticity and their large initial values make cooling faster; if the medium temperature is also high, 
cooling becomes even faster. As all these mentioned conditions are fulfilled in Fig.~\ref{fig10}, the cooling for T$_{SO-\Phi}^{\omega}$ becomes very 
fast that medium gets exhausted much before T$_\text{Ideal}$. The combined effect of large viscosity and the high temperature does not let the evolving medium 
change the direction of the vorticity, as shown in the $\omega-\tau$ plot of Fig.~\ref{fig10}, where $\omega$ is always positive and vanishes when T$\rightarrow 0$. 
Due to this coupling, $\Phi$ gets dissipated earlier.\\

Fig.~\ref{fig11} depicts that decrease of $\omega_{0}$ and increase of $\tau_0$, makes the variation of T$_{SO-\Phi}^{\omega}$ and T$_\text{Ideal}$ similar. 
However, T$_{SO-\Phi}^{\omega}$  remains faster in the region, which represents faster cooling than T$_{SO}$. While $\omega_{0}$ and $\Phi_{0}$ are small, the high $T_0$  and $\Phi_{0}$ together support vorticity to sustain its initial direction till temperature and viscosity become inefficient to restrict the change. The change in the cooling rate corresponding to very small vorticity and very large thermalization time and temperature
is depicted in Fig.~\ref{fig12}.  
In this scenario, the variation of T$_{SO-\Phi}^{\omega}$ and T$_{SO}$ turns out to be similar. Here the impact of viscosity is minimal on vorticity. The 
high initial temperature is a dominating factor in this case. Therefore a small rise in $\omega$ for a short duration is observed in Fig.~\ref{fig12}. 
Later it diffuses in the opposite direction; as a result, the cooling rate corresponding to $T_{SO-\Phi}^{\omega}$  $\sim$  $T_{SO}$ and it becomes slightly slower than $T_{SO}$ around $\tau >$ 7.0 fm.\\


\begin{figure*}[ht!]
\centering
\includegraphics[scale = 0.32]{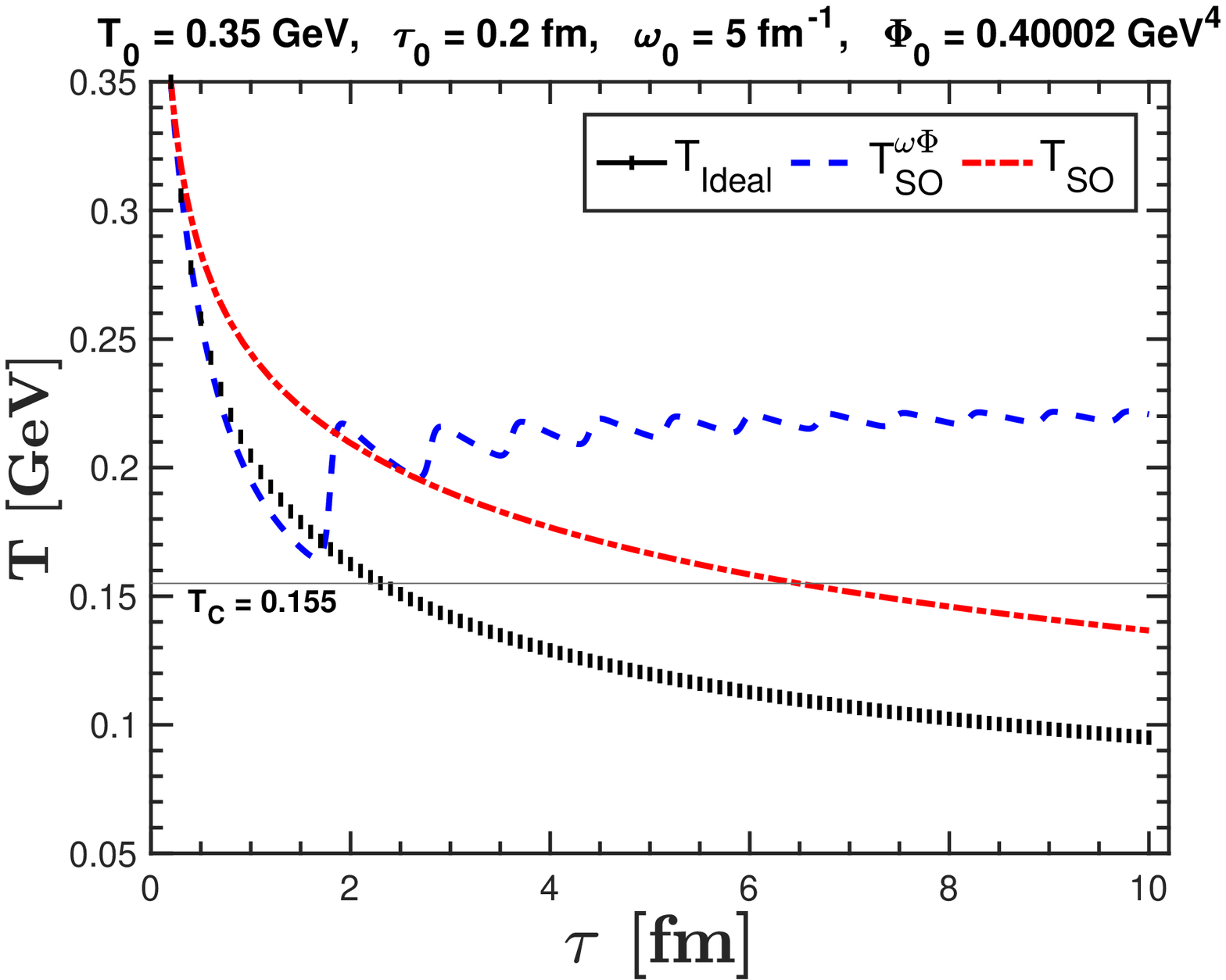}
\includegraphics[scale = 0.32]{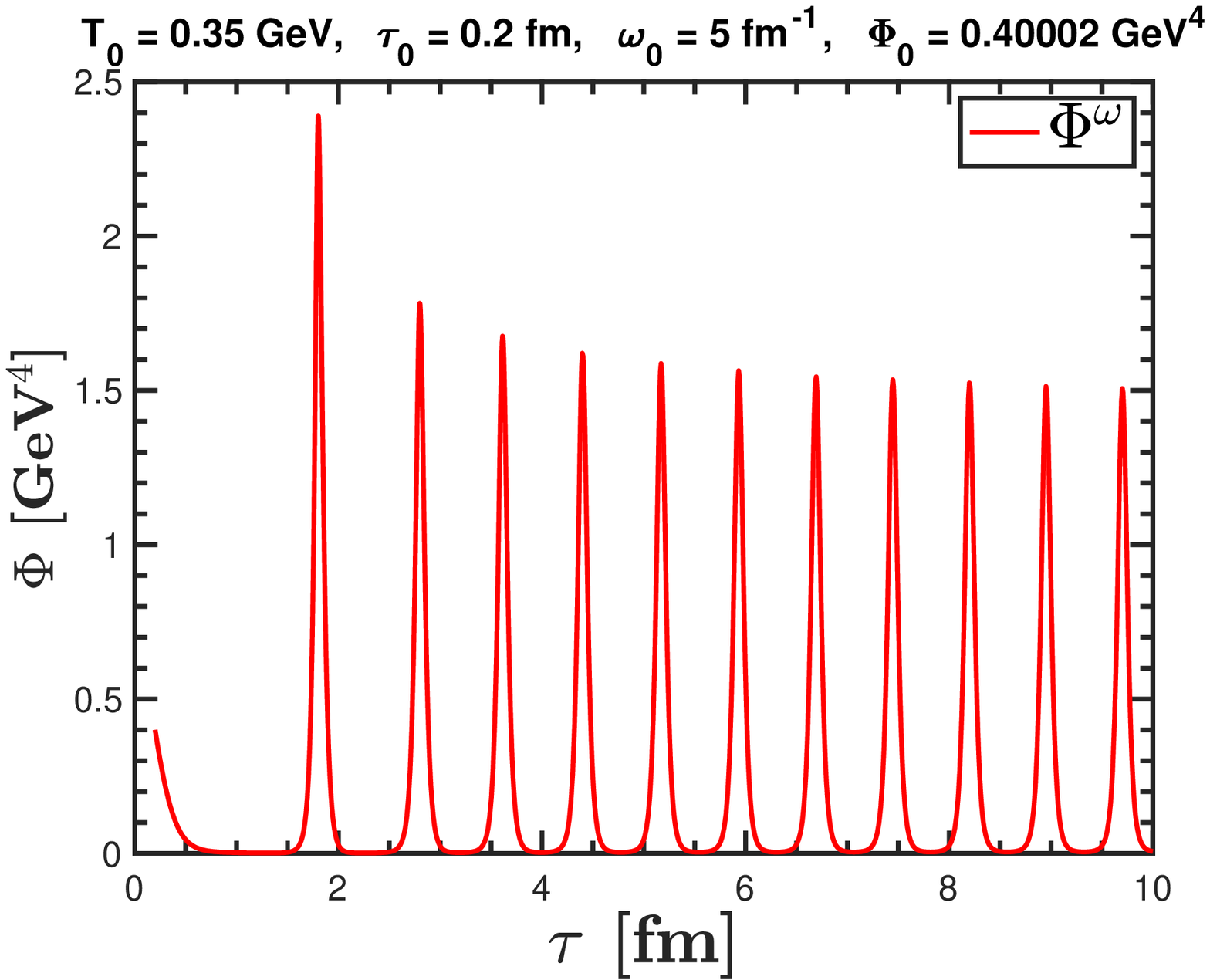}
\includegraphics[scale = 0.32]{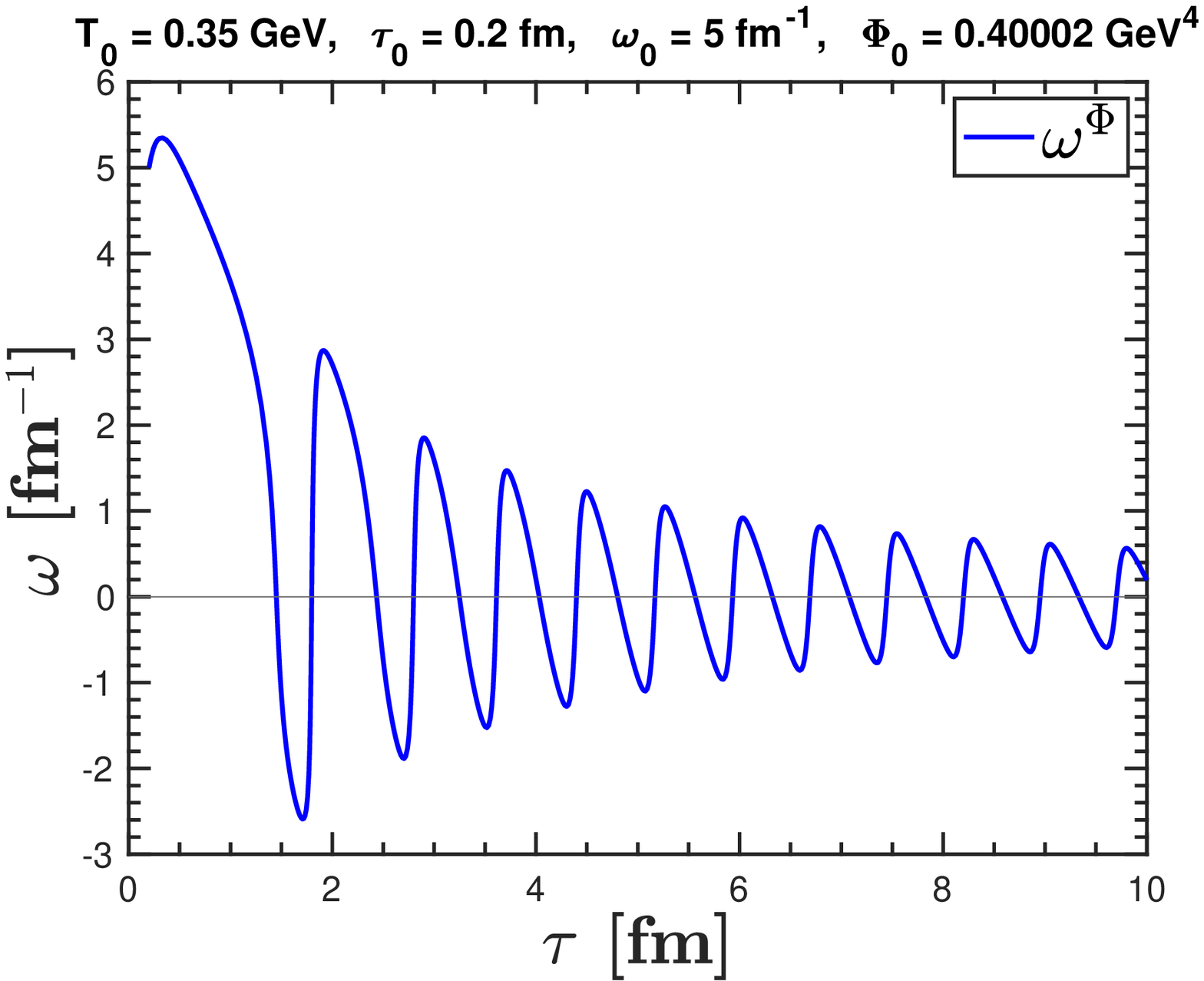}
\caption{(Color Online) {\bf Left to Right:} Temperature (T), viscous term ($\Phi$) and vorticity ($\omega$) are plotted, respectively, against time $\tau$ with the initial conditions: {\bf T = 0.35 GeV, $\tau_{0}$ = 0.2 fm, $\omega_{0}$ = 5.0 fm$^{-1}$, $\Phi_{0}$ = 0.40002 GeV$^4$}.}
\label{fig13}
\end{figure*}
\begin{figure*}[ht!]
\centering
\includegraphics[scale = 0.32]{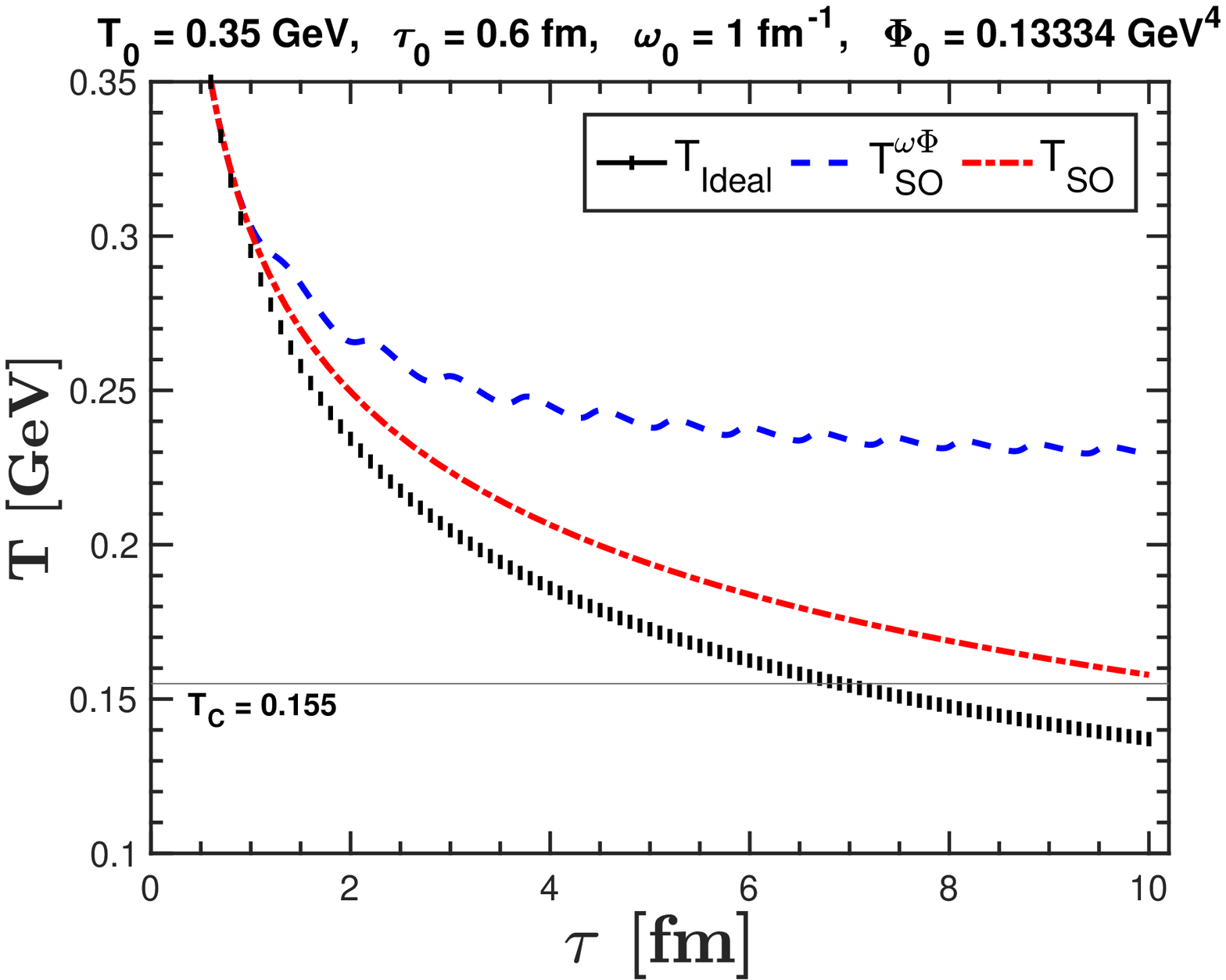}
\includegraphics[scale = 0.32]{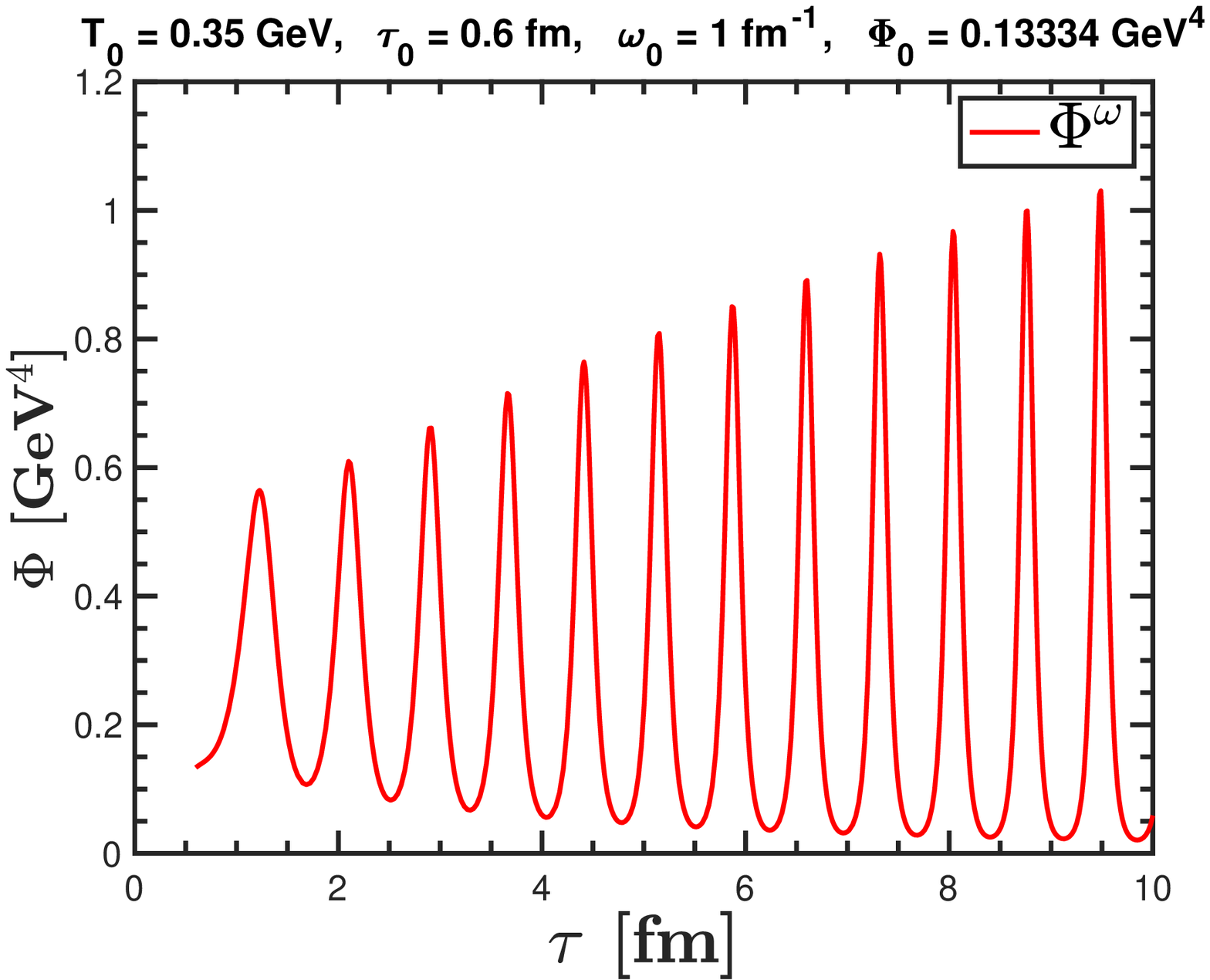}
\includegraphics[scale = 0.32]{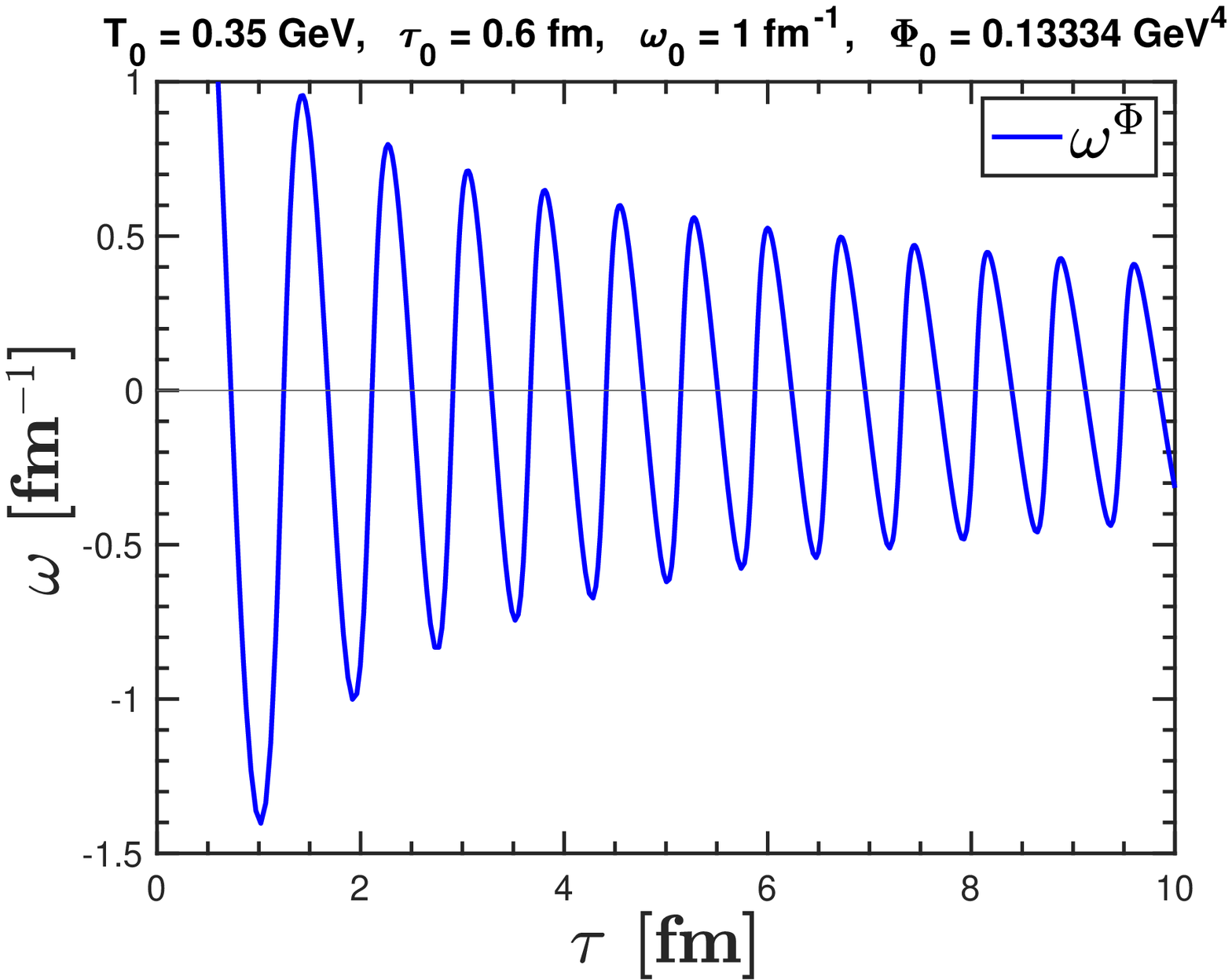}
\caption{(Color Online) {\bf Left to Right:} Temperature (T), viscous term ($\Phi$) and vorticity ($\omega$) are plotted, respectively, against time $\tau$ with the initial conditions: {\bf T = 0.35 GeV, $\tau_{0}$ = 0.6 fm, $\omega_{0}$ = 1.0 fm$^{-1}$, $\Phi_{0}$ = 0.13334 GeV$^4$}.}
\label{fig14}
\end{figure*}
\begin{figure*}[ht!]
\centering
\includegraphics[scale = 0.32]{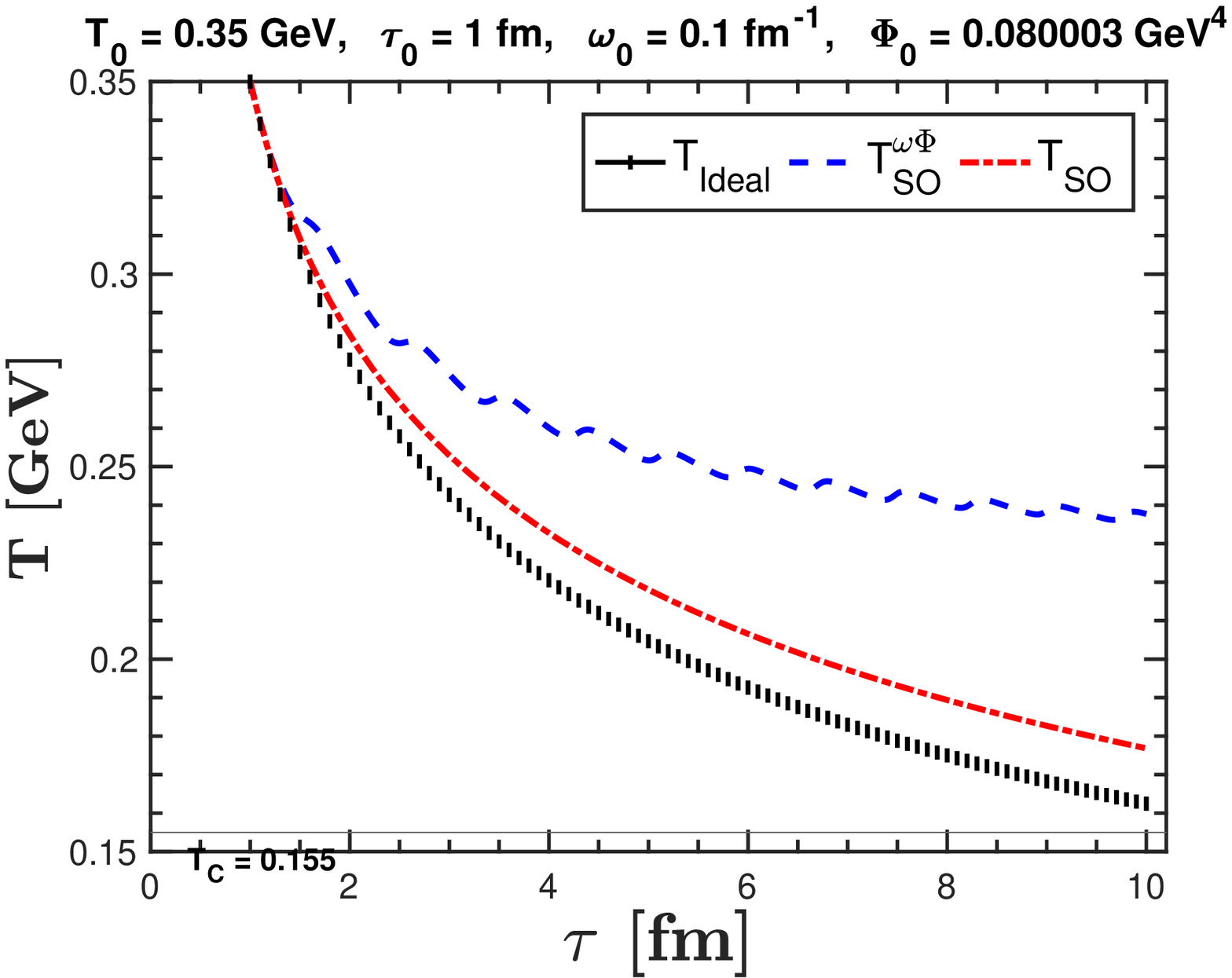}
\includegraphics[scale = 0.32]{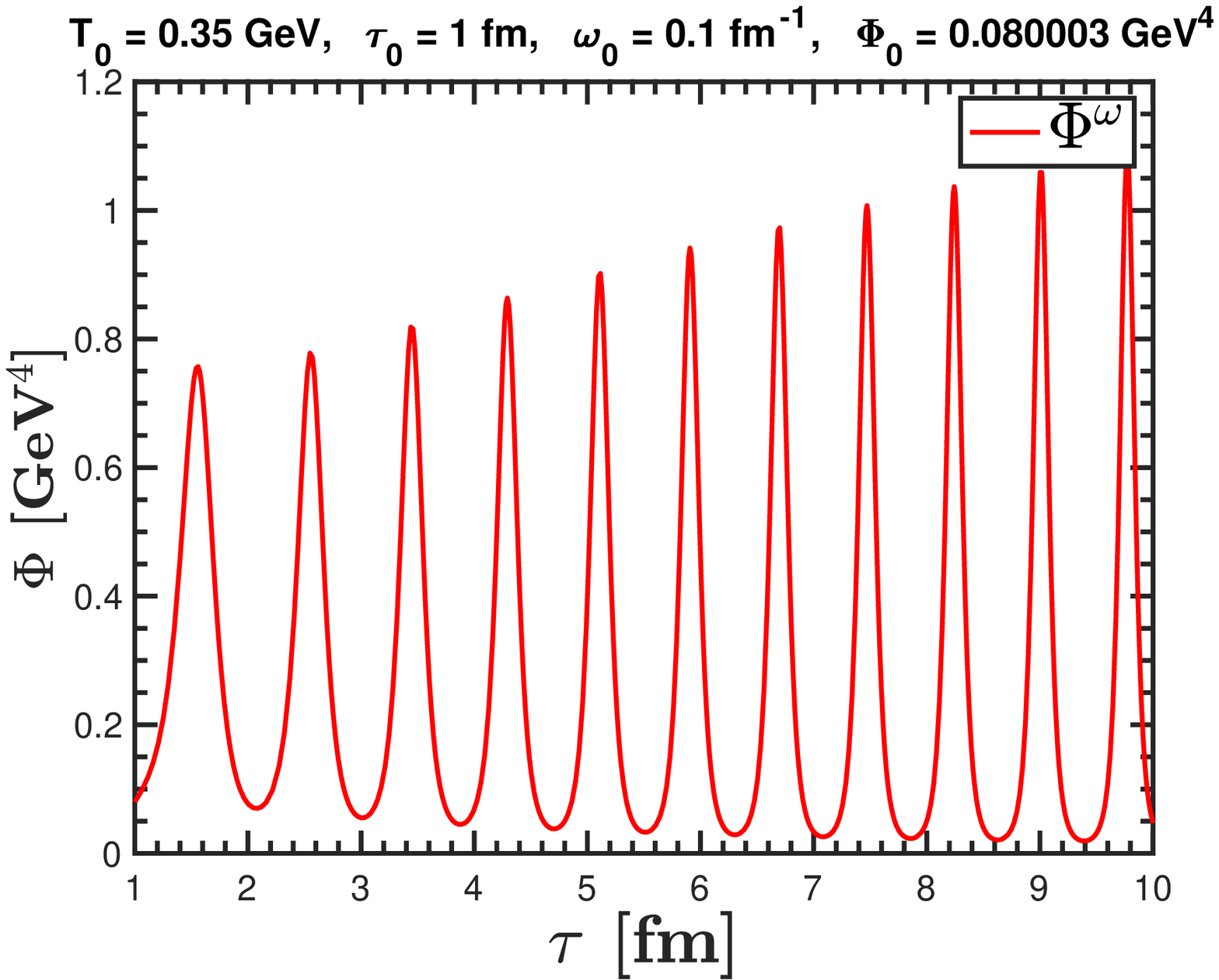}
\includegraphics[scale = 0.32]{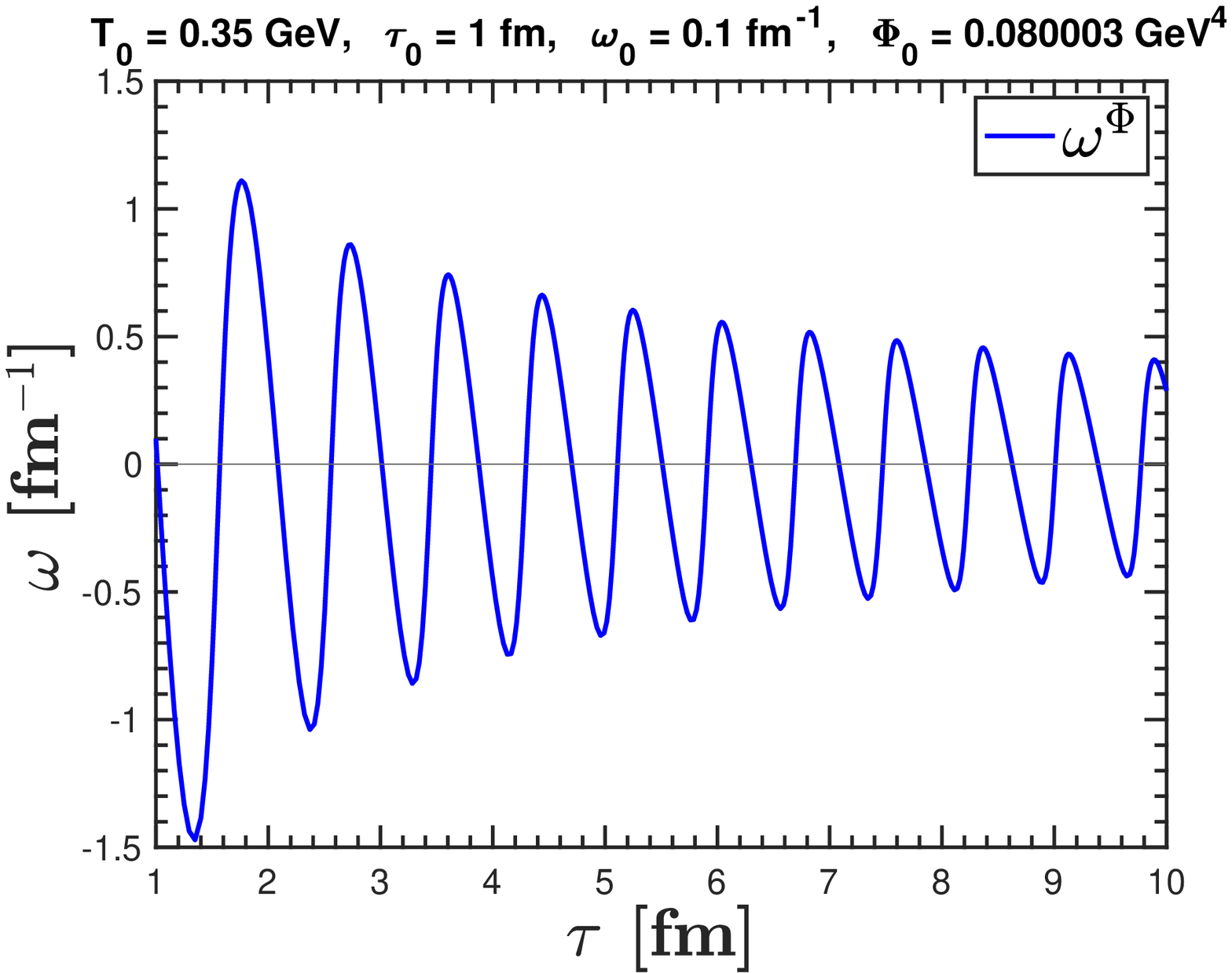}
\caption{(Color Online) {\bf Left to Right:} Temperature (T), viscous term ($\Phi$) and vorticity ($\omega$) are plotted, respectively, against time $\tau$ with the initial conditions: {\bf T = 0.35 GeV, $\tau_{0}$ = 1.0 fm, $\omega_{0}$ = 0.1 fm$^{-1}$, $\Phi_{0}$ = 0.080003 GeV$^4$}.}
\label{fig15}
\end{figure*}

\begin{figure*}[ht!]
\centering
\includegraphics[scale = 0.32]{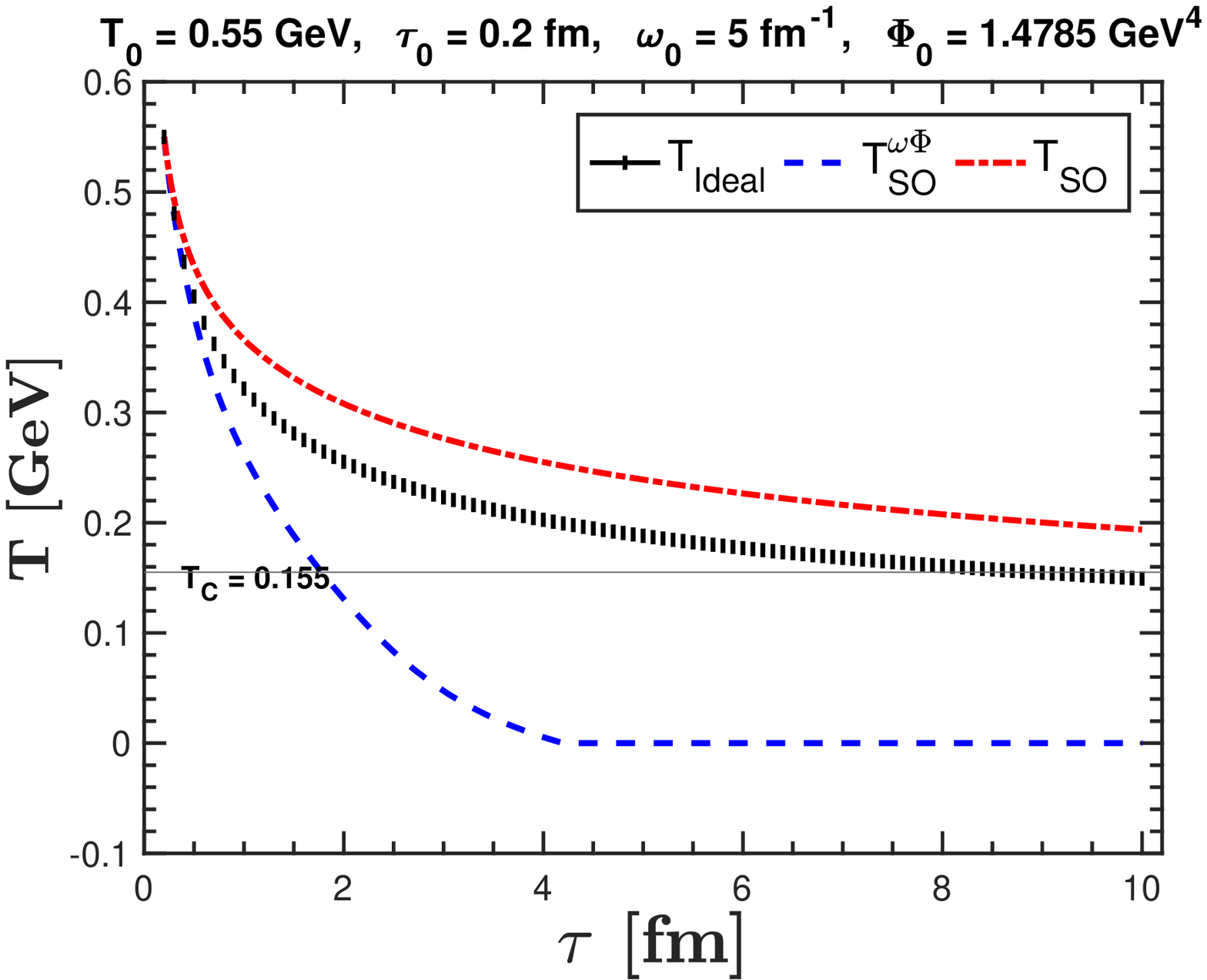}
\includegraphics[scale = 0.32]{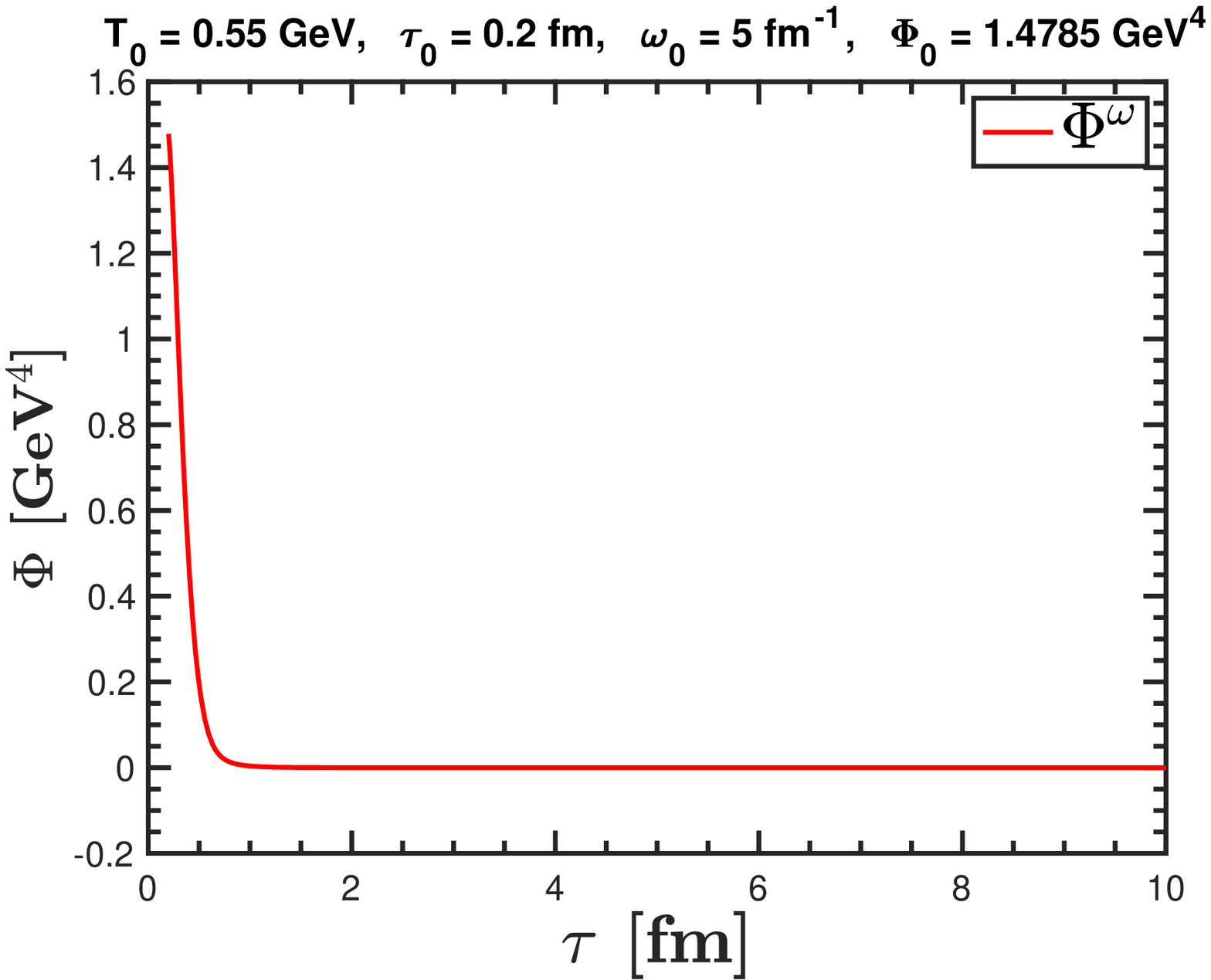}
\includegraphics[scale = 0.32]{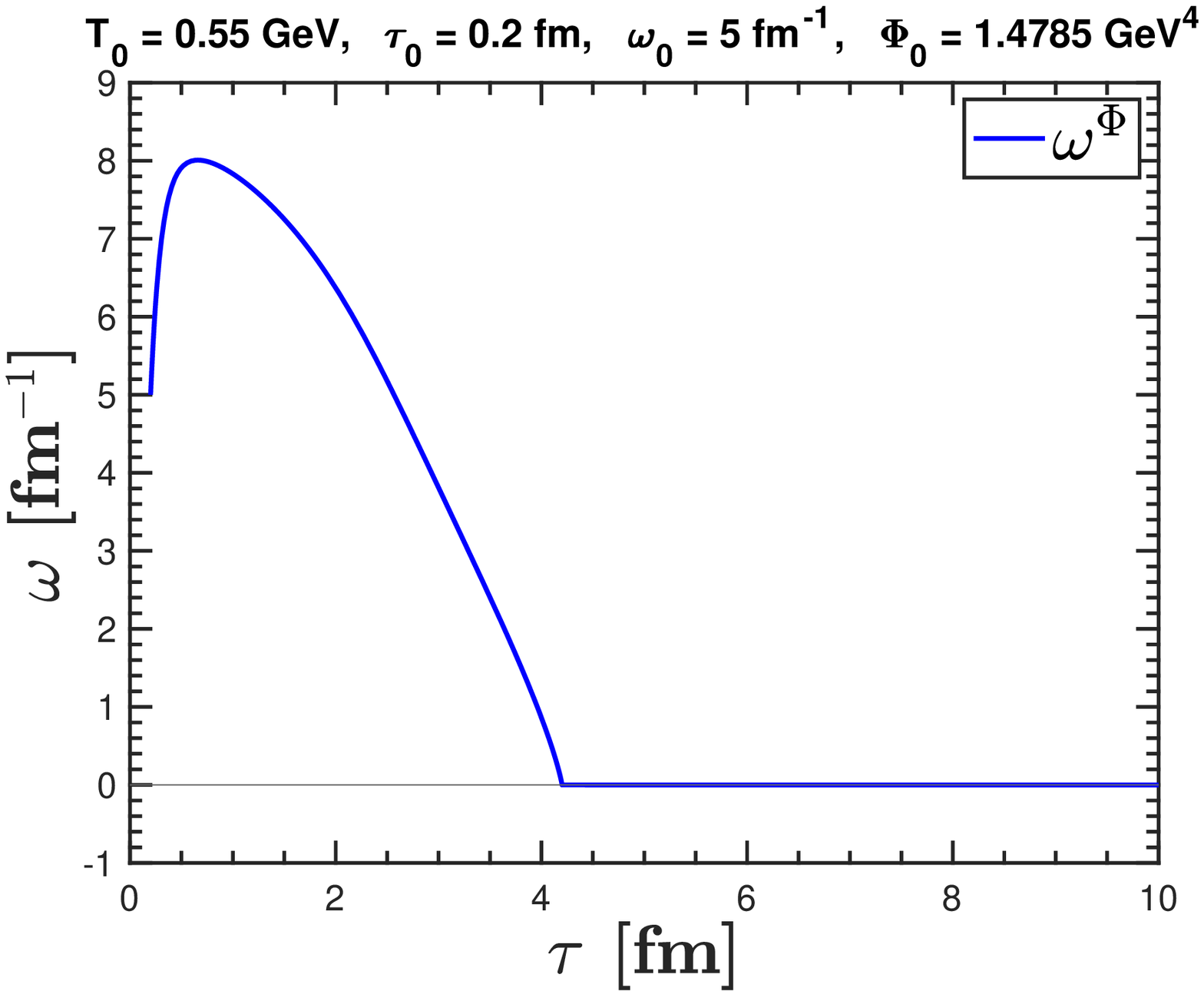}
\caption{(Color Online) {\bf Left to Right:} Temperature (T), viscous term ($\Phi$) and vorticity ($\omega$) are plotted, respectively, against time $\tau$ with the initial conditions: {\bf T = 0.55 GeV, $\tau_{0}$ = 0.2 fm, $\omega_{0}$ = 5.0 fm$^{-1}$, $\Phi$ = 1.4785 GeV$^4$}.}
\label{fig16}
\end{figure*}
\begin{figure*}[ht!]
\centering
\includegraphics[scale = 0.32]{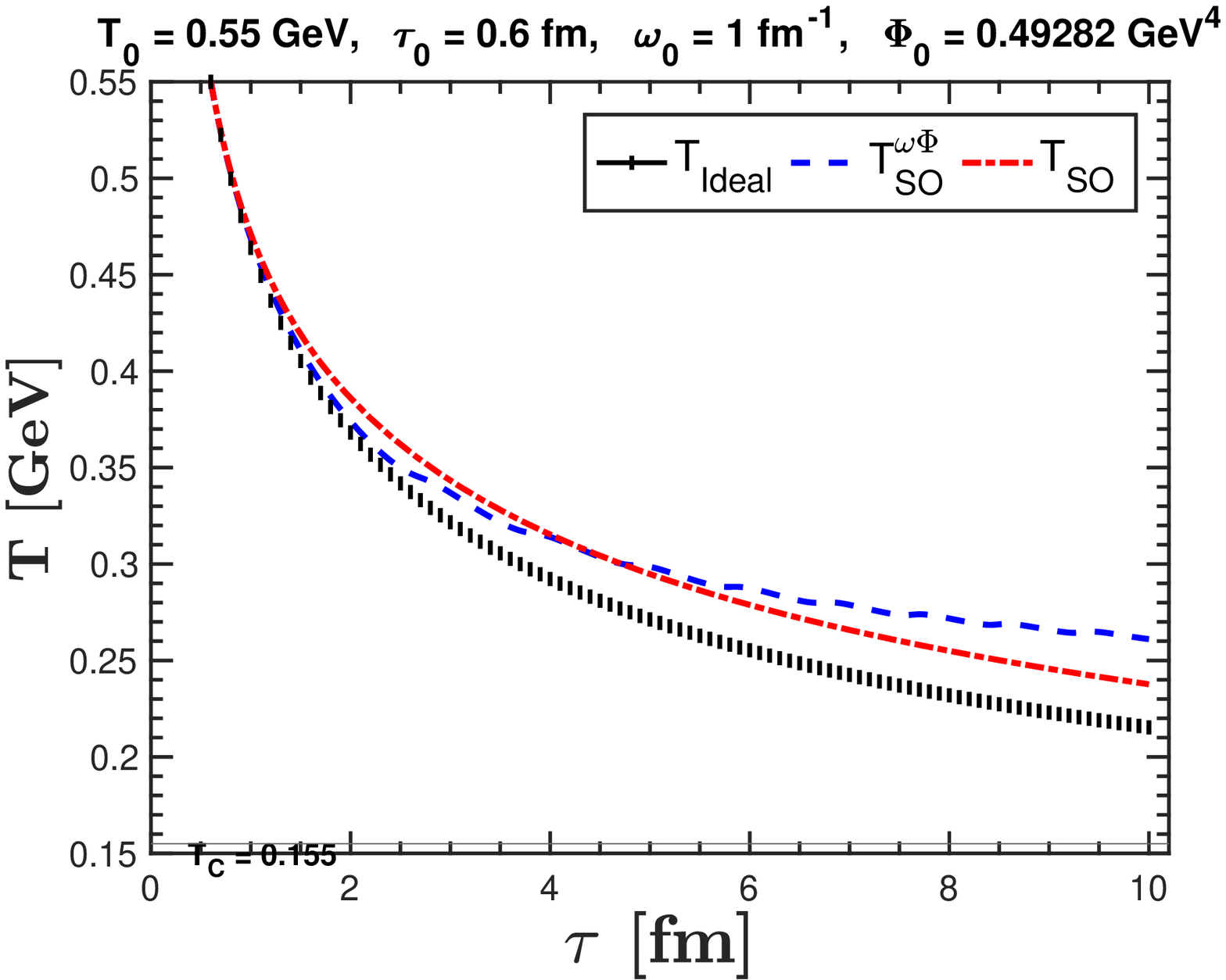}
\includegraphics[scale = 0.32]{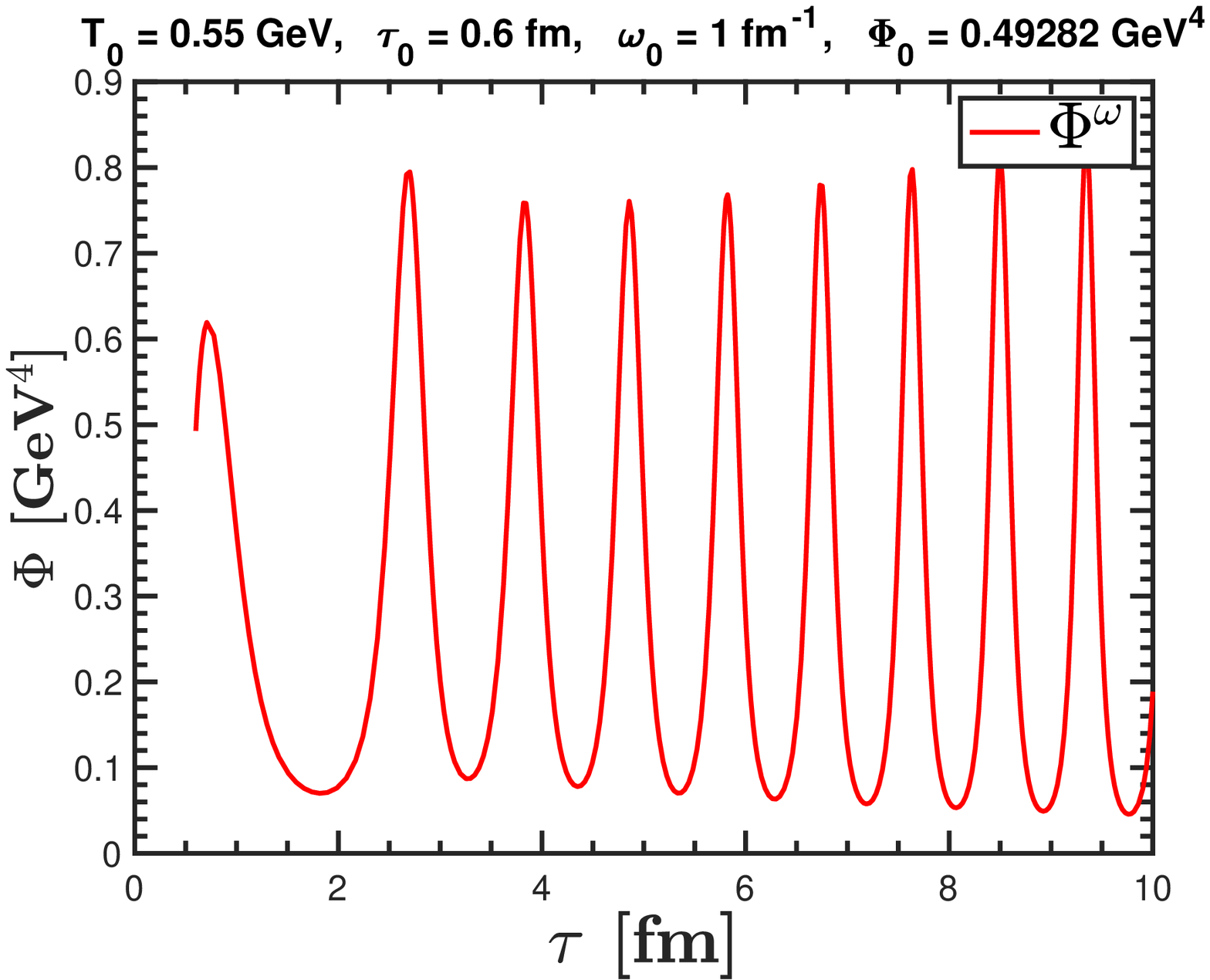}
\includegraphics[scale = 0.32]{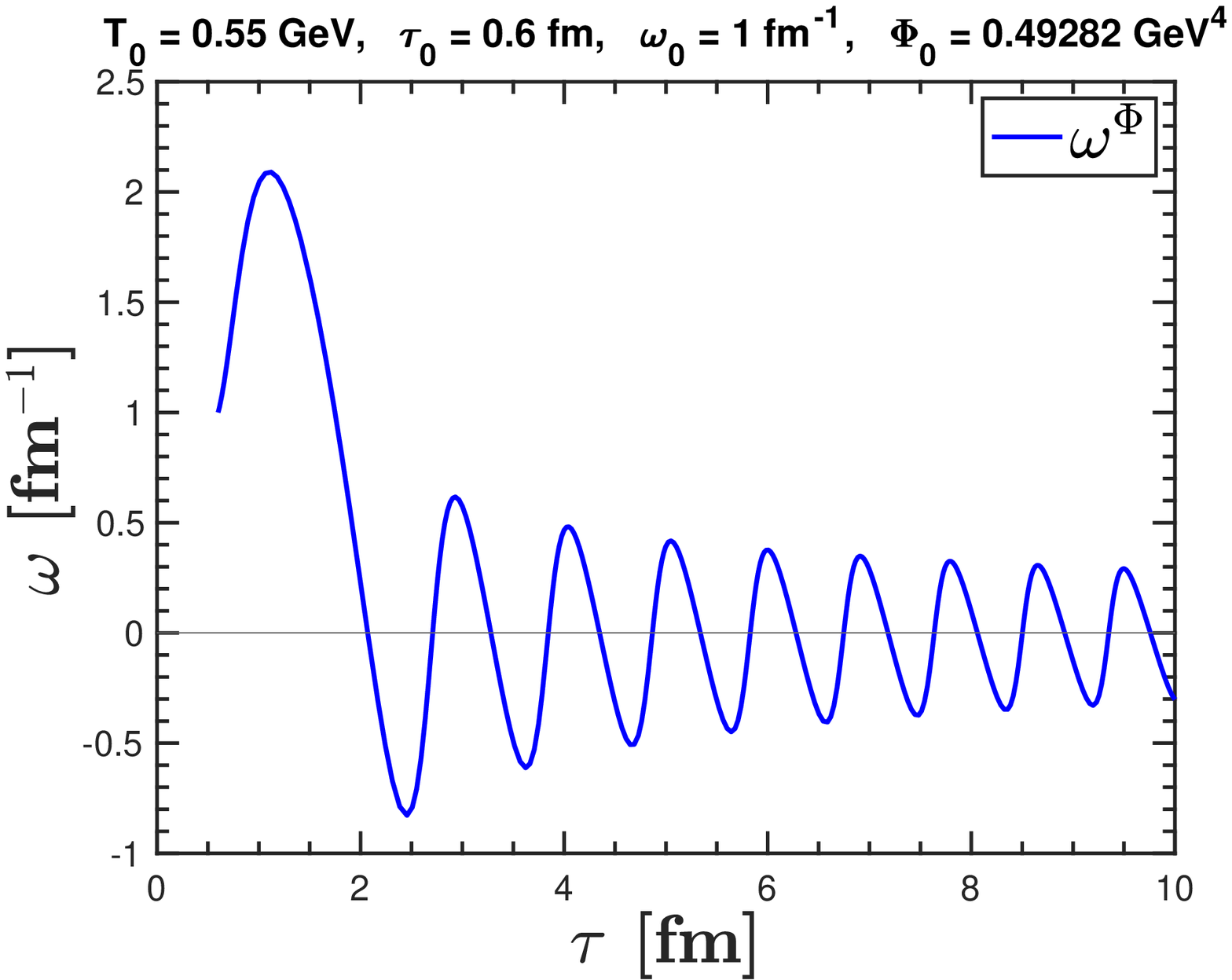}
\caption{(Color Online) {\bf Left to Right:} Temperature (T), viscous term ($\Phi$) and vorticity ($\omega$) are plotted, respectively, against time $\tau$ with the initial conditions: {\bf T = 0.55 GeV, $\tau_{0}$ = 0.6 fm, $\omega_{0}$ = 1.0 fm$^{-1}$, $\Phi$ = 0.49282 GeV$^4$}.}
\label{fig17}
\end{figure*}
\begin{figure*}[ht!]
\centering
\includegraphics[scale = 0.32]{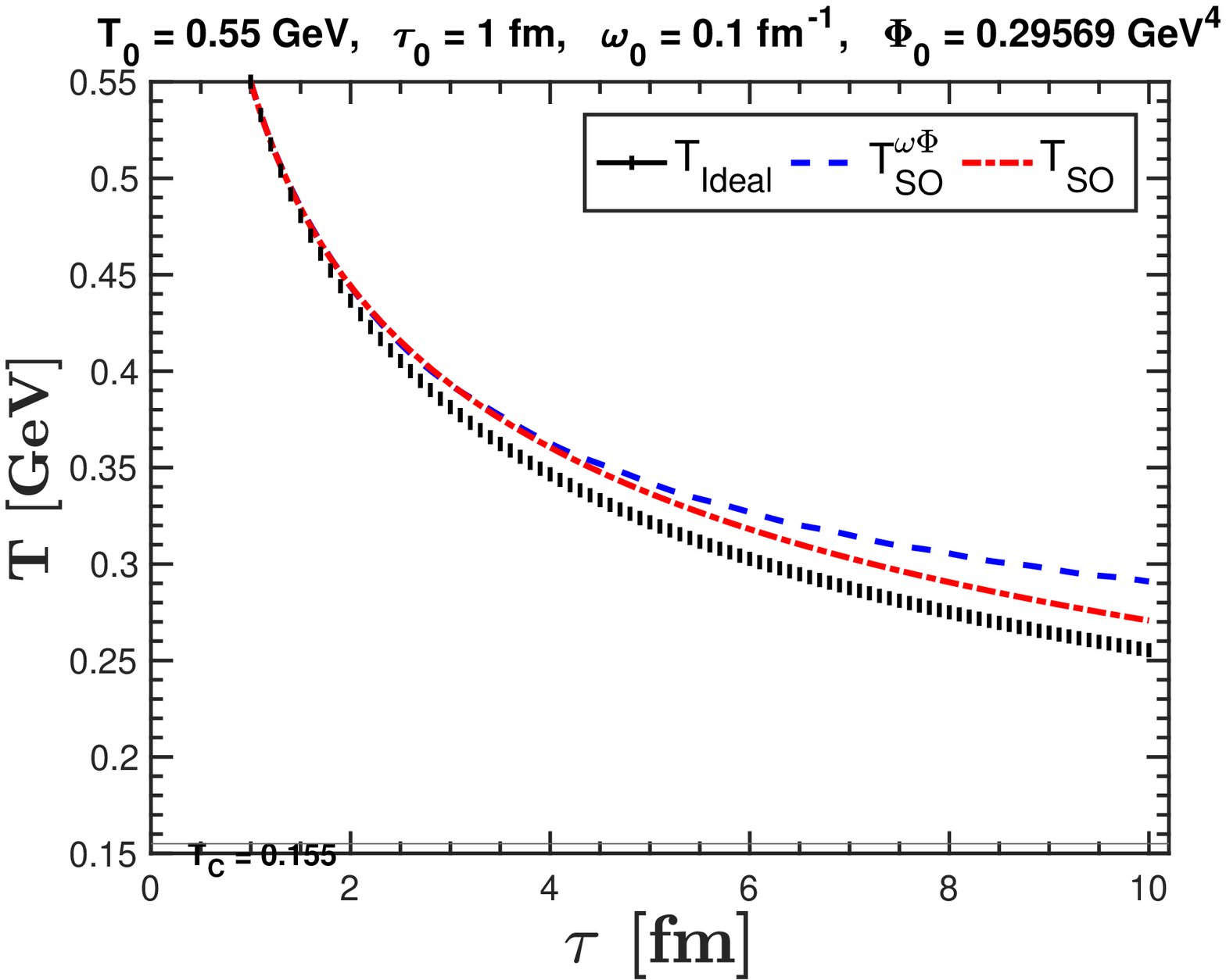}
\includegraphics[scale = 0.32]{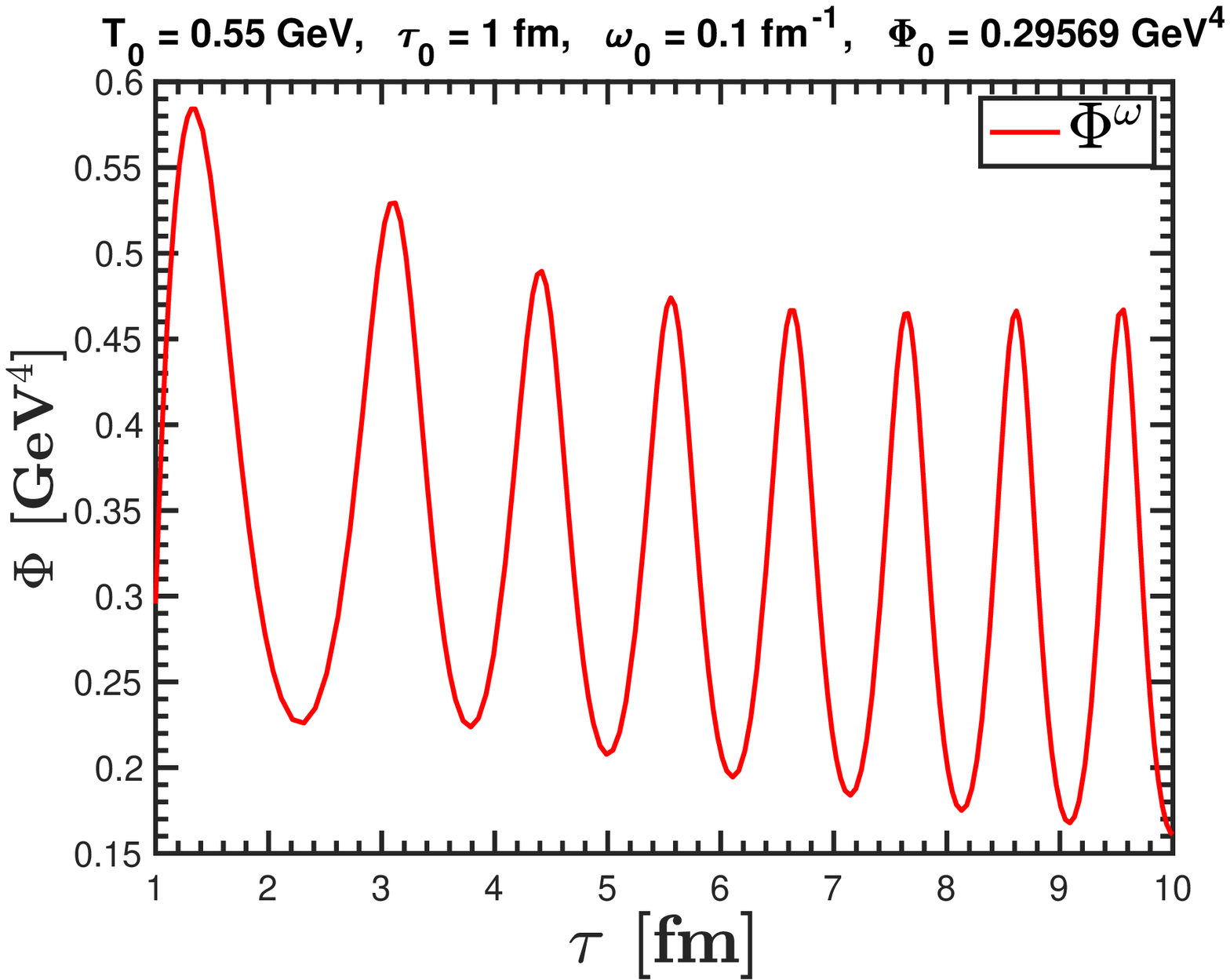}
\includegraphics[scale = 0.32]{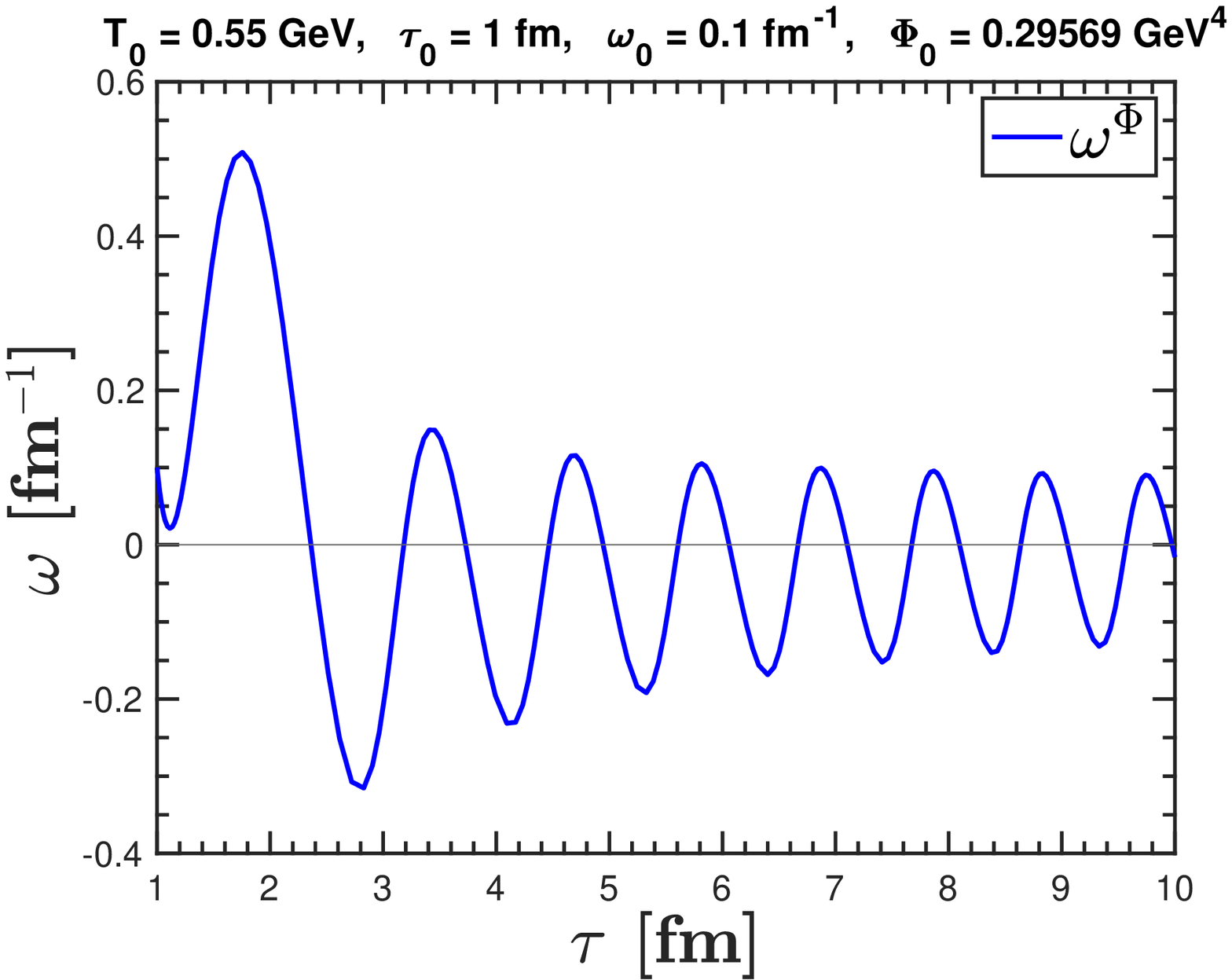}
\caption{(Color Online) {\bf Left to Right:} Temperature (T), viscous term ($\Phi$) and vorticity ($\omega$) are plotted, respectively, against time $\tau$ with the initial conditions: {\bf T = 0.55 GeV, $\tau_{0}$ = 1.0 fm, $\omega_{0}$ = 0.1 fm$^{-1}$, $\Phi$ = 0.29569 GeV$^4$}.}
\label{fig18}
\end{figure*}


\subsection*{Case-III: Direct coupling of $\omega$ with $\Phi$}
In earlier cases, $\omega$ was not directly contributing in the viscous term as the last term of Eq.~(\ref{eq26}) was taken as $\frac{\omega \Phi}{\gamma T\tau} = 0$. 
Similar to the previous case, viscosity induces a rotational motion in the fluid. In the same way, rotating fluid induces an additional viscosity in the medium due to the velocity gradient between rotating fluid cells. This coupling between $\omega$ and $\Phi$ plays a complementary role in the medium evolution. The temperature variation of  $\omega-\Phi$ coupling is presented by T$_{SO}^{\omega\Phi}$ which corresponds to the solution of the coupled rate equations; Eq.~(\ref{eq23}), Eq.~(\ref{eq24}) and Eq.~(\ref{eq26}). 
The rate of change in $\Phi$ due to its direct coupling with $\omega$ is shown by $\Phi^{\omega}$. \\

The results for a fixed initial temperature, $T_0=0.350$ GeV, are presented in Fig.\ref{fig13} to Fig.\ref{fig15}. In Fig.\ref{fig13}, it can be observed that a higher value of $\omega_{0}$ causes $\Phi$ to decrease to zero, resulting in T$_{SO}^{\omega\Phi}$ cooling being equivalent to T$_\text{Ideal}$. Conversely, a large negative value of $\omega$ leads to a sharp increase in $\Phi$, causing an abrupt change in T$_{SO}^{\omega\Phi}$ around $\tau = 2$ fm. This sudden rise in $\Phi$ changes the direction of $\omega$. On the whole, $+\omega$ decreases the $\Phi$ and $-\omega$ increases it. Similarly, Non-zero $\Phi$ generates 
the vorticity in the opposite direction of the existing vorticity. This cyclic process produces oscillations in $\omega$ and $\Phi$ cooling, as observed in Fig.~\ref{fig13}. Consequently, T$_{SO}^{\omega\Phi}$ cooling becomes very slow, resembling a damped step function over time. For diluted initial conditions, as depicted in Fig.~\ref{fig14}, T$_{SO}^{\omega\Phi}$ does not exhibit any abrupt changes in cooling. However, the cooling process becomes very slow in this case due to the increasing oscillation of $\Phi$ over time, caused by the relatively small initial vorticity ($\omega_{0}=1.0$ fm$^{-1}$) and viscosity ($\Phi_{0}=0.13334$ GeV$^4$).  The low viscosity accelerates the evolution, generating significant vorticity in the opposite direction, which increases $-\omega$. Due to the small $\omega_{0}$, $\Phi$ is not completely dissipated and instead adds up to the viscosity induced by $-\omega$. Consequently, the peak of $\Phi$ increases with each oscillation, establishing a self-sustaining system that does not dissipate over time. Fig.~\ref{fig15} follows the same trend as Fig.~\ref{fig14}, with the magnitude differences reflecting the use of different initial conditions.\\

We adopt the same initial conditions for $\tau_0$ and $\omega_{0}$ at a higher initial temperature T$_0$ = 0.550 GeV. Fig.~\ref{fig16} demonstrates that at the high initial temperature, vorticity and viscosity coupling ($\omega-\Phi$) enable fluid rotation in one direction, leading to a sudden drop in $\Phi$. Consequently, all these systems cool down at a faster rate than the ideal case, and vorticity also diminishes over time. This behavior is reflected in T$_{SO}^{\omega\Phi}$ as shown in Fig.~\ref{fig16}. Furthermore, when $\Phi_{0}$ and $\omega_{0}$ are small, the $\omega-\Phi$ coupling induces oscillations in vorticity and $\Phi$ over time. At high initial temperatures, the damping of $\omega$ and $\Phi$ is more pronounced compared to 
the results displayed in Fig.~\ref{fig14}, resulting in smaller oscillation amplitudes, as depicted in Fig.~\ref{fig17}. Consequently, T$_{SO}^{\omega\Phi}$ dissipates faster in Fig.~\ref{fig17} compared to the case shown in Fig.~\ref{fig14}. However, T$_{SO}^{\omega\Phi}$ cooling remains slow and oscillatory compared to its cooling rate depicted in Fig.~\ref{fig16}. The slight oscillation in T$_{SO}^{\omega\Phi}$ in Fig.~\ref{fig17} arises from finite $\omega$ and $\Phi$ oscillations. If we further decrease $\omega_0$ at high $\tau_0$, the oscillation in T$_{SO}^{\omega\Phi}$ disappears. In this scenario, we observe an opposite damped shift in $\omega$ and $\Phi$ oscillations, as shown in Fig.~\ref{fig18}. Here, the positive $\omega$ phase increases slowly, while the negative $\omega$ phase decreases at a faster rate, resulting in damped oscillations in $\Phi$. Overall, in this case, $\omega-\Phi$ compensate each other in a way that T$_{SO}^{\omega\Phi}$ exhibits continuous and slow cooling compared to T$_{SO}$, as depicted in Fig.~\ref{fig18}.


\begin{figure*}[ht!]
\centering
\includegraphics[scale = 0.32]{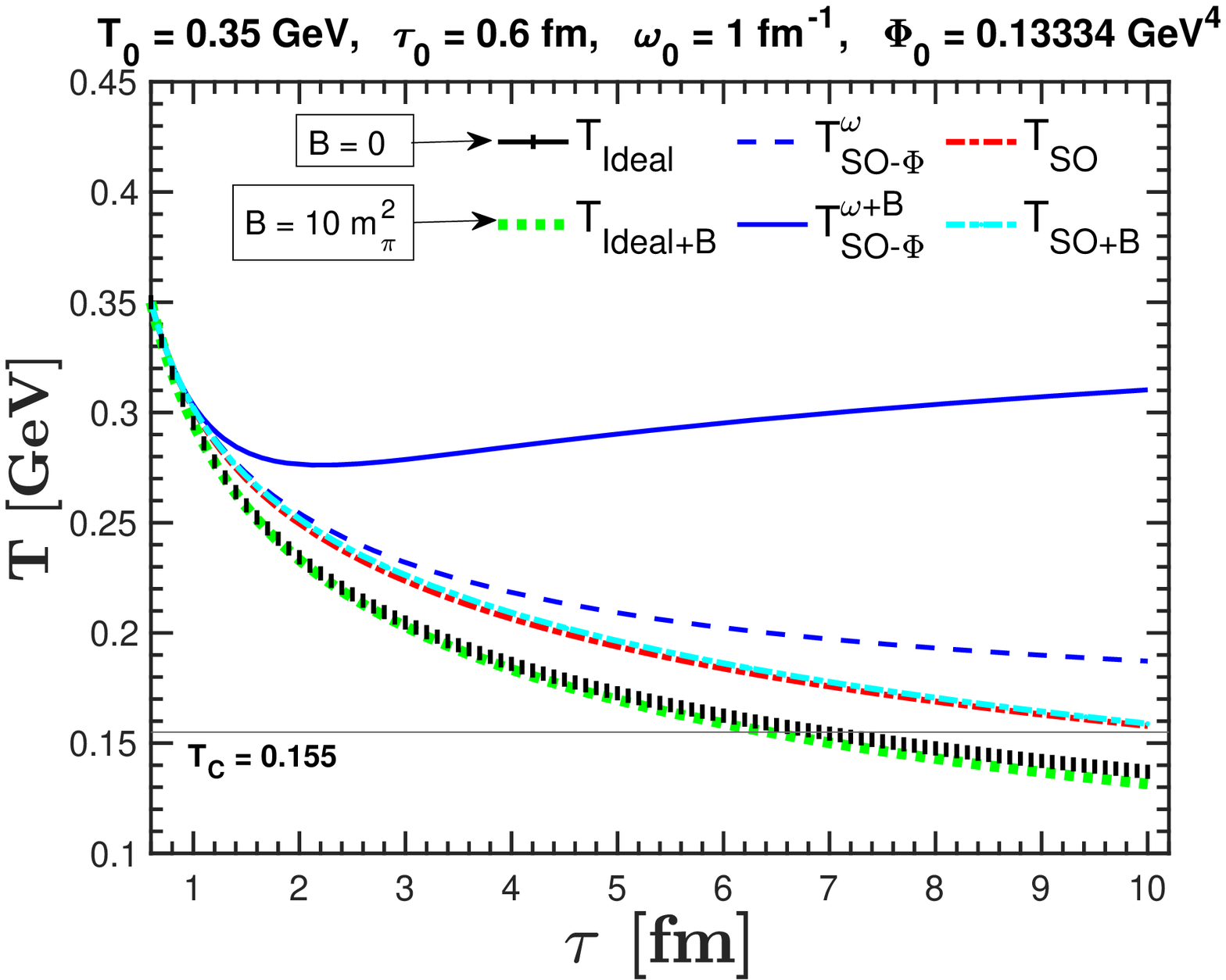}
\includegraphics[scale = 0.32]{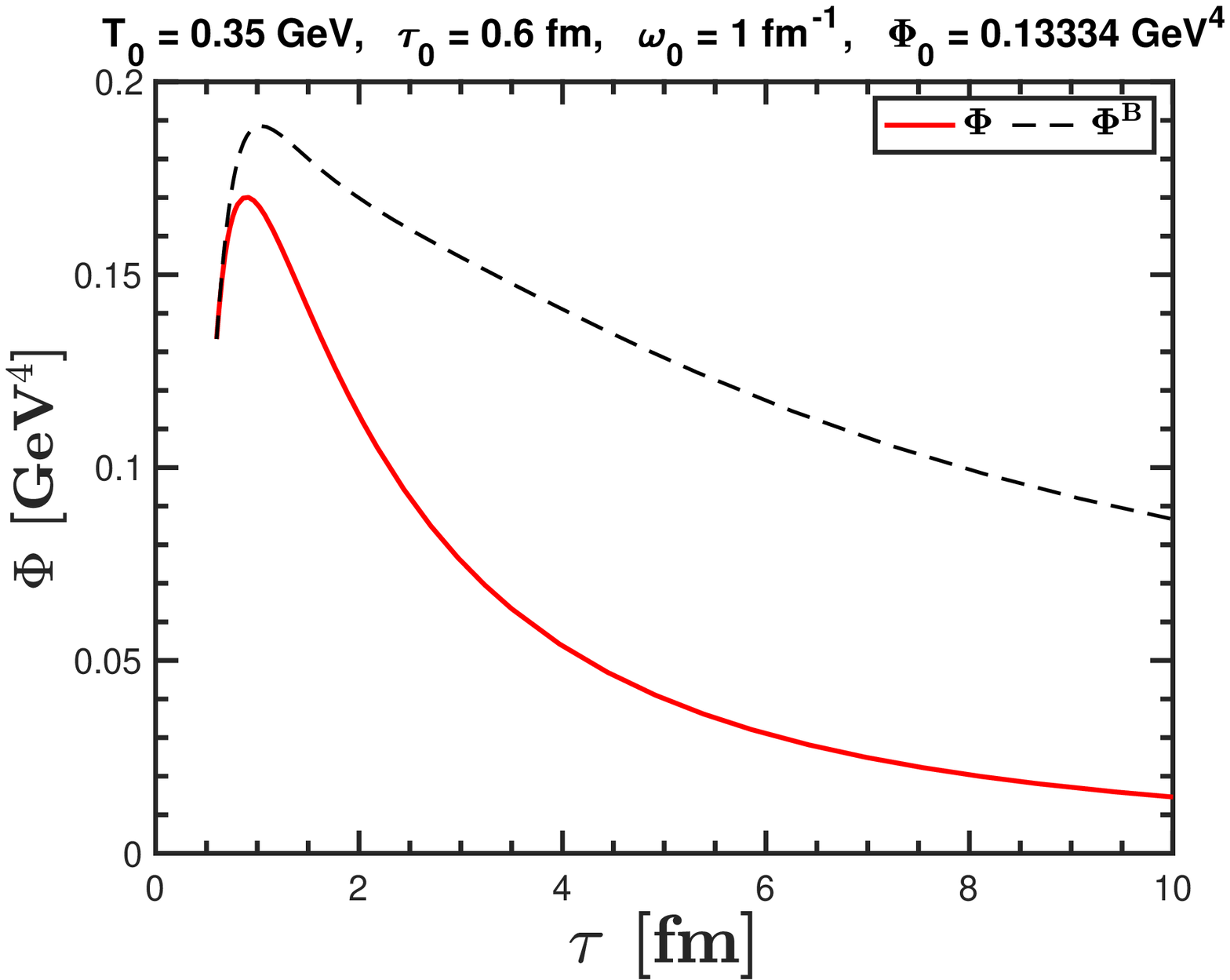}
\includegraphics[scale = 0.32]{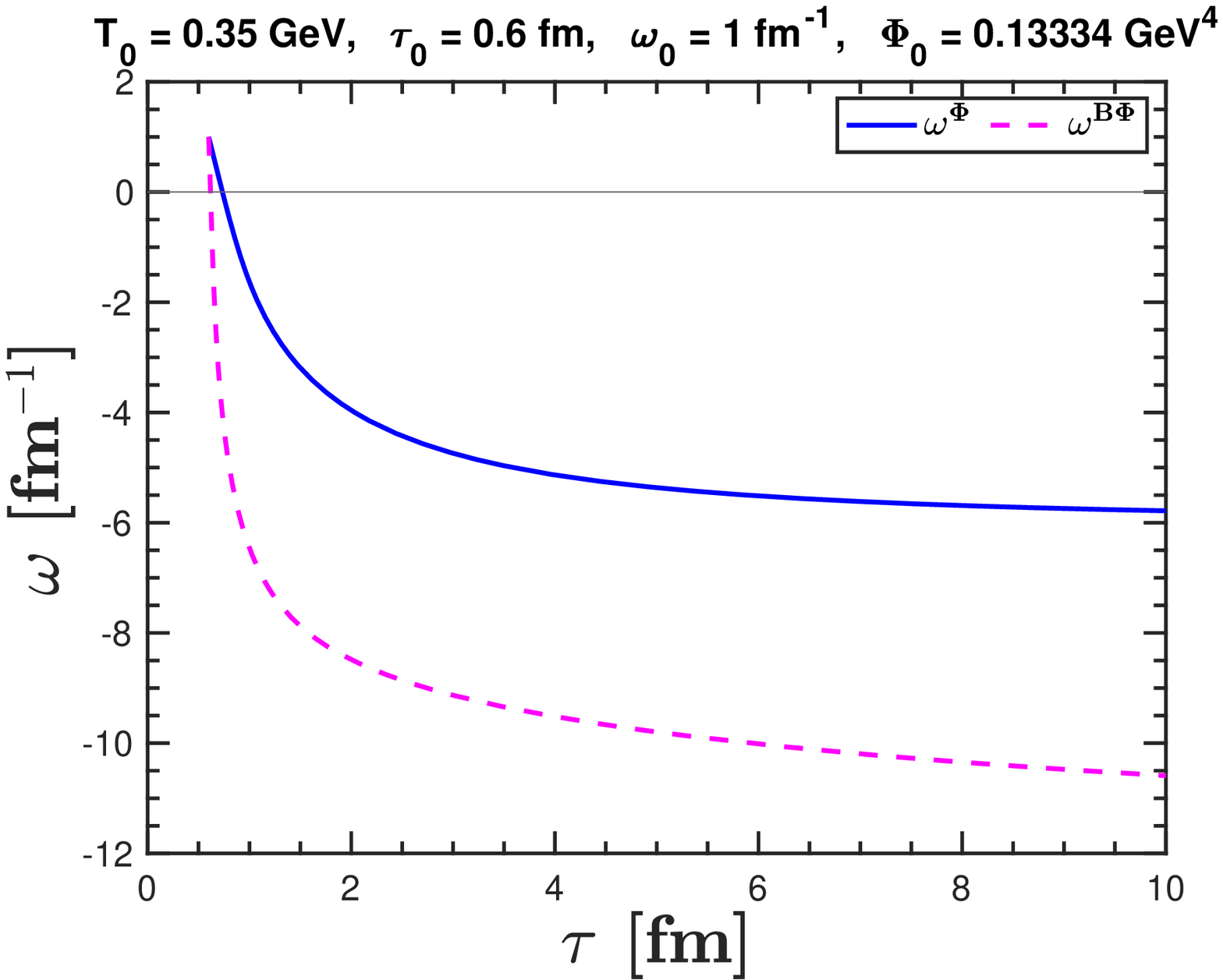}
\caption{(Color Online) {\bf Left to Right:} Temperature (T), viscous term ($\Phi$) and vorticity ($\omega$) are plotted, respectively, against time $\tau$ with the initial conditions: {\bf T = 0.35 GeV, $\tau_{0}$ = 0.6 fm, $\omega_{0}$ = 1.0 fm$^{-1}$, $\Phi$ = 0.13334  GeV$^4$}.}
\label{fig19}
\end{figure*}
\begin{figure*}[ht!]
\centering
\includegraphics[scale = 0.32]{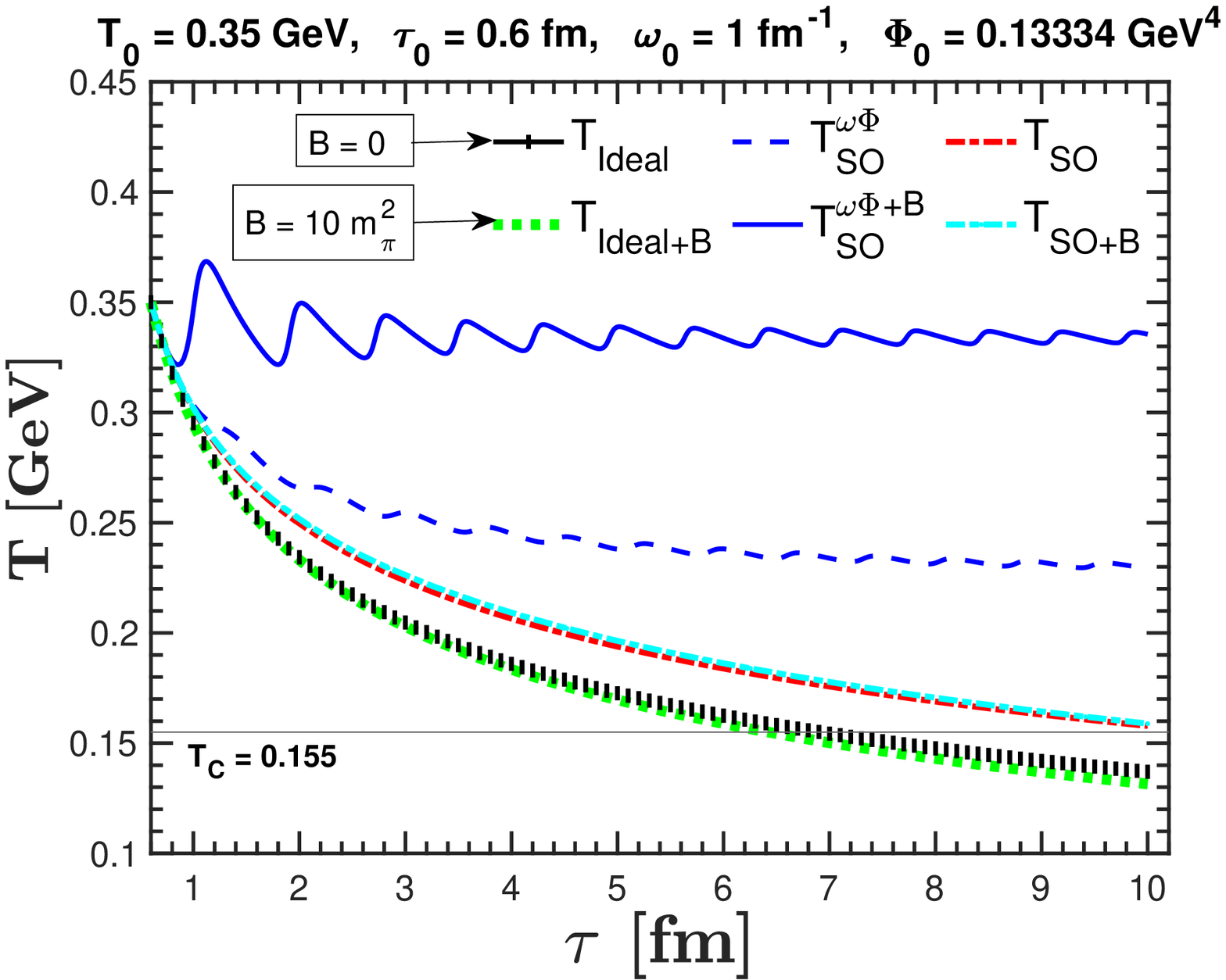}
\includegraphics[scale = 0.32]{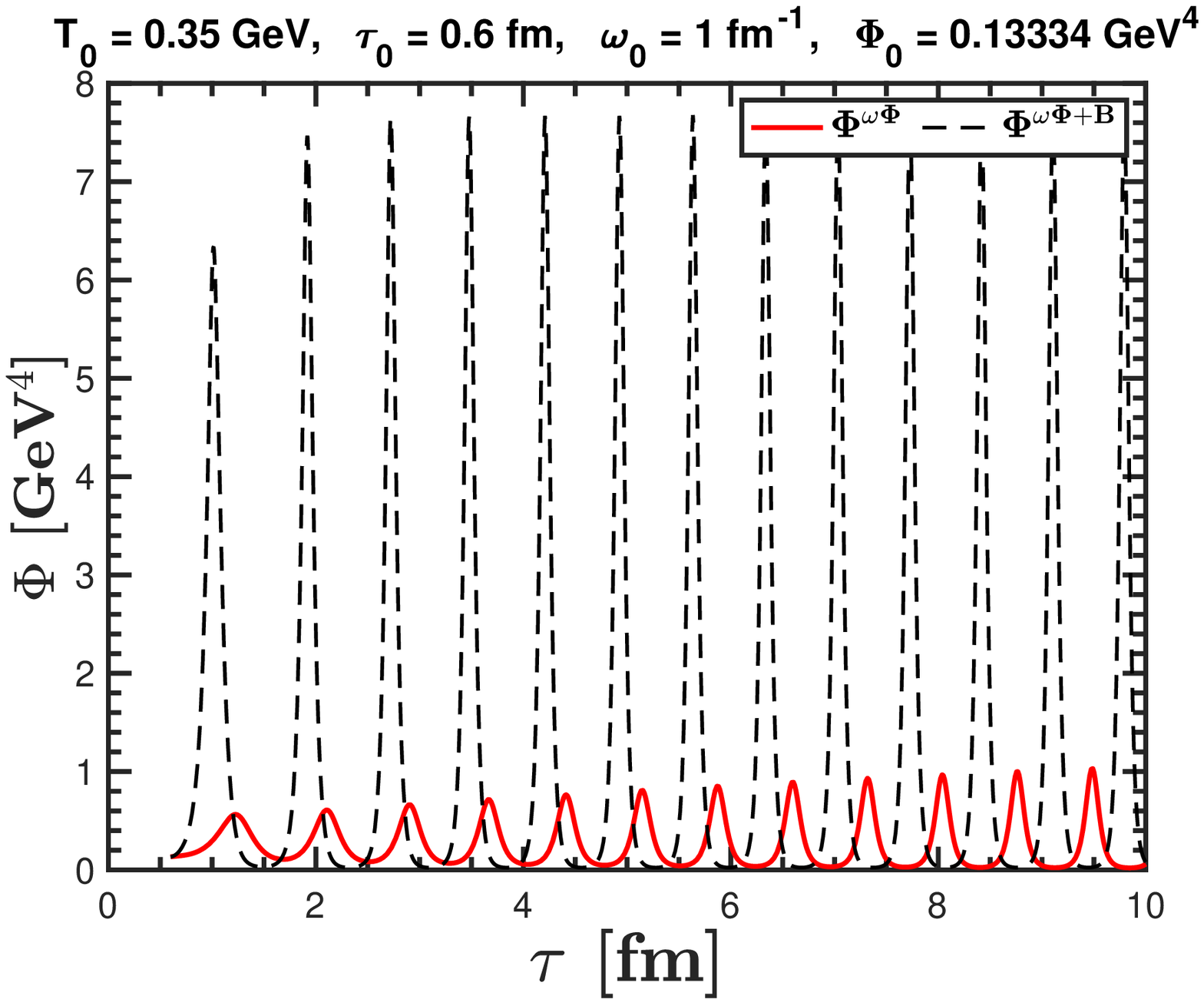}
\includegraphics[scale = 0.32]{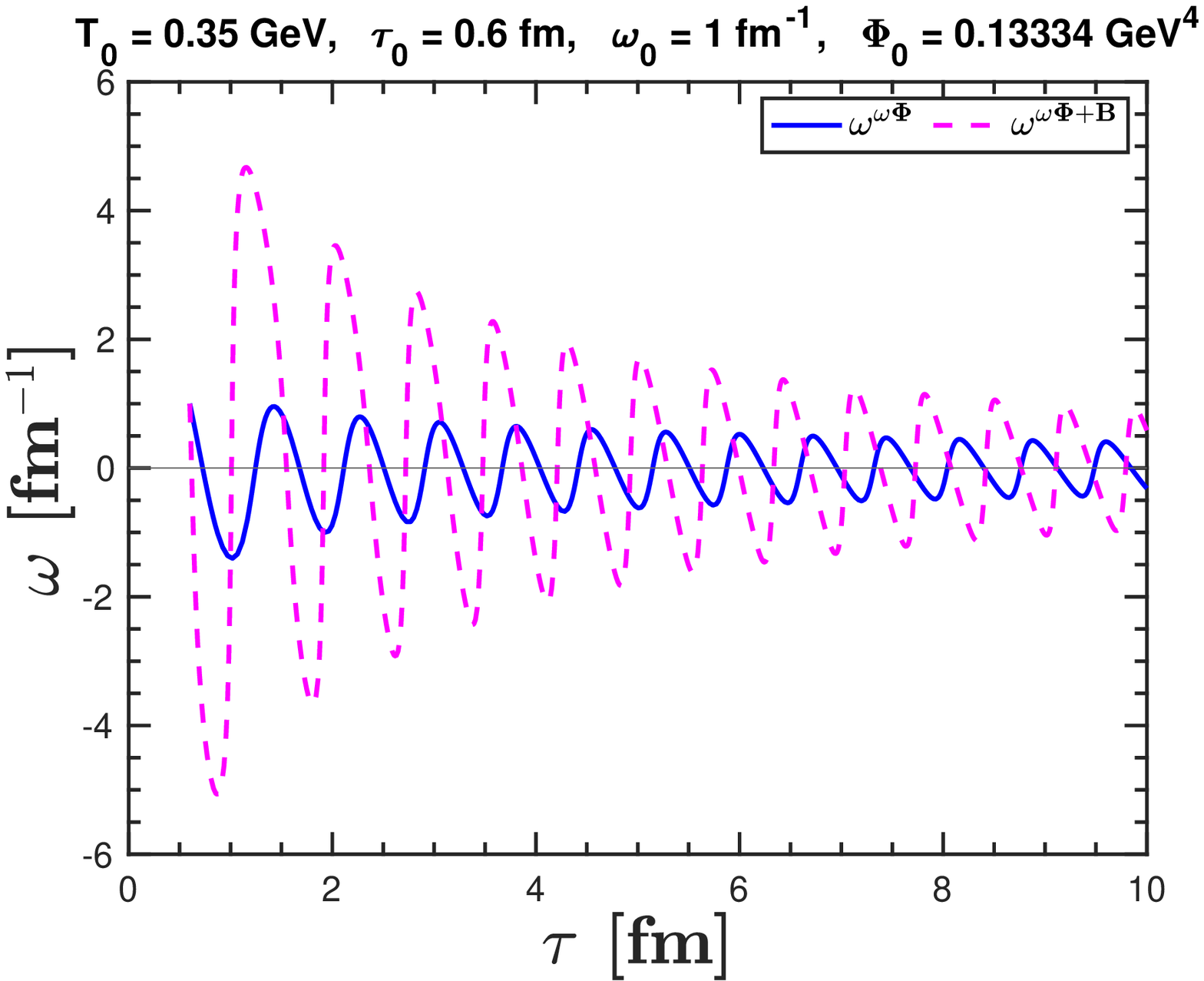}
\caption{(Color Online) {\bf Left to Right:} Temperature (T), viscous term ($\Phi$) and vorticity ($\omega$) are plotted, respectively, against time $\tau$ with the initial conditions: {\bf T = 0.35 GeV, $\tau_{0}$ = 0.6 fm, $\omega_{0}$ = 1.0 fm$^{-1}$, $\Phi$ = 0.13334  GeV$^4$}.}
\label{fig20}
\end{figure*}

\subsection*{Case IV: Change in the medium evolution due to the static magnetic field (B)}

Considering an external static magnetic field (B)  
along with vorticity and viscosity, changes the hydrodynamical evolution of the medium. 
Here are a few scenarios for combining the magnetic field with non-viscous, viscous, and vorticity. 
We have considered the impact of the static magnetic field ($B\neq 0$) in the following cases:
\begin{itemize}
 \item At, $\omega = 0$, $\Phi = 0$; we get the temperature evolution for an ideal case in the presence of the static magnetic field as, 
\begin{equation}
    \frac{dT}{d\tau} = - \frac{T}{3\gamma}\bigg(1 +\frac{\chi_{m}eB^{2}}{Ts}\bigg) \partial_{\mu} u^{\mu} \nonumber 
\end{equation}
We call the solution of this equation as $T_{Ideal+B}$.
 Here $\chi_{m}$ is magnetic sucesptibility, in our calculation we have taken $\chi_{m} = 0.03$ ~\cite{Pu:2016ayh} and $eB = 10 m_{\pi}^{2}$. The net electric charge is considered taking the sum over the electric charges of $u$, $d$, and $s$ quarks to obtain the magnetic field; $eB = \sum_{f} |q_{f}| B$.

 \item Now we consider that $\omega = 0$, but medium has finite viscosity,  $\Phi \ne 0$.
 \begin{equation}
    \frac{dT}{d\tau} = \left[- \frac{T}{3\gamma}\bigg( 1 +\frac{\chi_{m}eB^{2}}{Ts}\bigg) +  \frac{\Phi T^{-3}}{12a\gamma} \right]\partial_{\mu} u^{\mu}  \nonumber 
\end{equation}
The solution of this equation is denoted as $T_{SO+B}$
 
\item Next, we assume that medium is viscous and has vorticity as well, s.t. $\omega \ne 0$, $\Phi \ne 0$. However, in this case, $\Phi$ does not 
arise due to vorticity, while vorticity gets induced due to viscosity. So the cooling respective to the mentioned condition is defined in 
Eq.~(\ref{eq31}) is represented here as; T$_{SO+\Phi}^{\omega+B}$, and corresponding vorticity and viscosity dissipation with time are depicted 
by $\omega^{B\Phi}$ and $\Phi^{B}$, respectively.

\item Further, we consider the case when vorticity and viscosity play a complementary relation, 
i.e., $\omega(\Phi)$ and $\Phi(\omega)$. The change of viscosity and vorticity dissipation under $\omega-\Phi$ 
coupling in the presence of magnetic field (B) is denoted as T$_{SO}^{\omega\Phi+B}$, 
$\Phi^{\omega\Phi+B}$ and $\omega^{\omega\Phi+B}$, respectively.

\end{itemize}

Fig.~\ref{fig19} shows that inclusion of static magnetic field along with vorticity and viscosity does not let the medium cool down. As seen in the T vs. $\tau$ plot, 
the solid blue line initially decreases and slowly increases with time. While the magnetic field for the ideal and viscous case slightly increases 
the cooling. It can be interpreted in this way that the magnetic field separates the +ve and -ve charge particles in opposite directions 
to create charge polarization 
in the medium. This charge polarization gives a boost to the cooling. Therefore inclusion of a magnetic field makes cooling faster in the absence of vorticity. 
The vorticity or rotation in the medium disturbs the charge polarization while the magnetic field works to retain it. In this process, the magnetic field drastically 
increases the vorticity in the opposite direction, as depicted in the $\omega$ vs. $\tau$ plot in Fig.~\ref{fig19}. Because of this, the viscous term $\Phi$ also gets 
altered, and its dissipation rate gets reduced, as shown by the dashed black line in $\Phi$ vs. $\tau$ plot in Fig.~\ref{fig19}.\\

Fig.~\ref{fig20} shows the changes brought in the medium evolution due to the $\omega-\Phi$ coupling in the presence of the static magnetic field. 
The oscillations in dissipation rates follow a similar explanation as the previous results of $\omega-\Phi$ coupling where $B=0$. Here, the non-zero 
static magnetic field enhances the amplitude of the damped oscillatory solutions for $\omega$, which can be witnessed in Fig.~\ref{fig20}. 
The magnetic field along with $\omega-\Phi$ coupling also largely enhances the fluctuation in viscosity, $\Phi^{\omega\Phi+B}$, dissipation rate as 
compared with $B = 0$ case, i.e. $\Phi^{\omega\Phi}$. This $\omega-\Phi$ coupling, along with $B$, produces an additional heat which raises the temperature 
$T>T_{0}$ when vorticity is maximum ($\omega \approx -5.0$ fm$^{-1}$) as depicted in Fig.~\ref{fig20}. It also shows that the cooling of the medium 
becomes stagnant if $\omega-\Phi$ coupling occurs in the presence of the static magnetic field. Fig.~\ref{fig19} and Fig.~\ref{fig20} suggest that 
a non-zero static magnetic field induces a shift in the temperature (T), viscosity ($\Phi$) and vorticity ($\omega$) dissipation rate if the medium has finite initial vorticity.   

\section{Summary}
\label{sum}

Within the ambit of second-order causal dissipative hydrodynamics, we have 
investigated the impact of vorticity on the evolution of  viscous QGP
 and compared the results with the evolution of an ideal QGP. 
We have found that the medium evolution is very sensitive to the initial temperature ($T$), the viscous term 
($\Phi$) as well as on vorticity ($\omega$). 
These initial conditions significantly modify the medium evolution and QGP lifetime. Evolution becomes more 
complex with the coupling of vorticity and viscosity. 
Such a complementary relation between $\omega$ and $\Phi$ generates oscillations or fluctuations 
in the medium dissipation.
In addition, the inclusion of a static magnetic field vastly reduces the cooling rate.
We have adopted a simplified approach to a complex system with some conjectured velocity profiles 
to describe the medium created in ultra-relativistic collisions. However, considering a coupled system of vorticity, 
viscosity, and time-dependent magnetic field along with its associated electric field in a (3+1)D hydrodynamics is a more realistic picture of QGP medium evolution. 
The evolution of QGP incorporates the interplay between various physical phenomena, which makes its cooling very complex.

\section*{Acknowledgement}
Raghunath Sahoo and Captain R. Singh acknowledge the financial support under DAE-BRNS, the Government of India, Project No. 58/14/29/2019-BRNS. Bhagyarathi Sahoo acknowledges the Council of Scientific and Industrial Research, Govt. of India, for financial support. The authors acknowledge the Tier-3 computing facility in the experimental high-energy physics laboratory of IIT Indore, supported by the ALICE project.


\appendix

\section{Calculation of $u^{\mu}\partial_{\mu}$}
\label{appe1}
The convective derivative  $u^{\mu}\partial_{\mu}$ is obtained as;
\begin{equation}   
 u^{\mu}\partial_{\mu} = u^{0}\partial_{0} + u^{x}\partial_{x} + u^{y}\partial_{y} + u^{z}\partial_{z}
\end{equation}
Using the four velocity $u^{\mu} =\gamma(1, v_{x}, 0, v_{z} )$, we have
\begin{equation}   
 u^{\mu}\partial_{\mu} = \gamma \bigg( \frac{\partial}{\partial t} + v_{x} \frac{\partial}{\partial x}+ v_{z} \frac{\partial}{\partial z} \bigg)
\end{equation}

In the present scenario, we approximate a constant temperature and viscosity along the x and z directions. In mid-rapidity region (in the limit $\eta \rightarrow0$) the above expression becomes, 

\begin{equation}   
 u^{\mu}\partial_{\mu} = \gamma \frac{\partial}{\partial \tau}
\end{equation}

\section{Calculation of $\partial_{\mu}u^{\mu}$}
\label{appe2}

The expansion rate $\partial_{\mu}u^{\mu}$ explore the velocity grandient, as given by;
\begin{equation}   
 \partial_{\mu}u^{\mu} = \partial_{0}u^{0} + \partial_{x}u^{x} + \partial_{y}u^{y} + \partial_{z}u^{z}
\end{equation}

With the help of four velocity $u^{\mu}$, we get;

\begin{equation}   
 \partial_{\mu}u^{\mu} = \bigg( \frac{\partial \gamma}{\partial t} + \frac{\partial  ( \gamma v_{x} )}{\partial x}+ \frac{\partial (\gamma v_{z})}{\partial z} \bigg)
\end{equation}

Solving the above expression by using the velocity profile mentioned in Eq.~(\ref{eq15}) and  Eq.~(\ref{eq16}), we obtain; 

\begin{equation}   
 \partial_{\mu}u^{\mu} = \bigg[  \frac {\gamma}{\tau} + \frac{\gamma^{3}}{2} \bigg( \frac{\omega^2x^2}{2\tau} - \frac{\omega xz}{\tau^2} \bigg) \bigg]
\end{equation}
In the limit $\eta \rightarrow0$, 

\begin{equation}   
 \partial_{\mu}u^{\mu} = \frac{\gamma}{\tau} \bigg( 1 + \frac{\gamma^{2}\omega^2x^2}{4} 	 \bigg)
\end{equation}

For irrotational fluid ($\omega = 0$) the divergence of velocity profile  $\partial_{\mu}u^{\mu}$ is equal to $\frac{\gamma}{\tau}$.

\section{Calculation of $\bigtriangledown^{<0}u^{0>} - \bigtriangledown^{<z}u^{z>} $ }
\label{appe3}

We know, 

\begin{equation}
\label{Ceq11}
 \bigtriangledown^{<\mu}u^{\nu>} \equiv 2\bigtriangledown^{(\mu}u^{\nu)} - \frac{2}{3}\Delta^{\mu\nu}\bigtriangledown^{\alpha} u_{\alpha} 
\end{equation} 
where 
\begin{equation}
    \bigtriangledown^{(\mu}u^{\nu)}  = \frac{1}{2} \left( \bigtriangledown^{\mu}u^{\nu} + \bigtriangledown^{\nu}u^{\mu} \right)    
    \label{Ceq12}
\end{equation}
and \hspace{2mm}
$\Delta^{\mu\nu} = g^{\mu\nu} -u^{\mu}u^{\nu},\;\;\;\;\;\; \bigtriangledown^{\mu} = \partial^{\mu} -u^{\mu} D$.\\

Now $\bigtriangledown^{<0}u^{0>}$ and $\bigtriangledown^{<z}u^{z>}$ can be defined as, 

\begin{equation}
    \bigtriangledown^{<0}u^{0>} = 2\bigtriangledown^{0}u^{0} - \frac{2}{3}(g^{00} - u^0u^0)\bigtriangledown^{\alpha} u_{\alpha} 
    \label{Ceq13}
\end{equation}
\begin{equation}
  \bigtriangledown^{<z}u^{z>}   = 2\bigtriangledown^{z}u^{z} - \frac{2}{3}(g^{zz} - u^zu^z)\bigtriangledown^{\alpha} u_{\alpha} 
    \label{Ceq14}
\end{equation}

Taking the difference of Eq.~(\ref{Ceq13}) and Eq.~(\ref{Ceq14}), we have

\begin{align}
  \bigtriangledown&^{<0}u^{0>} - \bigtriangledown^{<z}u^{z>}\nonumber \\
  &= 2(\bigtriangledown^{0}u^{0} - \bigtriangledown^{z}u^{z}) - \frac{2}{3}\left(2+\gamma^2(v_{z}^2 -1)\right)\bigtriangledown^{\alpha} u_{\alpha} 
    \label{Ceq15}
\end{align}

Solving the Eq.~(\ref{Ceq15}) in the limit $\eta \rightarrow0$, finally we get;
\begin{align}
\label{Ceq16}
\nonumber
    &\bigtriangledown^{<0}u^{0>} - \bigtriangledown^{<z}u^{z>} = \frac{2\gamma}{\tau} \bigg[  \gamma^{4} \frac{\omega x}{2}\left(\frac{\omega^3x^3}{8} +1\right) - 1\\ &- \frac{1}{3}\bigg(1+\frac{\gamma^{2}\omega^2x^2}{4} \bigg)^2 -\frac{1}{6}(1-\gamma^{2}) \bigg( \frac{\gamma^2\omega^2x^2}{2} +2 \bigg) \bigg]
\end{align}

\section{Coupling vorticity with viscosity}
\label{appe4}

Let's solve $\lambda \pi^{(\mu}_{\alpha}\omega^{\nu)\alpha}$
\begin{equation}
    \nonumber
    \pi^{(\mu}_{\alpha}\omega^{\nu)\alpha}
= \frac{1}{2}\left(\pi_{\alpha}^{\mu}\omega^{\nu \alpha} + \pi_{\alpha}^{\nu}\omega^{\mu \alpha}\right)
\end{equation}
 By using the relation 
\begin{equation}
 \nonumber
    A^{(\mu}B^{\nu)}
= \frac{1}{2} \left(A^{\mu}B^{\nu} + A^{\nu}B^{\mu} \right)
\end{equation}
With the help of metric tensor, we get;
\begin{equation}
\nonumber
    \pi^{(\mu}_{\alpha}\omega^{\nu)\alpha} = \frac{1}{2}\left(g_{\alpha \rho}\pi^{\rho \mu} \omega^{\nu \alpha} + g_{\alpha \sigma}\pi^{\sigma \nu} \omega^{\mu \alpha}\right)        
\end{equation}

From Eq.~(\ref{eq20}), we have 
\begin{align}
\label{Deq29}
\nonumber
    D\pi^{\mu\nu} = - \frac{1}{\tau_{\pi}}\pi^{\mu\nu}- \frac{1}{2\beta_{2}}\pi^{\mu\nu}\left[\beta_{2}\theta + TD\left(\frac{\beta_{2}}{T}\right)\right] \nonumber\\
    +\frac{1}{\beta_{2}} \bigtriangledown^{<\mu}u^{\nu>} + \lambda \pi^{(\mu}_{\alpha}\omega^{\nu)\alpha}
\end{align}

For $\mu=0$ and $\nu=0$,
\begin{align}
\label{Deq30}
\nonumber
    D\pi^{00} = - \frac{1}{\tau_{\pi}}\pi^{00}- \frac{1}{2\beta_{2}}\pi^{00}\left[\beta_{2}\theta + TD\left(\frac{\beta_{2}}{T}\right)\right] \nonumber\\
    +\frac{1}{\beta_{2}} \bigtriangledown^{<0}u^{0>} + \lambda \pi^{(0}_{\alpha}\omega^{0)\alpha}
\end{align}
For $\mu=z$ and $\nu=z$, we have
\begin{align}
\label{Deq31}
\nonumber 
    D\pi^{zz} = - \frac{1}{\tau_{\pi}}\pi^{zz}- \frac{1}{2\beta_{2}}\pi^{zz}\left[\beta_{2}\theta + TD\left(\frac{\beta_{2}}{T}\right)\right] \nonumber\\
    +\frac{1}{\beta_{2}} \bigtriangledown^{<z}u^{z>} + \lambda \pi^{(z}_{\alpha}\omega^{z)\alpha}
\end{align}

Writing $ D = \gamma \frac{d}{d\tau}$ and $\theta = \partial_{\mu}u^{\mu}$ and subtracting Eq.~(\ref{Deq31}) from Eq.~(\ref{Deq30}), we get;

\begin{align}
\label{Deq32}
    \nonumber
    \frac{d}{d\tau}\left( \pi^{00}-\pi^{zz} \right) = - \frac{1}{\gamma \tau_{\pi}} \left(\pi^{00}- \pi^{zz}\right)- \frac{1}{2\gamma \beta_{2}} \left(\pi^{00}-\pi^{zz}\right) \nonumber\\
    \left[\beta_{2} \partial_{\mu}u^{\mu}+ T \gamma \frac{d}{d \tau}\left(\frac{\beta_{2}}{T}\right)\right] +\frac{1}{\gamma \beta_{2}} \left(\bigtriangledown^{<0}u^{0>} - \bigtriangledown^{<z}u^{z>}\right) \nonumber\\
    + \frac{\lambda}{\gamma} \left( \pi^{(0}_{\alpha}\omega^{0)\alpha} - \pi^{(z}_{\alpha}\omega^{z)\alpha} \right)
\end{align}

Choosing $\omega^{0z} = -\omega^{z0} = \frac{\omega}{T}$ and writing  $\pi^{00}-\pi^{zz}$  = $\Phi$, \\
The last term of Eq.~(\ref{Deq32}) transformed as; \\
\begin{align}
 \left( \pi^{(0}_{\alpha}\omega^{0)\alpha} - \pi^{(z}_{\alpha}\omega^{z)\alpha} \right) &= \left( \pi^{00}\omega^{z0} - \pi^{zz}\omega^{0z}\right)\nonumber \\
&= -\frac{\omega}{T}\left( \pi^{00} - \pi^{zz} \right) \nonumber \\
&= -\frac{\omega \Phi}{T}
\end{align}

Using second-order transport coefficient $\lambda$ = $\frac{1}{\tau},  \tau_{\pi} = 2\eta\beta_{2}, \beta_{2} = 3/4P ,\eta = bT^{3}$,  Eq.~(\ref{eq26}) extends to the following form;

\begin{align}
\label{Deq33}
    \frac{d\Phi}{d\tau} = - \frac{2aT\Phi}{3\gamma b} - \frac{\Phi}{2\gamma}\left( \partial_{\mu}u^{\mu} - \frac{5\gamma}{T}\frac{dT}{d\tau} \right) \nonumber \\ +\frac{4aT^{4}}{3\tau\gamma}\left(\bigtriangledown^{<0}u^{0>} - \bigtriangledown^{<z}u^{z>}\right) - \frac{\omega \Phi}{\gamma T \tau}. 
\end{align}

 \end{document}